\documentclass[12pt,a4paper]{article}
\usepackage{amsmath,amssymb,color,graphicx,amsthm}
\usepackage{natbib}
\usepackage{bibentry}
\usepackage{times}
\usepackage{multirow}
\usepackage{diagbox}
\usepackage{cancel}
\usepackage{color, colortbl}
\usepackage[list=off]{caption}
\usepackage{setspace}
\usepackage[hmargin=2.2cm, vmargin=2.85cm]{geometry}
\nobibliography*

\usepackage{floatrow}
\DeclareFloatFont{tiny}{\tiny}
\usepackage{hyperref}
\hypersetup{
    colorlinks,
    linktocpage,
    citecolor=blue,
    filecolor=blue,
    linkcolor=blue,
    urlcolor=blue
}

\newcolumntype{x}[1]{%
>{\centering\arraybackslash}m{#1}}%

\newcommand{\Gammainv}[1]{\mathcal{G}^{-1} \left(#1\right)}
\newcommand{\Gammad}[1]{ \mathcal{G}\left(#1\right)}

\newcommand{\Normal}[1]{\mathcal{N}\!\left(#1\right)}

\newcommand{\Normult}[2]{N _{#1}\left(#2\right)}

\newcommand{\Betafun}[1]{B (#1)}
\newcommand{\Betadis}[1]{\mathcal{B}\left(#1\right)}

\newcommand{\ym}{{\mathbf y}}             
\newcommand{\error}{\epsilon}             
\newcommand{\hsv}{h}             
\newcommand{\hsvm}{{\mathbf h}}             

\newcommand{\mupar}{\mu}
\newcommand{\phipar}{\phi}
\newcommand{\sigmapar}{\sigma}
\newcommand{\etat}{\eta}
\newcommand{\omegapar}{\omega}

\theoremstyle{plain}

\theoremstyle{definition}
\newtheorem{alg}{Algorithm}

\DeclareMathOperator{\e}{e}
\newcommand{\R}{R}

{\begin{figure}
\begin{center}
\scalebox{#4}{\includegraphics{#3}}
\caption{#1}\label{#2}}
{\end{center}\end{figure}}

{\begin{table}\begin{center}
\caption{#1}\label{#2}}
{\end{center}\end{table}}

\begin{document}


\title{Ancillarity-Sufficiency Interweaving Strategy (ASIS) for Boosting MCMC Estimation of Stochastic Volatility Models}

\author{Gregor Kastner, Sylvia Fr\"uhwirth-Schnatter\\[1.4em]Institute for Statistics and Mathematics\\WU Vienna University of Economics and Business\\Welthandelsplatz 1 / D4 / level 4, 1020 Vienna, Austria}


\date{With minor editorial changes, this article is published as:\\[1em] \bibentry{kas-fru:anc}.}

\maketitle

\hrule
\vspace*{-1.2em}
\begin{abstract}
Bayesian inference for stochastic volatility models using MCMC methods highly depends on actual parameter values in terms of sampling efficiency. While draws from the posterior utilizing the standard centered parameterization break down when the volatility of volatility parameter in the latent state equation is small, non-centered versions of the model show deficiencies for highly persistent latent variable series. The novel approach of ancillarity-sufficiency interweaving has recently been shown to aid in overcoming these issues for a broad class of multilevel models. In this paper, we demonstrate how such an interweaving strategy can be applied to stochastic volatility models in order to greatly improve sampling efficiency for all parameters and throughout the entire parameter range. Moreover, this method of ``combining best of different worlds'' allows for inference for parameter constellations that have previously been infeasible to estimate without the need to select a particular parameterization beforehand.
\end{abstract}
\hrule

\vspace{2em}

\newpage

\section{Introduction}

Returns of financial and economic time series often exhibit time-varying volatilities. To account for this behavior, \cite{tay:fin} suggests in his pioneering paper to model the logarithm of the squared volatilities by latent autoregressive processes of order one. This specification, commonly referred to as the \emph{stochastic volatility (SV)} model, presents itself as a competitive alternative to GARCH-type designs by modeling the volatilities non-deterministically. Also, it arises naturally as a discretization of continuous-time models frequently appearing in the mathematical finance literature \citep[see e.g.][]{hul-whi:pri}.

Following e.g.\ \cite{jac-etal:bayJBES} or \cite{kim-etal:sto}, observed log-returns are denoted $\mathbf y = (y_1, y_2, \dots, y_T)'$ and the SV model is specified as
\begin{eqnarray}  
 \label{c1}
 y_{t} &=&  \e^{\hsv_{t}/2}\error_{t},
 \\
 \label{c2}
  \hsv_{t}&=& \mupar +  \phipar (\hsv_{t-1}- \mupar )  + \sigmapar \etat_{t},
\end{eqnarray}
where it is assumed that the iid standard normal innovations $\error_t$ and $\etat_s$ are independent for $t, s \in \{1,\dots,T\}$. The unobserved process $\hsvm=(\hsv_{0}, \hsv_{1}, \ldots,\hsv_{T})$ appearing in state equation (\ref{c2}) is usually interpreted as the latent time-varying \emph{volatility process} with initial state distributed according to the stationary distribution, i.e.
$
\hsv_0|\mu,\phi,\sigma \sim \Normal{\mu,\sigma^2/(1-\phi^2)}.
$
From now on, we will refer to equations (\ref{c1}) and (\ref{c2}) as the SV model in its \emph{centered parameterization (C)}.

Simulation efficiency in state-space models can often be improved through \emph{model reparameterization}. Papers related to this matter include \cite{gel-etal:eff}, \cite{pit-she:ana}, \cite{fru:eff}, \cite{rob-etal:bay04}, \cite{fru-soe:bayhes}, and \cite{str-etal:par}. The pioneering paper by~\cite{tay:fin} as well as several other papers like \cite{kim-etal:sto} or \cite{lie-ric:cla} consider a \emph{partially non-centered parameterization}, where the level $\mu$ of $\hsv_{t}$ -- which defines  the scale of $y_{t}$ -- is shifted from the state equation (\ref{c2}) to the observation equation (\ref{c1}) by setting $\bar h_t = h_t - \mu$.
\cite{kim-etal:sto} compare both parameterizations within a Bayesian inference. They show that the partially non-centered  parameterization leads to very high inefficiency when sampling $\mupar$ and recommend choosing the centered parameterization in any case. Nevertheless, the centered parameterization has several disadvantages. Firstly, inefficiency when drawing $\sigmapar$ is still high, see e.g.~Table~1 in \cite{kim-etal:sto}. Secondly, the conclusions are only valid if  $\phipar$ is close to one, which is commonly the case when the  SV model is applied to capture conditional heteroskedasticity of observed financial times series. However, this is not necessarily true when the SV model is applied in more general contexts such as capturing conditional heteroskedasticity in latent variables or regression residuals.

For the purpose of this paper, the (fully) \emph{non-centered parameterization (NC)}, given through
\begin{eqnarray} 
 &&  y_{t}   \sim \Normal{0, \omegapar   e ^{\sigmapar \tilde \hsv_{t}}}, \label{nc1}\\
&& \tilde \hsv_{t}=  \phipar \tilde \hsv_{t-1}  + \etat_{t}, \qquad  \etat_{t} \sim \Normal{0,1}, \label{nc2}
\end{eqnarray}
where $\omega=\e^\mu$, is of particular importance. The initial value of $\tilde \hsv_{0}|\phi$ is once again drawn from the stationary distribution of the latent process, i.e.\ $\tilde \hsv_{0}|\phi \sim \Normal{0, 1/(1- \phipar^2)}$. Note that $\tilde h_t = (h_t-\mu)/\sigma$. 
For a moderate parameter range where $\phi_\text{true} \in \{0.8, 0.9, 0.95\}$ and $\sigma_\text{true} \in \{0.2, 0.3, 0.4\}$, \cite{str-etal:par} illustrate that this type of non-centering typically yields lowest inefficiency factors when estimating stochastic volatility and stochastic conditional duration models with randomly sized block updating. Also, in similar contexts, there are several papers showing that MCMC sampling improves a lot by considering a non-centered version of a state space model, see e.g.~\cite{fru:eff} and \cite{fru-wag:sto}. These authors show that non-centering is especially useful if the error variance in the state equation is considerably smaller than the error variance in the observation equation.
\cite{pit-she:ana} show for linear Gaussian state space models that the speed of convergence in the centered parameterization decreases as $|\phi|$ increases when the signal to noise ratio is fixed.

No matter which parameterization is chosen, the likelihood in the SV model has an intractable form. Thus, Bayesian estimation commonly relies on sampling the latent states $\mathbf{h}$ and treat these as known for updating the parameters $\mu$, $\phi$, and $\sigma$. In their seminal paper, \cite{jac-etal:bayJBES} propose a \emph{single-move} Metropolis-Hastings (MH) algorithm. Each individual $h_t$ is sampled conditional on past and future, i.e.~drawn from
 $
 p(h_t|\mathbf{h}_{[-t]},\sigma,\phi,\mu,\ym),
 $
 where $\mathbf{h}_{[-t]}$ denotes all elements of $\mathbf{h}$ except $h_t$. Due to the commonly high persistence of the latent process, \cite{she-kim:bay} note that the draws obtained from this sampler are also highly correlated and thus only slowly converging to the stationary distribution.
Alternatively, \cite{she-pit:lik} propose a \emph{multi-move} sampler, where volatility blocks of random length are updated at a time, while \cite{she:par} and \cite{omo-etal:sto} propose a method to draw directly from 
$
p(\mathbf{h}|\sigma,\phi,\mu,\ym).
$
This becomes possible through a normal mixture approximation of $\log(\error_t^2)$ and requires forward filtering backward sampling (FFBS) methods \citep{car-koh:ong,fru:dat,dur-koo:sim}. Within a more general Gaussian state-space framework, \cite{rue:fas} and \cite{mcc-etal:sim} propose sampling the latent volatilities through Cholesky-factorization of the precision matrix by exploiting its band-diagonal structure. We adopt this method to sample the latent volatilities ``all without a loop'' (AWOL).
For a more extensive review of both Bayesian and non-Bayesian SV estimation methods, see \cite{bos:rel}.

The contribution of this paper is threefold.
Firstly, we explore the impact of alternative parameterizations for a wide parameter range including empirically plausible values and more extreme ones that can be relevant for applications of SV models within more general frameworks such as SV factor models or regression analysis. It turns out that simulation efficiency heavily depends on the true parameter values of the data generating process, thus no single ``best'' parameterization exists.
Secondly, we provide a strategy to overcome this deficiency
by \emph{interweaving} C and NC utilizing an ancillarity-sufficiency interweaving strategy (ASIS) introduced by \cite{yu-men:cen}. This results in a robustly efficient sampler that always outperforms the more efficient parameterization with respect to all parameters at little extra cost in terms of design and computation.
Thirdly, we provide evidence that empirical sampling efficiency heavily depends on the realization of the data generating process by massively parallel simulation experiments.

The paper is structured as follows: Section~\ref{sec:est} gives insight into the estimation procedure for each of the two selected parameterizations. Section~\ref{sec:int} explains how ASIS can be applied in order to interweave these parameterizations. Extensive simulation results presented in Section~\ref{sec:simres} compare sampling efficiency for all parameters amongst the different parameterizations, Section~\ref{sec:app} provides real-data results for several daily exchange rates, and Section~\ref{sec:con} concludes.

\section{Bayesian Inference in the SV Model}
\label{sec:est}

\subsection{Prior Distributions}
\label{sec:priors}

To perform Bayesian inference, a prior distribution $p(\mu,\phi,\sigma)$ needs to be specified. For both parameterizations, we choose the same independent components for each parameter. The level $\mupar \in \mathbb{R}$ is equipped with the usual normal prior $\mu \sim \Normal{b_{\mu}, B_{\mu}}$. For the persistence parameter $\phipar \in (-1,1)$, we choose $(\phi+1)/2 \sim \Betadis{a_0, b_0}$ as in \cite{kim-etal:sto}, implying 
\begin{eqnarray}
 \label{beta_transformed}
 p(\phi) = \frac{1}{2\Betafun{a_0,b_0}}\left ( \frac{1+\phi}{2}\right ) ^{a_0-1}\left ( \frac{1-\phi}{2}\right ) ^{b_0-1}.
\end{eqnarray}
Clearly, the support of this distribution is the unit ball and thus guarantees stationarity of the autoregressive volatility process.
For the volatility of volatility $\sigmapar \in \mathbb{R}^+$, we choose $\sigma^2 \sim B_\sigma\cdot \chi^2_1=\Gammad{\frac{1}{2},\frac{1}{2B_\sigma}}$.
Note that this specification differs from the commonly employed conjugate Inverse-Gamma prior $\sigma^2 \sim \Gammainv{c_0,C_0}$ and is motivated by \cite{fru-wag:sto}, who equivalently stipulate the prior for $\pm \sqrt { \sigma^2}$ to follow a centered normal distribution, i.e.\ $\pm \sqrt { \sigma^2} \sim \Normal{0,B_\sigma}$. It turns out that this choice is less influential when the true volatility of volatility is small because $\sigma$ is not bound away from zero a priori.

\subsection{MCMC Methodology}

Observation equation (\ref{c1}) can easily be rewritten as
\begin{eqnarray}
 \label{lin}
 \tilde y_t&=&h_t + \log(\error_t^2),
\quad \error_t \sim \Normal{0,1},
\end{eqnarray}
where $\tilde y_t$ denotes $\log y_t^2$. Alternatively, $\tilde y_t$ can be interpreted as the transformed de-meaned returns $\log (y_t-\bar y)^2$, but might just as well be taken $\log ( (y_t-\bar y)^2 + c)$ with a fixed offset constant $c=10^{-3}$ as in \cite{kim-etal:sto} or $\log ( y_t^2 + c)$ with $c= 10^{-4}$ as in \cite{omo-etal:sto} in order to avoid values equal to zero. Equation (\ref{lin}) now takes the form of a linear but non-Gaussian state space model. Moreover, one can approximate the distribution of $\log(\error_t^2)$ by a mixture of normal distributions, i.e.\
$
\log(\error_t^2)|r_t \sim \Normal{m_{r_t},s_{r_t}^2}.
$
Here, $r_t\in\{1,\ldots,10\}$ defines the mixture component indicator at time $t$, while $m_{r_t}$ and $s^2_{r_t}$ denote mean and variance of the $r_t$th mixture component as tabulated in \cite{omo-etal:sto}. This representation allows rewriting (\ref{lin}) as a linear and conditionally Gaussian state space model,
\begin{eqnarray}
 \label{mixcenter1} \tilde y_t&=&m_{r_t}+h_t+\error_t,
\quad \error_t \sim \Normal{0,s^2_{r_t}},
\end{eqnarray}
where speedy MCMC sampling becomes possible in three steps.

\begin{alg}[AWOL Sampler]

Choose appropriate starting values for the parameters $\mu$, $\phi$, $\sigma$ and the indicators $\mathbf{r}=(r_1,r_2,\dots, r_T)'$ -- e.g.~start with components with high weights -- and repeat the following steps:
 \begin{enumerate}
  \item[(a)] Sample the latent volatilities AWOL by drawing from $\hsvm_{[-0]}|\ym, \mathbf{r}, \mu, \phi, \sigma^2$ or $\tilde \hsvm_{[-0]}|\ym, \mathbf{r}, \mu, \phi, \sigma^2$, respectively.
   The initial value is drawn from $h_0|h_1,\mu,\phi,\sigma^2$ or from $\tilde h_0|\tilde h_1,\phi$.
  \item[(b)] Sample $\mupar, \phipar, \sigmapar^2$ via Bayesian regression.
  \begin{itemize}
   \item[-] For C, we investigate a 1-block sampler, drawing from $\mupar, \phipar, \sigmapar^2|\hsvm$, a 2-block sampler, where $\sigma^2$ is drawn from $\sigma^2|\hsvm,\mu,\phi$, while $\mu$ and $\phi$ are sampled jointly from $\mu,\phi|\hsvm,\sigma^2$, and a 3-block sampler, where all parameters are individually drawn from the full conditionals.
    Due to non-conjugacy of the chosen priors, MH updates are used in all variants.
   \item[-] In NC, MH is needed only for updating $\phi$ by drawing from $\phi|\tilde \hsvm$, while $\mu$ and $\sigma^2$ can be Gibbs-updated jointly from $\mu,\sigma^2|\mathbf{y}, \tilde \hsvm,\mathbf{r}$ (2-block) or individually from $\mu|\mathbf{y}, \tilde \hsvm,\mathbf{r},\sigma^2$ and $\sigma^2|\mathbf{y}, \tilde \hsvm,\mathbf{r},\mu$ (3-block).
 \end{itemize}
\item[(c)] Update the indicators $\mathbf{r}$ from $\mathbf{r}|\mathbf{y}, \hsvm$ in C, or $\mathbf{r}|\mathbf{y},\tilde \hsvm,\mu,\sigma^2$ in NC, via inverse transform sampling.
\end{enumerate}
\end{alg}

\subsection{Step (a): Sampling the Latent Volatilities AWOL}
\label{sub:sampleh}

Conditional on all other variables, the joint density for $\hsvm$ (and $\tilde \hsvm$) is multivariate normal. Due to the order-one autoregressive nature of the latent volatility process, this distribution can be written in terms of the tridiagonal precision matrix $\mathbf{\Omega}$, giving rise to sampling
\emph{all without a loop (AWOL)}.
This method is employed in \cite{rue:fas} and \cite{mcc-etal:sim} and does not require the ``end-user'' to implement any loops -- hence the name. Thus, it is very convenient in terms of implementation and fast in terms of computation. No FFBS methods are needed, there is no need to invert the tridiagonal precision matrix $\mathbf{\Omega}$ and it is fast due to the availability of band back-substitution already implemented in practically all widely used programming libraries.

In the centered parameterization, we draw from $\hsvm_{[-0]}|\mu, \sigma, \phi, \mathbf{r}, \ym \sim \Normult{T}{\mathbf{\Omega}^{-1}\mathbf{c}, \mathbf{\Omega}^{-1}}$ with
\[
\mathbf{\Omega}=\begin{bmatrix}
\frac{1}{s_{r_1}^2}+\frac{1}{\sigma^2} & \frac{-\phi}{\sigma^2}& 0 &\hdots &0 \\
\frac{-\phi}{\sigma^2} & \frac{1}{s_{r_2}^2}+\frac{1+\phi^2}{\sigma^2} & \frac{-\phi}{\sigma^2} &\ddots &\vdots \\ 
0 & \frac{-\phi}{\sigma^2} & \ddots & \ddots &0 \\ 
\vdots & \ddots & \ddots &\frac{1}{s_{r_{T-1}}^2}+\frac{1+\phi^2}{\sigma^2} & \frac{-\phi}{\sigma^2}\\ 
0 & \hdots& 0 & \frac{-\phi}{\sigma^2} &\frac{1}{s_{r_T}^2}+\frac{1}{\sigma^2}
\end{bmatrix},
\]
and
\[
\mathbf{c}=
\begin{bmatrix}
\frac{1}{s_{r_1}^2}(\tilde y_1-m_{r_1})+\frac{\mu(1-\phi)}{\sigma^2} \\
\frac{1}{s_{r_2}^2}(\tilde y_2-m_{r_2})+\frac{\mu(1-\phi)^2}{\sigma^2} \\
\vdots \\
\frac{1}{s_{r_{T-1}}^2}(\tilde y_{T-1}-m_{r_{T-1}})+\frac{\mu(1-\phi)^2}{\sigma^2} \\
\frac{1}{s_{r_T}^2}(\tilde y_T-m_{r_T})+\frac{\mu(1-\phi)}{\sigma^2}
\end{bmatrix}.
\]

Analogously, in the noncentered case, we draw from $\tilde \hsvm_{[-0]}|\mu, \sigma, \phi, \mathbf{r}, \ym \sim \Normult{T}{\mathbf{\Omega}^{-1}\mathbf{c}, \mathbf{\Omega}^{-1}}$ with
\[
\mathbf{\Omega}=\begin{bmatrix}
\frac{\sigma^2}{s_{r_1}^2}+1 & -\phi& 0 &\hdots &0 \\
-\phi & \frac{\sigma^2}{s_{r_2}^2}+1+\phi^2 & -\phi &\ddots &\vdots \\ 
0 & -\phi & \ddots & \ddots &0 \\ 
\vdots & \ddots & \ddots &\frac{\sigma^2}{s_{r_{T-1}}^2}+1+\phi^2 & -\phi\\ 
0 & \hdots& 0 & -\phi &\frac{\sigma^2}{s_{r_T}^2}+1
\end{bmatrix},
\]
and
\[
\mathbf{c}=
\begin{bmatrix}
\frac{\sigma}{s_{r_1}^2}(\tilde y_1-m_{r_1}-\mu) \\
\frac{\sigma}{s_{r_2}^2}(\tilde y_2-m_{r_2}-\mu) \\
\vdots \\
\frac{\sigma}{s_{r_{T-1}}^2}(\tilde y_{T-1}-m_{r_{T-1}}-\mu) \\
\frac{\sigma}{s_{r_T}^2}(\tilde y_T-m_{r_T}-\mu)
\end{bmatrix}.
\]

For both parameterizations, this is accomplished by first computing the Cholesky decomposition $\mathbf{\Omega}=\mathbf{LL}'$. Due to the band structure of $\mathbf{\Omega}$, this is computationally inexpensive and can either be implemented directly or via the LAPACK-routine \verb.dpbtrf. \citep{lapack}, to name but one of the many widely available (and thoroughly tested) linear algebra routines designed for this task. Note that only main diagonal and lower first off-diagonal elements of $\mathbf{L}$ will be nonzero. Next, we draw $\boldsymbol{\error} \sim \Normult{T}{\boldsymbol{0},\mathbf{I}_T}$ and then
efficiently solve $\mathbf{L}\mathbf{a}=\mathbf{c}$ for $\mathbf{a}$ and $\mathbf{L}'\hsvm=\mathbf{a}+\boldsymbol{\error}$ for $\hsvm$ by using band back-substitution instead of actually calculating $\mathbf{L}^{-1}$.
Finally, the initial value can be sampled from $h_0|h_1,\mu,\phi,\sigma \sim \Normal{\mu+\phi(h_{1}-\mu), \sigma^2}$ in C and from
$\tilde h_0|\tilde h_1,\phi \sim 
N(\tilde h_1\phi,1)$
in NC.

\subsection{Step (b)-C: Sampling of $\mu$, $\phi$ and $\sigma$ in C} 
\label{sub:sampleparaC}

For sampling $\boldsymbol{\theta} = (\mu, \phi, \sigma^2)$, it is helpful to rewrite the conditional AR(1) model as a conditional regression model with the lagged latent variables as regressors,
\[
 h_t=\gamma + \phi h_{t-1} + \eta_t, \quad \eta_t \sim \Normal{0,\sigma^2},
\]
via $\gamma=(1-\phi)\mu$. Note that the implied conditional prior $p(\gamma|\phi)$ follows a normal distribution with mean $b_{\mu}(1-\phi)$ and variance $B_\mu(1-\phi)^2$.
In this Subsection, we will discuss three common blocking strategies for sampling $\boldsymbol{\theta}$.

For a \emph{one block} update of $\boldsymbol{\theta}$, we use a single MH step. The posterior arising from an auxiliary regression model with conjugate priors is used as the proposal density:
\[
p_\text{aux}(
\boldsymbol{\theta}_\text{new}
|\hsvm) =
 p_\text{aux}(\gamma_\text{new}, \phi_\text{new} |\hsvm, \sigma^2_\text{new})
p_\text{aux}(\sigma^2_\text{new}|\hsvm).
\]
We choose $p_\text{aux}(\sigma^2) \propto\sigma^{-1} $ to denote the density of an auxiliary improper conjugate  $\Gammainv{-\frac{1}{2}, 0}$ prior, and $p_\text{aux}(\gamma,\phi|\sigma^2)
$ to denote the density of an auxiliary conjugate $\Normult{2}{\mathbf{0}, \sigma^2\mathbf{B}_0}$ prior with $\mathbf{B}_0=\text{diag}(B_0^{11}, B_0^{22})$. 
More specifically, $p_\text{aux}(\gamma|\sigma) \sim \Normal{0, \sigma^2B_0^{11}}$ and $p_\text{aux}(\phi|\sigma) \sim \Normal{0, \sigma^2B_0^{22}}$.
In order to avoid collinearity problems when $\sigma^2$ is close to zero (and thus $h_t$ almost constant for all $t$), we pick slightly informative variances, i.e.~$B_0^{11} = 10^{12}$ and $B_0^{22} = 10^{8}$.
This yields
\begin{eqnarray}
 \label{regression1}
\gamma,\phi|\hsvm, \sigma^2 \sim \Normult{2}{\mathbf{b}_T,\sigma^2\mathbf{B}_T},
\end{eqnarray}
with $\mathbf{B}_T = (\mathbf{X'X} + \mathbf{B}_0^{-1})^{-1}$ and $ \mathbf{b}_T = \mathbf{B}_T\mathbf{X}'\mathbf{h}_{[-0]}$, where $\mathbf{X}$ is the $T \times 2$ design matrix with ones in the first column and $\mathbf{h}_{[-T]}$ in the second. The marginalized auxiliary posterior distribution for $\sigma^2$ is given through
$
\sigma^2 | \hsvm \sim \Gammainv{c_T, C_T},
$
with $c_T = (T-1)/2$ and
$C_T =  \frac{1}{2}\left( \sum_{i=1}^T \hsv_i^2 - \mathbf{b}_T' \mathbf{X}' \hsvm_{[-0]} \right)$.
The acceptance probability is given through $\min(1,R)$, with
\begin{eqnarray*}
R
=
 \frac{p(\hsv_0|\boldsymbol{\theta}_\text{new})p(\gamma_\text{new}|\phi_\text{new})p(\phi_\text{new})p(\sigma^2_\text{new})}
 	 {p(\hsv_0|\boldsymbol{\theta}_\text{old})p(\gamma_\text{old}|\phi_\text{old})p(\phi_\text{old})p(\sigma^2_\text{old})}
    \times
    \frac{p_\text{aux}(\phi_\text{old}, \gamma_\text{old}|\sigma^2_\text{old})p_\text{aux}(\sigma^2_\text{old})}
         {p_\text{aux}(\phi_\text{new}, \gamma_\text{new}|\sigma^2_\text{new})p_\text{aux}(\sigma^2_\text{new})}.
\end{eqnarray*}

In the \emph{two-block} sampler, we draw the first block from the full conditional distribution $\gamma,\phi|\hsvm, \sigmapar^2$ given in (\ref{regression1}) and accept with probability $\min(1,R)$, where
\[
R=
\frac
 {p(h_0|\gamma_\text{new},\phi_\text{new})p(\gamma_\text{new}|\phi_\text{new})p(\phi_\text{new})}
 {p(h_0|\gamma_\text{old},\phi_\text{old})p(\gamma_\text{old}|\phi_\text{old})p(\phi_\text{old})}
\times
\frac
 {p_\text{aux}(\gamma_{\text{old}},\phi_{\text{old}})}
 {p_\text{aux}(\gamma_{\text{new}},\phi_{\text{new}})}.
\]

In order to construct a suitable proposal for the -- now full conditional -- density $p(\sigma^2|\hsvm,\mu,\phi)$, we again use the 
auxiliary conjugate prior $p_\text{aux}(\sigma^2) \propto\sigma^{-1} $, under which we straightforwardly obtain 
\begin{eqnarray}
 \label{IGprop}
\sigma^2|\hsvm,\mu,\phi \sim \Gammainv{c_T, C_T},
\end{eqnarray}
where $c_T=T/2$ and
$
 C_T=\frac{1}{2}\left(\sum_{t=1}^T ((h_t-\mu)-\phi(h_{t-1}-\mu))^2 +(h_0-\mu)^2(1-\phi^2)\right).
$
The acceptance probability simplifies to $\min(1,R)$ with
\begin{eqnarray*}
 R=\frac
 {p(\sigma^2_\text{new})}
 {p(\sigma^2_\text{old})}
 \times
 \frac
 {p_\text{aux}(\sigma^2_\text{old})}
 {p_\text{aux}(\sigma^2_\text{new})}=
 \exp \left \{\frac{\sigma^2_\text{old}-\sigma^2_\text{new}}{2B_\sigma}\right \}.
\end{eqnarray*}


In the \emph{three-block} sampler, each individual parameter is drawn from the full conditional distribution $\mupar|\cdot$, $\phipar|\cdot$, and $\sigmapar^2|\cdot$, respectively. Thus, $\sigmapar^2$ is drawn from (\ref{IGprop}). For sampling $\phipar$, we obtain a proposal from
\[
\phipar|\hsvm, \gamma,\sigmapar^2 \sim \Normal{
\frac{\left[\sum_{t=1}^Th_{t-1}h_t\right]-\gamma\sum_{t=0}^{T-1}h_t}{\sum_{t=0}^{T-1}h_t^2 + 1/B_0^{22}},
\frac{\sigma^2}{\sum_{t=0}^{T-1}h_t^2 + 1/B_0^{22}}
}.
\]
The acceptance probability is equal to $\min(1,R)$ with
\[
R=\frac{p(\hsv_0|\phipar_\text{new}, \mu, \sigmapar^2)p(\phipar_\text{new})}{p(\hsv_0|\phipar_\text{old}, \mu, \sigmapar^2)p(\phipar_\text{old})}\times
\frac{p_\text{aux}(\phipar_\text{old}|\sigmapar^2)}{p_\text{aux}(\phipar_\text{new}|\sigmapar^2)}.
\]
For sampling $\gamma$ from the full conditional posterior distribution, we obtain a proposal from
\[
\gamma|\hsvm, \phipar,\sigmapar^2 \sim
\Normal{
\frac{\sum_{t=1}^T h_t-\phi\sum_{t=0}^{T-1}h_t}{T + 1/B_0^{11}},
\frac{\sigma^2}{T + 1/B_0^{11}}}
\]
and an acceptance probability equaling $\min(1,R)$ with
\[
R=\frac{p(\hsv_0|\gamma_\text{new}, \phipar, \sigmapar^2)p(\gamma_\text{new}|\phi)}{p(\hsv_0|\gamma_\text{old}, \phipar, \sigmapar^2)p(\gamma_\text{old}|\phi)}\times
\frac{p_\text{aux}(\gamma_\text{old}|\sigmapar^2)}{p_\text{aux}(\gamma_\text{new}|\sigmapar^2)}.
\]

\subsection{Step (b)-NC: Sampling of $\mu$, $\phi$ and $\sigma$ in NC} 
\label{sub:sampleparaNC}
In the noncentered parameterization, only $\phipar$ is left in the state equation. To sample this parameter, we employ a flat auxiliary prior $p_\text{aux}(\phi)\propto c$, yielding the proposal
\[
\phi|\tilde \hsvm \sim 
\Normal{
 \frac{\sum_{t=0}^{T-1}\tilde h_t\tilde h_{t+1}}{\sum_{t=0}^{T-1}\tilde h_t^2},
 \frac{1}{\sum_{t=0}^{T-1}\tilde h_t^2}
},
\]
and an acceptance probability of $\min(1,R)$, where
$
 R=
 {p(\tilde h_0|\phi^\text{new})p(\phi^\text{new})}/
 { p(\tilde h_0|\phi^\text{old})p(\phi^\text{old})}.
$

For sampling $\mupar$ and $\sigmapar$, one can straightforwardly rewrite the conditional observation equation (\ref{mixcenter1}) as a regression model with homoskedastic errors, i.e.\
\begin{eqnarray}
 \label{ncreg}
 \mathbf{\breve y} = \mathbf{X} \begin{bmatrix} \mupar \\ \sigmapar \end{bmatrix} +
  \boldsymbol{\error},
\end{eqnarray}
  where 
  $\boldsymbol{\error} \sim \Normult{K}{\mathbf{0},\mathbf{I}_K}$,
and
\[
\mathbf{\breve y}=
\begin{bmatrix}
(\tilde y_1-m_{r_1})/s_{r_1} \\
\vdots \\
(\tilde y_T-m_{r_T})/s_{r_T}
\end{bmatrix},
\quad
\mathbf{X}=
\begin{bmatrix}
\tilde h_{1}/ s_{r_{1}} & 1/ s_{r_{1}}\\
\vdots & \vdots \\
\tilde h_T / s_{r_T}& 1 / s_{r_T}
\end{bmatrix}.
\]


The joint posterior distribution is again bivariate Gaussian with variance-covariance matrix $\mathbf{B}_T=(\mathbf{B}_0^{-1}+\mathbf{X}'\mathbf{X})^{-1}$ and mean $\mathbf{b}_T = \mathbf{B}_T(\mathbf{B}_0^{-1}\mathbf{b}_0+\mathbf{X}'\mathbf{\breve y})$, where $\mathbf{b}_0=(b_\mu,0)'$ and $\mathbf{B}_0=\text{diag}(B_\mupar, B_\sigmapar)$ denote mean and variance of the joint prior density $p(\mu,\sigma)$, respectively.

Alternatively, one could sample both parameters from the full conditional posteriors (\emph{three-block} sampling), yielding
$
\mu|\ym, \tilde \hsvm,\mathbf{r},\sigma \sim \Normal{b_{T,\mu}, B_{T,\mu}}
$
with
\[
b_{T,\mu} = B_{T,\mu}\left( \sum_{t=1}^T \frac{\tilde y_t -m_{r_t}-\sigma \tilde h_t}{s^2_{r_t}} + \frac{b_\mu}{B_\mu} \right),
\qquad
B_{T,\mu} = 1/\left({\displaystyle \sum_{t=1}^T 1/s_{r_t}^2 +\frac{1}{B_\mu}}\right),
\]
and 
$
\sigma|\ym, \tilde \hsvm,\mathbf{r},\mu \sim \Normal{b_{T,\sigma}, B_{T,\sigma}}
$
with
\[
b_{T,\sigma} = B_{T,\sigma} \sum_{t=1}^T \frac{\tilde h_t(\tilde y_t-m_{r_t}-\mu)}{s^2_{r_t}},
\qquad
B_{T,\sigma} = 1/\left({\sum_{t=1}^T\frac{\tilde h_t^2}{s_{r_t}^2} +\frac{1}{B_\sigma}}\right).
\]

\subsection{Step (c): Sampling the Indicators $\mathbf{r}$}
\label{sub:sampleind}

We proceed exactly as \cite{omo-etal:sto}. Observing that $\tilde y_t-h_t=\error_t^*$ with $\error_t^* \sim \Normal{m_{r_t},s^2_{r_t}}$,
one easily obtains the posterior probabilities $\mathbb{P}(r_t=k|\cdot)$ for $k \in \{1,\dots, 10\}$ and $t \in \{1,\dots, T\}$ according to 
\[
\mathbb{P}(r_t=k|\cdot) \propto \mathbb{P}(r_t=k)\frac{1}{s_k}\exp \left\{-\frac{(\error_t^*-m_k)^2}{2s_k^2} \right\},
\]
where $\mathbb{P}(r_t=k)$ denotes the mixture weights of the $k$th component. In our implementation, we do the calculations on a $\log$-scale and normalize with respect to the maximum as required. The actual drawing is then conducted via inverse transform sampling. Note that due to $T\times10$ exponential function calls, this step is computationally rather expensive but can easily be parallelized.

\section{Interweaving C and NC by ASIS}
\label{sec:int}

To provide some intuition about the sampling efficiency in C, let $\phi=0$ for a moment. This implies that the state equation (\ref{c2}) reduces to $h_t \sim \Normal{\mu,\sigma^2}$ iid for all $t \in \{1,\dots,T\}$.
In this setting, $\hsvm$ becomes more informative about $\mu$ when the conditional variance $\sigma^2$ gets smaller. Thus, \emph{more} information is missing when treating $\hsvm$ as latent data.
Consequently, when sampling $\mu$ under the assumption that $\phi = 0$ and $\sigma^2$ is small, we expect C to be inefficient.
On the other hand, if $\phi$ approaches $1$, the latent process converges towards a random walk and $\hsvm$ will be very uninformative about $\mu$. Thus, only little information is lost when treating $\hsvm$ as latent data and C has better chances to work fine.
In NC, no major troubles are to be expected if $\phi=0$, since the state equation (\ref{nc2}) reduces to $\tilde h_t \sim \Normal{0,1}$ iid for all $t \in \{1,\dots,T\}$, which is obviously independent of the value of $\sigma$. Thus, sampling $\mu$ and $\sigma$ in the linearized equation (\ref{ncreg}) reduces to simple linear regression with independent regressors. If, however, $\phi$ goes towards $1$, we are prone to running into spurious regression problems. Certainly these arguments rely on massive oversimplification (e.g.\ by not taking into account the impact of the mixture approximation or spillover effects by inefficient proposal densities and different blocking strategies) and can only provide a faint idea of what is going on in the general case.

Nevertheless, due to the fact that in the context of the model at hand, the latent variables $\hsvm$ in C form a sufficient statistic for $\mu$ and $\sigma$, while the transformed volatilities $\tilde \hsvm$ in NC form an ancillary statistic for these parameters, there is hope that interweaving C and NC helps to increase sampling efficiency. \cite{yu-men:cen} propose an \emph{ancillary-sufficiency interweaving strategy (ASIS)} which, in certain situations, converges geometrically even when C and/or NC fail to do so.
They explain this ``seemingly magical property'' by relating to Basu's theorem \citep{bas:ons} on the independence of complete sufficient and ancillary statistics and show in a quite general context that the geometric convergence rate of the sampler interweaving $\hsvm$ and $\tilde \hsvm$ is always bound by $R\sqrt{r_\text{C}r_\text{NC}}$, where $R$ is the maximal correlation between $\hsvm$ and $\tilde \hsvm$ in their joint posterior distribution $p(\hsvm, \tilde \hsvm|\mathbf{y})$ and $r_\text{C},r_\text{NC}$ denote the geometric rate of convergence of C and NC, respectively. This means that the rate of convergence of the interwoven sampler is mainly governed by the individual convergence rates and the posterior correlation $R$, implying that ancillary-sufficiency pairs of latent variables are likely to be good candidates for reducing sampling inefficiency. It is worth noting that the original ASIS notation $Y_{obs}$ for the observed data directly transforms to $\ym$ for the model at hand, while $Y_{mis}$ -- denoting the ``missing'' part of the data -- equals $\mathbf{h}$.

The idea of interweaving is surprisingly simple. It is based on sampling the parameters in question -- in our case $\mu$, $\sigma$ (and $\phi$) -- twice: once utilizing C and again utilizing NC.
Ad hoc, it is not clear whether one should start C and redraw NC (``baseline C'') or vice versa (``baseline NC''). We will discuss both strategies and assess their performance individually.
Algorithm~\ref{gisc} below describes the former, i.e.\ both the latent volatilities and the indicators are sampled \emph{once with baseline C}, while the parameters $\mu, \phi, \sigma$ are sampled \emph{once in each parameterization} within each iteration of the sampler.
It is termed ``GIS-C'', where the first three letters are borrowed from \cite{yu-men:cen} and stand for \emph{global interweaving strategy}. C simply denotes the fact that we use the centered baseline.

\begin{alg}[GIS-C]
 \label{gisc}
Choose appropriate starting values and repeat the following steps:
\begin{enumerate}
 \item[(a)] Draw $\mathbf{h}$ (C).
 \item[(b)] Draw $\mu, \phi, \sigma$ (C).
 \item[(b*)] Move to NC by the simple deterministic transformation $\tilde h_t = \frac{h_t-\mu}{\sigma}$ for all $t.$
 \item[(b**)] Redraw $\mu, \phi, \sigma$ (NC).
 \item[(b***)] Move back to C by calculating $h_t = \mu + \sigma \tilde h_t$ for all $t$.
 \item[(c)] Draw the indicators $\mathbf{r}$ (C).
\end{enumerate}
\end{alg}

The individual sampling steps are implemented exactly as described in subsections \ref{sub:sampleh} to \ref{sub:sampleind}. Note that since $\phi$ is not involved in the reparameterization, in step (b**), one might as well redraw $\mu$ and $\sigma$ only; the difference concerning sampling efficiency is however negligible.
Also note that although additional sampling steps are introduced as (b*) to (b***), overall sampling time is only affected minimally because these steps are very cheap in terms of computation cost.

The sampler with noncentered baseline is of course very similar. As before, for each iteration the parameters $\mu$, $\phi$, and $\sigma$ are sampled twice (once in C and once in NC), while the latent volatilities and the indicators are sampled in NC only.

\begin{alg}[GIS-NC]
Choose appropriate starting values and repeat the following steps:
\begin{enumerate}
 \item[(a)] Draw $\mathbf{\tilde h}$ (NC).
 \item[(b)] Draw $\mu, \phi, \sigma$ (NC).
 \item[(b*)] Move to C by the simple deterministic transformation $h_t = \mu + \sigma \tilde h_t$ for all $t$.
 \item[(b**)] Redraw $\mu, \phi, \sigma$ (C).
 \item[(b***)] Move back to NC by transforming back: $\tilde h_t = \frac{h_t-\mu}{\sigma}$ for all $t$.
 \item[(c)] Draw the indicators $\mathbf{r}$ (NC).
\end{enumerate}
\end{alg}


To conclude, note that the strategy of interweaving is intrinsically different to \emph{alternating} the parameterizations, for instance by (randomly) choosing one parameterization and running a complete MCMC cycle within that parameterization. Also, it is distinct from \emph{compromising} between two parameterizations, e.g.\ by partial noncentering.

\section{Simulation Results}
\label{sec:simres}

In order to assess simulation efficiency of our algorithms, we simulate data from the model specified in equations (\ref{c1}) and (\ref{c2}).
For the sake of simplicity and readability, $\mu_\text{true}$ is set to $-10$ for all runs. Results not reported here show that this choice is of minor influence. The parameters $\phi_\text{true}$ and $\sigma_\text{true}$ vary on a $\{0,0.5,0.8,0.9,0.95,0.96,0.97,0.98,0.99\}\times\{0.5,0.4,0.3,0.2,0.1\}$ grid, resulting in $45$ distinct parameter settings. This choice includes previously investigated and empirically plausible values, see e.g.\ \cite{jac-etal:bayJBES}, \cite{kim-etal:sto}, \cite{lie-ric:cla}, and \cite{str-etal:par}. Moreover, the range is chosen to also include more extreme values that frequently arise when univariate SV is applied to capture conditional heteroskedasticity in latent variables such as factors or residuals of regression-type problems. We repeat this exercise for $500$ data sets and apply four sampling schemes (C, NC, GIS-C, GIS-NC) by using $M=100\,000$ MCMC draws after a burn-in of $10\,000$ for each data set.
Time series length is fixed to $T=5000$, which corresponds to just above 20 years of daily data.
Overall, this results in $90\ 000$ chains of length $110\ 000$, or a total of around $50$ trillion latent instantaneous volatility draws. 
Nevertheless, due to parallel implementation of native C code on our local computer cluster using 500 cores, sampling can easily be done overnight.
Throughout all simulations we use priors with means equaling the true values, more specifically $b_\mu=\mu_{\text{true}}$, $B_\mu = 10$, $a_0=40$, $b_0=80/(1+\phi_\text{true})-40$, $B_\sigma = \sigma_\text{true}^2$, and starting values are set to true values to avoid values outside the stationary distribution after the burn-in period.

Computation of parallel MCMC chains for each parameter constellation was conducted on a cluster of workstations consisting of 44 IBM dx360M3 nodes with a total of 544 cores running \R\ 2.15.1 \citep{r:r} and OpenMPI 1.4.3 \citep{gab-etal:ope}. For high-level-parallelization and parallel random number generation according to \cite{lec-etal:obj}, the R packages \verb.parallel. (part of \R{}) and \verb.snow. \citep{r:sno} were used. Ex-post analysis and timing was done on a Laptop with a 2.67GHz Intel i5 M560 CPU running the same R version. For the actual sampling, the \R{} package \verb.stochvol. \citep{r:sto}, available on CRAN, was created. The core implementation is written in C, interfaced to R via \verb.Rcpp. \citep{edd-fra:rcp}. Inefficiency factors and effective sample sizes were computed with the \R{} package \verb.coda.~\citep{plu-etal:cod}.

The mean time for running $1000$ simulation draws varies between $2.3$ seconds for C and $2.4$ seconds for GIS-NC on a Laptop with a 2.67GHz Intel i5 M560 CPU using one core. Note that these numbers are fairly constant for all true parameter values and grow linearly with $T$. As an example, the time to run $1000$ simulations for $T=500$ varies between $0.23$ and $0.24$ seconds.

\subsection{To Center Or Not to Center?}

\begin{figure}[h!t]
\centering
  \includegraphics[width=\textwidth]{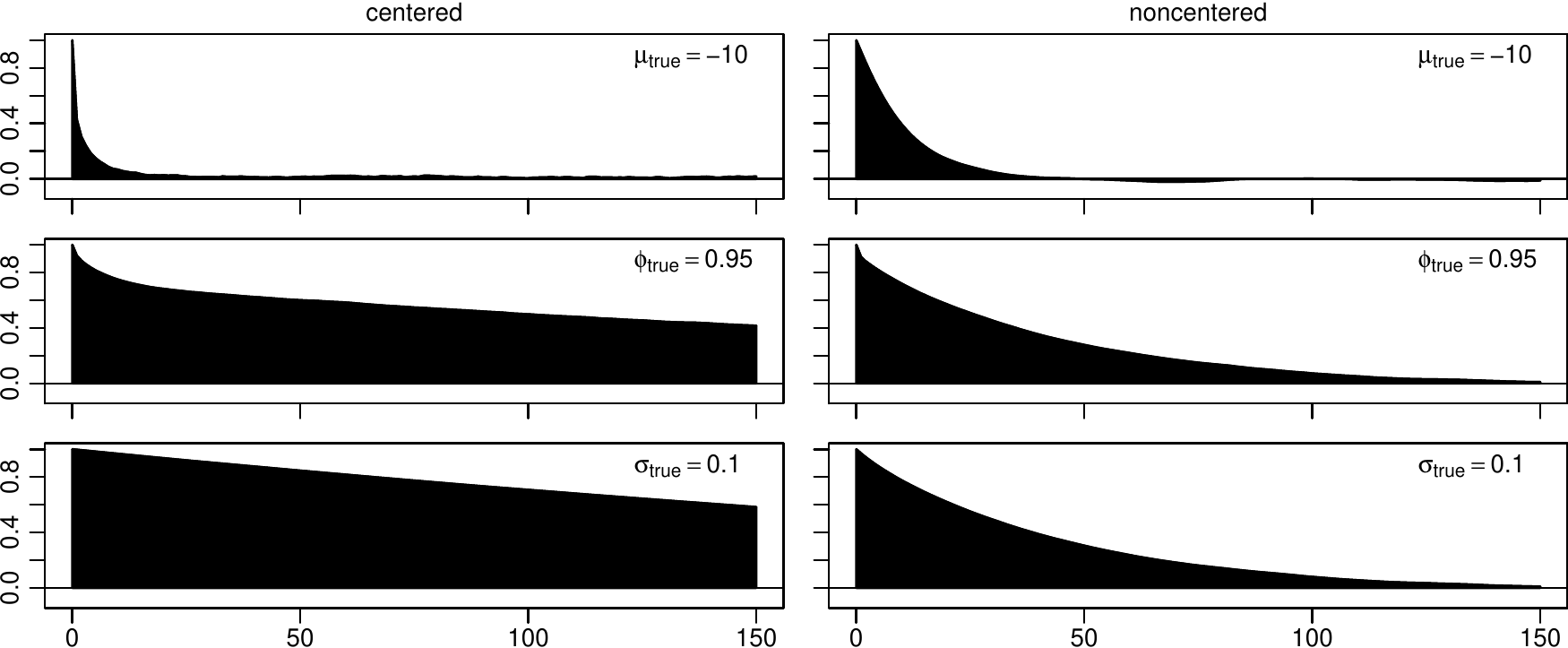}
  \caption{Sample autocorrelations of 100\,000 MCMC draws obtained from C (left hand side) and NC (right hand side) for a \emph{small volatility of volatility} setup.}
 \label{acf1}
\end{figure}

Simulation efficiency of the two raw parameterizations mainly depends on the values of the parameters $\phi$ (persistence) and $\sigma$ (volatility of volatility). To illustrate the latter, Figure~\ref{acf1} shows autocorrelations of an exemplary parameter setup with small volatility of volatility $\sigma_\text{true}=0.1$ for a single time series that has been randomly selected from the pool of all $500$ time series. Here, C ``fails'' in the sense that the draws from $p(\mu|\mathbf{y})$ and $p(\sigma|\mathbf{y})$ exhibit large autocorrelation, while NC performs substantially better. This observation is in line with findings of \cite{pit-she:ana} and \cite{fru:eff}, who observe that simulation efficiency in the centered parameterization decreases with decreasing $\sigma_\text{true}$ for linear Gaussian state space models. 

\begin{figure}[h!t]
\centering
  \includegraphics[width=\textwidth]{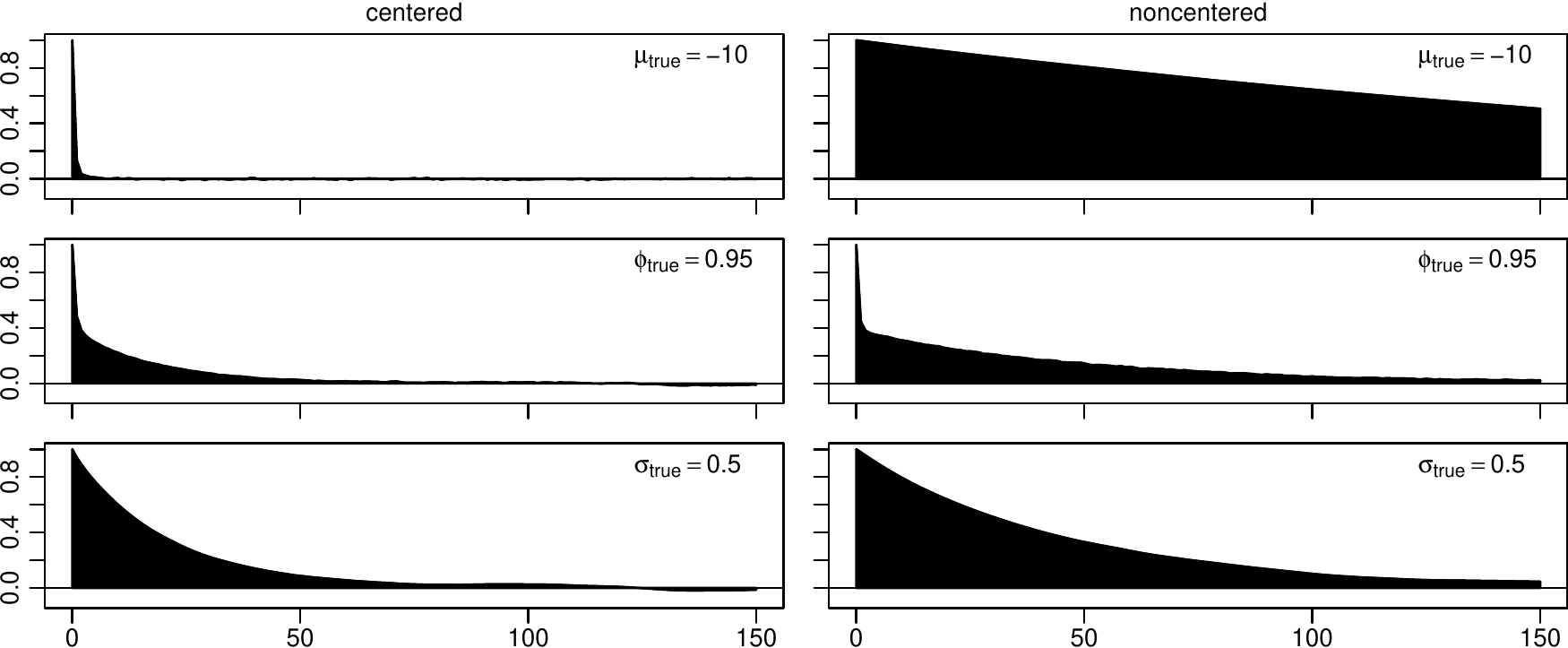}
 \caption{Sample autocorrelations of 100\,000 MCMC draws obtained from C (left hand side) and NC (right hand side) for a \emph{large volatility of volatility} setup.}
 \label{acf2}
\end{figure}

On the other hand, Figure~\ref{acf2} portraits a parameter setup with larger volatility of volatility $\sigma_\text{true}=0.5$, while persistence $\phi_\text{true}$ and level $\mu_\text{true}$ are the same as before. Here, we see that draws from C show little autocorrelation, while MCMC chains obtained from NC do not mix well.

\subsection{Sampling Efficiency}
\label{sec:eff}
For assessing simulation efficiency, the inefficiency factor (IF) is employed as a benchmark. It is an estimator for the integrated autocorrelation time $\tau$ of a stochastic process given through
$
\tau = 1+2\sum_{s=1}^\infty\rho(s),
$
where $\rho(s)$ is the autocorrelation function for lag $s$. We estimate $\tau$ through the spectral density of the Markov chain, i.e.\ $\text{IF}=\gamma_0/s^2$, where $\gamma_0$ denotes the estimated spectral density evaluated at zero and $s^2$ denotes the sample variance of the MCMC draws.
%
The inefficiency factor is directly proportional to the squared Monte Carlo standard error MCSE$^2$ through the relationship $\text{MCSE}^2 = \frac{s^2}{M}\times \text{IF}$. In other words, $100\,000$ draws from a Markov chain with an IF of $100$ have roughly the same MCSE as $1000$ draws from an independent sample. Consequently, the effective sample size ESS is given by $M/\text{IF}$.
Clearly, the aim is to provide samplers with small IFs, thus large ESSs, at smallest possible computational cost.

\begin{figure}[h!t]
\centering
  \includegraphics[width=\textwidth]{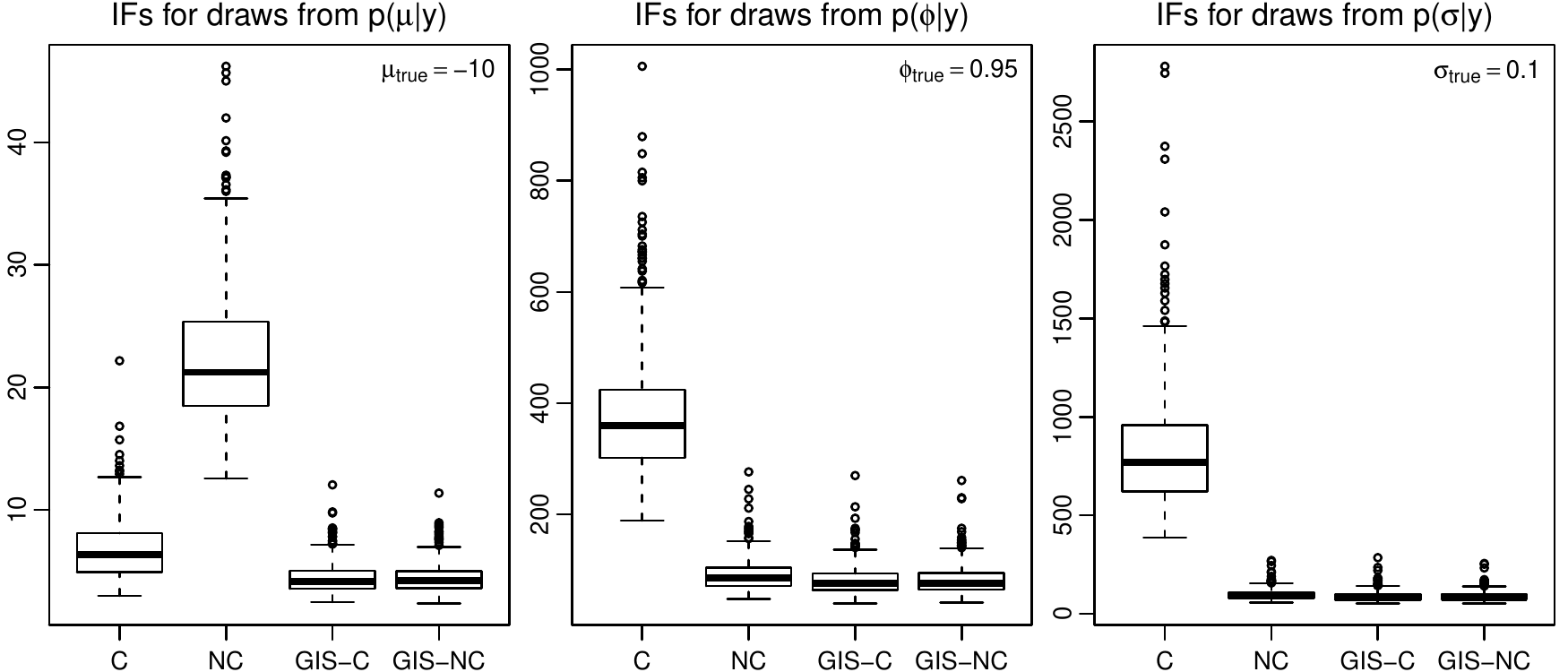}
  \caption{Boxplots of $500$ repeated measurements of inefficiency factors of $100\,000$ draws from the marginal densities. The underlying latent volatility process exhibits \emph{small volatility of volatility}.}
 \label{boxplots1}
\end{figure}

\begin{figure}[h!t]
 \centering 
 \includegraphics[width=\textwidth]{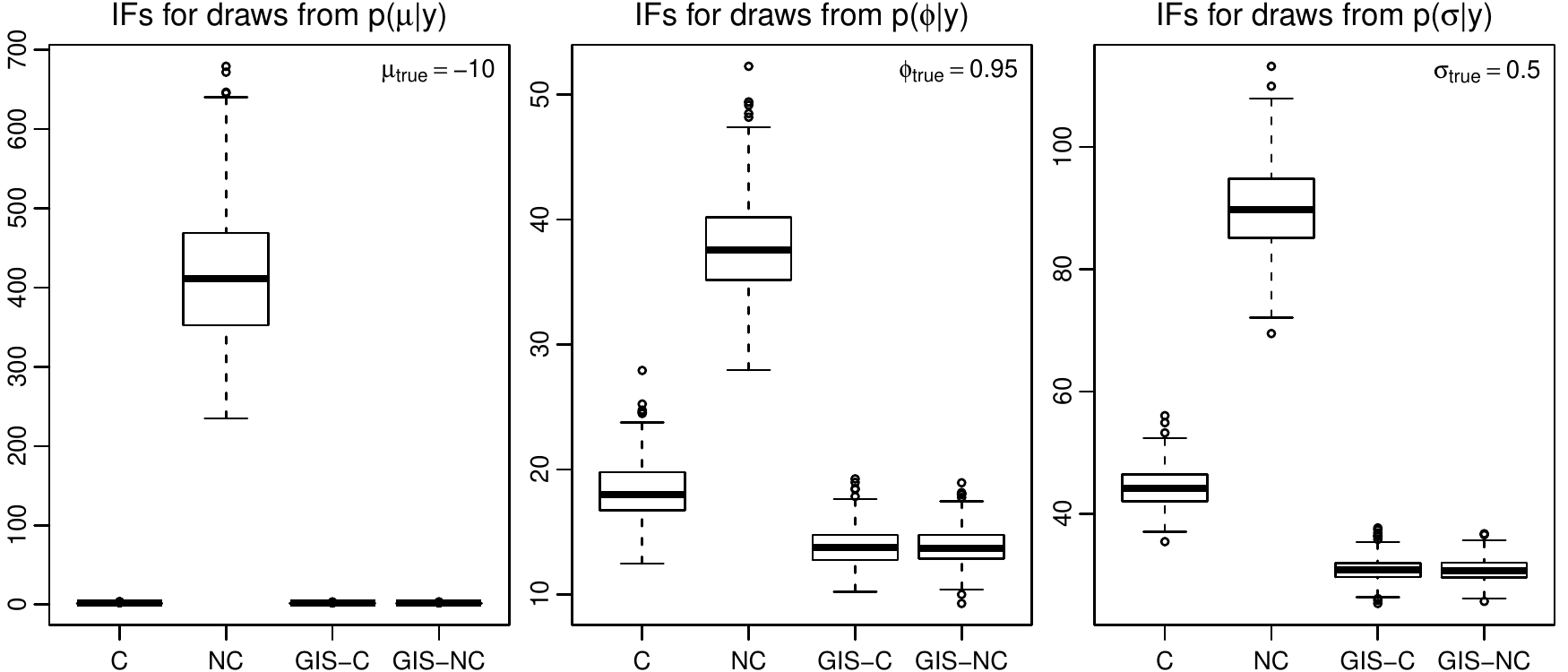}
  \caption{Boxplots of $500$ repeated measurements of inefficiency factors of $100\,000$ draws from the marginal densities. The underlying latent volatility process exhibits \emph{large volatility of volatility}.}
 \label{boxplots2}
\end{figure}

Even for artificially created datasets of length $T=5000$ or larger, estimation results may depend substantially on the actual realization of the underlying process. Also, other factors -- most importantly the initial seed for drawing pseudo random variables in the individual MCMC steps -- can influence both sample statistics from the posterior distribution as well as sample statistics for evaluating simulation efficiency. To compensate for this fact, we repeat each simulation with $500$ independently generated artificial data sets. The boxplots provided in Figure~\ref{boxplots1} and Figure~\ref{boxplots2} illustrate the variation of IFs for the same parameter constellations as above. While the overall picture about non-centering remains the same, we can now observe some substantial deviation from the median for certain realizations. 
Furthermore, the plots show that both interweaving strategies GIS-C and GIS-NC help to avoid woe by working well no matter which raw parameterization fails.

\subsection{Efficiency Overview}

In order to gain insight into the entire parameter range of interest, Tables~\ref{ifmu}, \ref{ifphi} and \ref{ifsigma} provide a summary of median inefficiency factors across all $45$ parameter constellations.

\begin{table}[ht!]
\caption{Inefficiency factors for $100\,000$ draws from $p(\mu|\mathbf{y})$ in various parameterizations using different blocking strategies. Time series length $T=5000$, the values reported are medians of $500$ repetitions and $T_\text{CPU}$ denotes the median time to complete $1000$ iterations.\\[.5em] Shading: \begin{tabular}{|x{ 0.84 cm} x{0.84cm}x{0.84cm}x{0.84cm}x{0.84cm}x{0.84cm}x{0.84cm}x{0.84cm}x{0.84cm}x{0.84cm} x{ 0.84 cm}|}
   \hline
\cellcolor[rgb]{1, 1 , 1 } 0&\cellcolor[rgb]{1, 0.968 , 0.968 } 50&\cellcolor[rgb]{1, 0.936 , 0.936 } 100&\cellcolor[rgb]{1, 0.904 , 0.904 } 150&\cellcolor[rgb]{1, 0.872 , 0.872 } 200&\cellcolor[rgb]{1, 0.84 , 0.84 } 250&\cellcolor[rgb]{1, 0.808 , 0.808 } 300&\cellcolor[rgb]{1, 0.776 , 0.776 } 350&\cellcolor[rgb]{1, 0.744 , 0.744 } 400&\cellcolor[rgb]{1, 0.712 , 0.712 } 450&\cellcolor[rgb]{1, 0.68 , 0.68 } 500+ \tabularnewline\hline
   \end{tabular}}
\label{ifmu}
\begin{tabular}{ccrrrrrrrrr}
   \hline
$p(\mu|\mathbf{y})$ & \backslashbox{$\sigma_{\text{true}}$}{$\phi_{\text{true}}$} & 0 & 0.5 & 0.8 & 0.9 & 0.95 & 0.96 & 0.97 & 0.98 & 0.99 \\ 
   \hline
\hline
 & 0.1 & \cellcolor[rgb]{1,0.68,0.68}591 & \cellcolor[rgb]{1,0.85472,0.85472}227 & \cellcolor[rgb]{1,0.96032,0.96032}62 & \cellcolor[rgb]{1,0.98848,0.98848}18 & \cellcolor[rgb]{1,0.99616,0.99616}6 & \cellcolor[rgb]{1,0.9968,0.9968}5 & \cellcolor[rgb]{1,0.99808,0.99808}3 & \cellcolor[rgb]{1,0.99872,0.99872}2 & \cellcolor[rgb]{1,0.99808,0.99808}3 \\ 
   & 0.2 & \cellcolor[rgb]{1,0.8432,0.8432}245 & \cellcolor[rgb]{1,0.9424,0.9424}90 & \cellcolor[rgb]{1,0.98528,0.98528}23 & \cellcolor[rgb]{1,0.99424,0.99424}9 & \cellcolor[rgb]{1,0.99808,0.99808}3 & \cellcolor[rgb]{1,0.99872,0.99872}2 & \cellcolor[rgb]{1,0.99872,0.99872}2 & \cellcolor[rgb]{1,0.99872,0.99872}2 & \cellcolor[rgb]{1,0.99808,0.99808}3 \\ 
  {\bf C} (1-block) & 0.3 & \cellcolor[rgb]{1,0.92832,0.92832}112 & \cellcolor[rgb]{1,0.96928,0.96928}48 & \cellcolor[rgb]{1,0.9904,0.9904}15 & \cellcolor[rgb]{1,0.9968,0.9968}5 & \cellcolor[rgb]{1,0.99872,0.99872}2 & \cellcolor[rgb]{1,0.99872,0.99872}2 & \cellcolor[rgb]{1,0.99872,0.99872}2 & \cellcolor[rgb]{1,0.99872,0.99872}2 & \cellcolor[rgb]{1,0.99744,0.99744}4 \\ 
  $T_\text{CPU}= 2.30 $ & 0.4 & \cellcolor[rgb]{1,0.9552,0.9552}70 & \cellcolor[rgb]{1,0.97824,0.97824}34 & \cellcolor[rgb]{1,0.99296,0.99296}11 & \cellcolor[rgb]{1,0.99808,0.99808}3 & \cellcolor[rgb]{1,0.99872,0.99872}2 & \cellcolor[rgb]{1,0.99872,0.99872}2 & \cellcolor[rgb]{1,0.99872,0.99872}2 & \cellcolor[rgb]{1,0.99872,0.99872}2 & \cellcolor[rgb]{1,0.9968,0.9968}5 \\ 
   & 0.5 & \cellcolor[rgb]{1,0.96672,0.96672}52 & \cellcolor[rgb]{1,0.98336,0.98336}26 & \cellcolor[rgb]{1,0.99488,0.99488}8 & \cellcolor[rgb]{1,0.99808,0.99808}3 & \cellcolor[rgb]{1,0.99872,0.99872}2 & \cellcolor[rgb]{1,0.99872,0.99872}2 & \cellcolor[rgb]{1,0.99872,0.99872}2 & \cellcolor[rgb]{1,0.99872,0.99872}2 & \cellcolor[rgb]{1,0.9968,0.9968}5 \\ 
   \hline
 & 0.1 & \cellcolor[rgb]{1,0.68,0.68}641 & \cellcolor[rgb]{1,0.85728,0.85728}223 & \cellcolor[rgb]{1,0.95904,0.95904}64 & \cellcolor[rgb]{1,0.98848,0.98848}18 & \cellcolor[rgb]{1,0.99616,0.99616}6 & \cellcolor[rgb]{1,0.9968,0.9968}5 & \cellcolor[rgb]{1,0.99808,0.99808}3 & \cellcolor[rgb]{1,0.99872,0.99872}2 & \cellcolor[rgb]{1,0.99808,0.99808}3 \\ 
   & 0.2 & \cellcolor[rgb]{1,0.83808,0.83808}253 & \cellcolor[rgb]{1,0.94304,0.94304}89 & \cellcolor[rgb]{1,0.98592,0.98592}22 & \cellcolor[rgb]{1,0.99424,0.99424}9 & \cellcolor[rgb]{1,0.99808,0.99808}3 & \cellcolor[rgb]{1,0.99872,0.99872}2 & \cellcolor[rgb]{1,0.99872,0.99872}2 & \cellcolor[rgb]{1,0.99872,0.99872}2 & \cellcolor[rgb]{1,0.99808,0.99808}3 \\ 
  {\bf C} (2-block) & 0.3 & \cellcolor[rgb]{1,0.92768,0.92768}113 & \cellcolor[rgb]{1,0.96992,0.96992}47 & \cellcolor[rgb]{1,0.9904,0.9904}15 & \cellcolor[rgb]{1,0.9968,0.9968}5 & \cellcolor[rgb]{1,0.99872,0.99872}2 & \cellcolor[rgb]{1,0.99872,0.99872}2 & \cellcolor[rgb]{1,0.99872,0.99872}2 & \cellcolor[rgb]{1,0.99872,0.99872}2 & \cellcolor[rgb]{1,0.99744,0.99744}4 \\ 
  $T_\text{CPU}= 2.31 $ & 0.4 & \cellcolor[rgb]{1,0.95392,0.95392}72 & \cellcolor[rgb]{1,0.97888,0.97888}33 & \cellcolor[rgb]{1,0.99296,0.99296}11 & \cellcolor[rgb]{1,0.99808,0.99808}3 & \cellcolor[rgb]{1,0.99872,0.99872}2 & \cellcolor[rgb]{1,0.99872,0.99872}2 & \cellcolor[rgb]{1,0.99872,0.99872}2 & \cellcolor[rgb]{1,0.99872,0.99872}2 & \cellcolor[rgb]{1,0.9968,0.9968}5 \\ 
   & 0.5 & \cellcolor[rgb]{1,0.96672,0.96672}52 & \cellcolor[rgb]{1,0.98336,0.98336}26 & \cellcolor[rgb]{1,0.99488,0.99488}8 & \cellcolor[rgb]{1,0.99872,0.99872}2 & \cellcolor[rgb]{1,0.99872,0.99872}2 & \cellcolor[rgb]{1,0.99872,0.99872}2 & \cellcolor[rgb]{1,0.99872,0.99872}2 & \cellcolor[rgb]{1,0.99872,0.99872}2 & \cellcolor[rgb]{1,0.9968,0.9968}5 \\ 
   \hline
 & 0.1 & \cellcolor[rgb]{1,0.68,0.68}609 & \cellcolor[rgb]{1,0.84384,0.84384}244 & \cellcolor[rgb]{1,0.95712,0.95712}67 & \cellcolor[rgb]{1,0.9872,0.9872}20 & \cellcolor[rgb]{1,0.99616,0.99616}6 & \cellcolor[rgb]{1,0.9968,0.9968}5 & \cellcolor[rgb]{1,0.99744,0.99744}4 & \cellcolor[rgb]{1,0.99872,0.99872}2 & \cellcolor[rgb]{1,0.99872,0.99872}2 \\ 
   & 0.2 & \cellcolor[rgb]{1,0.84,0.84}250 & \cellcolor[rgb]{1,0.94304,0.94304}89 & \cellcolor[rgb]{1,0.98528,0.98528}23 & \cellcolor[rgb]{1,0.99488,0.99488}8 & \cellcolor[rgb]{1,0.99808,0.99808}3 & \cellcolor[rgb]{1,0.99872,0.99872}2 & \cellcolor[rgb]{1,0.99872,0.99872}2 & \cellcolor[rgb]{1,0.99936,0.99936}1 & \cellcolor[rgb]{1,0.99872,0.99872}2 \\ 
  {\bf C} (3-block) & 0.3 & \cellcolor[rgb]{1,0.92832,0.92832}112 & \cellcolor[rgb]{1,0.96992,0.96992}47 & \cellcolor[rgb]{1,0.99104,0.99104}14 & \cellcolor[rgb]{1,0.9968,0.9968}5 & \cellcolor[rgb]{1,0.99872,0.99872}2 & \cellcolor[rgb]{1,0.99872,0.99872}2 & \cellcolor[rgb]{1,0.99936,0.99936}1 & \cellcolor[rgb]{1,0.99936,0.99936}1 & \cellcolor[rgb]{1,0.99872,0.99872}2 \\ 
  $T_\text{CPU}= 2.31 $ & 0.4 & \cellcolor[rgb]{1,0.95584,0.95584}69 & \cellcolor[rgb]{1,0.97888,0.97888}33 & \cellcolor[rgb]{1,0.99296,0.99296}11 & \cellcolor[rgb]{1,0.99744,0.99744}4 & \cellcolor[rgb]{1,0.99872,0.99872}2 & \cellcolor[rgb]{1,0.99936,0.99936}1 & \cellcolor[rgb]{1,0.99936,0.99936}1 & \cellcolor[rgb]{1,0.99872,0.99872}2 & \cellcolor[rgb]{1,0.99872,0.99872}2 \\ 
   & 0.5 & \cellcolor[rgb]{1,0.96736,0.96736}51 & \cellcolor[rgb]{1,0.984,0.984}25 & \cellcolor[rgb]{1,0.99488,0.99488}8 & \cellcolor[rgb]{1,0.99808,0.99808}3 & \cellcolor[rgb]{1,0.99872,0.99872}2 & \cellcolor[rgb]{1,0.99936,0.99936}1 & \cellcolor[rgb]{1,0.99936,0.99936}1 & \cellcolor[rgb]{1,0.99872,0.99872}2 & \cellcolor[rgb]{1,0.99872,0.99872}2 \\ 
   \hline
\hline
 & 0.1 & \cellcolor[rgb]{1,0.99424,0.99424}9 & \cellcolor[rgb]{1,0.9936,0.9936}10 & \cellcolor[rgb]{1,0.99232,0.99232}12 & \cellcolor[rgb]{1,0.99168,0.99168}13 & \cellcolor[rgb]{1,0.98656,0.98656}21 & \cellcolor[rgb]{1,0.9808,0.9808}30 & \cellcolor[rgb]{1,0.968,0.968}50 & \cellcolor[rgb]{1,0.92768,0.92768}113 & \cellcolor[rgb]{1,0.68832,0.68832}487 \\ 
   & 0.2 & \cellcolor[rgb]{1,0.98464,0.98464}24 & \cellcolor[rgb]{1,0.98656,0.98656}21 & \cellcolor[rgb]{1,0.9904,0.9904}15 & \cellcolor[rgb]{1,0.98592,0.98592}22 & \cellcolor[rgb]{1,0.9552,0.9552}70 & \cellcolor[rgb]{1,0.93088,0.93088}108 & \cellcolor[rgb]{1,0.8784,0.8784}190 & \cellcolor[rgb]{1,0.73184,0.73184}419 & \cellcolor[rgb]{1,0.68,0.68}1729 \\ 
  {\bf NC} (2-block) & 0.3 & \cellcolor[rgb]{1,0.98528,0.98528}23 & \cellcolor[rgb]{1,0.98912,0.98912}17 & \cellcolor[rgb]{1,0.98912,0.98912}17 & \cellcolor[rgb]{1,0.97376,0.97376}41 & \cellcolor[rgb]{1,0.90528,0.90528}148 & \cellcolor[rgb]{1,0.85024,0.85024}234 & \cellcolor[rgb]{1,0.73952,0.73952}407 & \cellcolor[rgb]{1,0.68,0.68}922 & \cellcolor[rgb]{1,0.68,0.68}3743 \\ 
  $T_\text{CPU}= 2.34 $ & 0.4 & \cellcolor[rgb]{1,0.98848,0.98848}18 & \cellcolor[rgb]{1,0.98976,0.98976}16 & \cellcolor[rgb]{1,0.98528,0.98528}23 & \cellcolor[rgb]{1,0.9552,0.9552}70 & \cellcolor[rgb]{1,0.8304,0.8304}265 & \cellcolor[rgb]{1,0.73632,0.73632}412 & \cellcolor[rgb]{1,0.68,0.68}726 & \cellcolor[rgb]{1,0.68,0.68}1707 & \cellcolor[rgb]{1,0.68,0.68}6790 \\ 
   & 0.5 & \cellcolor[rgb]{1,0.98912,0.98912}17 & \cellcolor[rgb]{1,0.98912,0.98912}17 & \cellcolor[rgb]{1,0.97952,0.97952}32 & \cellcolor[rgb]{1,0.9328,0.9328}105 & \cellcolor[rgb]{1,0.73696,0.73696}411 & \cellcolor[rgb]{1,0.68,0.68}660 & \cellcolor[rgb]{1,0.68,0.68}1149 & \cellcolor[rgb]{1,0.68,0.68}2534 & \cellcolor[rgb]{1,0.68,0.68}9421 \\ 
   \hline
 & 0.1 & \cellcolor[rgb]{1,0.99424,0.99424}9 & \cellcolor[rgb]{1,0.9936,0.9936}10 & \cellcolor[rgb]{1,0.99168,0.99168}13 & \cellcolor[rgb]{1,0.99168,0.99168}13 & \cellcolor[rgb]{1,0.98656,0.98656}21 & \cellcolor[rgb]{1,0.98016,0.98016}31 & \cellcolor[rgb]{1,0.968,0.968}50 & \cellcolor[rgb]{1,0.92576,0.92576}116 & \cellcolor[rgb]{1,0.68,0.68}516 \\ 
   & 0.2 & \cellcolor[rgb]{1,0.98464,0.98464}24 & \cellcolor[rgb]{1,0.98656,0.98656}21 & \cellcolor[rgb]{1,0.99104,0.99104}14 & \cellcolor[rgb]{1,0.98592,0.98592}22 & \cellcolor[rgb]{1,0.95456,0.95456}71 & \cellcolor[rgb]{1,0.93024,0.93024}109 & \cellcolor[rgb]{1,0.87776,0.87776}191 & \cellcolor[rgb]{1,0.72032,0.72032}437 & \cellcolor[rgb]{1,0.68,0.68}1883 \\ 
  {\bf NC} (3-block) & 0.3 & \cellcolor[rgb]{1,0.98528,0.98528}23 & \cellcolor[rgb]{1,0.98912,0.98912}17 & \cellcolor[rgb]{1,0.98912,0.98912}17 & \cellcolor[rgb]{1,0.97376,0.97376}41 & \cellcolor[rgb]{1,0.90464,0.90464}149 & \cellcolor[rgb]{1,0.8464,0.8464}240 & \cellcolor[rgb]{1,0.73376,0.73376}416 & \cellcolor[rgb]{1,0.68,0.68}935 & \cellcolor[rgb]{1,0.68,0.68}3843 \\ 
  $T_\text{CPU}= 2.35 $ & 0.4 & \cellcolor[rgb]{1,0.98848,0.98848}18 & \cellcolor[rgb]{1,0.98976,0.98976}16 & \cellcolor[rgb]{1,0.98528,0.98528}23 & \cellcolor[rgb]{1,0.9552,0.9552}70 & \cellcolor[rgb]{1,0.82848,0.82848}268 & \cellcolor[rgb]{1,0.7376,0.7376}410 & \cellcolor[rgb]{1,0.68,0.68}735 & \cellcolor[rgb]{1,0.68,0.68}1722 & \cellcolor[rgb]{1,0.68,0.68}7152 \\ 
   & 0.5 & \cellcolor[rgb]{1,0.98912,0.98912}17 & \cellcolor[rgb]{1,0.98912,0.98912}17 & \cellcolor[rgb]{1,0.97952,0.97952}32 & \cellcolor[rgb]{1,0.93216,0.93216}106 & \cellcolor[rgb]{1,0.73824,0.73824}409 & \cellcolor[rgb]{1,0.68,0.68}661 & \cellcolor[rgb]{1,0.68,0.68}1151 & \cellcolor[rgb]{1,0.68,0.68}2589 & \cellcolor[rgb]{1,0.68,0.68}10355 \\ 
   \hline
\hline
 & 0.1 & \cellcolor[rgb]{1,0.99424,0.99424}9 & \cellcolor[rgb]{1,0.99424,0.99424}9 & \cellcolor[rgb]{1,0.99296,0.99296}11 & \cellcolor[rgb]{1,0.99488,0.99488}8 & \cellcolor[rgb]{1,0.99744,0.99744}4 & \cellcolor[rgb]{1,0.99808,0.99808}3 & \cellcolor[rgb]{1,0.99808,0.99808}3 & \cellcolor[rgb]{1,0.99872,0.99872}2 & \cellcolor[rgb]{1,0.99808,0.99808}3 \\ 
   & 0.2 & \cellcolor[rgb]{1,0.98528,0.98528}23 & \cellcolor[rgb]{1,0.9872,0.9872}20 & \cellcolor[rgb]{1,0.99296,0.99296}11 & \cellcolor[rgb]{1,0.9968,0.9968}5 & \cellcolor[rgb]{1,0.99872,0.99872}2 & \cellcolor[rgb]{1,0.99872,0.99872}2 & \cellcolor[rgb]{1,0.99872,0.99872}2 & \cellcolor[rgb]{1,0.99872,0.99872}2 & \cellcolor[rgb]{1,0.99808,0.99808}3 \\ 
  {\bf GIS-C} (2-block) & 0.3 & \cellcolor[rgb]{1,0.98592,0.98592}22 & \cellcolor[rgb]{1,0.9904,0.9904}15 & \cellcolor[rgb]{1,0.99424,0.99424}9 & \cellcolor[rgb]{1,0.99744,0.99744}4 & \cellcolor[rgb]{1,0.99872,0.99872}2 & \cellcolor[rgb]{1,0.99872,0.99872}2 & \cellcolor[rgb]{1,0.99872,0.99872}2 & \cellcolor[rgb]{1,0.99872,0.99872}2 & \cellcolor[rgb]{1,0.99808,0.99808}3 \\ 
  $T_\text{CPU}= 2.36 $ & 0.4 & \cellcolor[rgb]{1,0.98912,0.98912}17 & \cellcolor[rgb]{1,0.99168,0.99168}13 & \cellcolor[rgb]{1,0.99552,0.99552}7 & \cellcolor[rgb]{1,0.99808,0.99808}3 & \cellcolor[rgb]{1,0.99872,0.99872}2 & \cellcolor[rgb]{1,0.99872,0.99872}2 & \cellcolor[rgb]{1,0.99872,0.99872}2 & \cellcolor[rgb]{1,0.99872,0.99872}2 & \cellcolor[rgb]{1,0.99744,0.99744}4 \\ 
   & 0.5 & \cellcolor[rgb]{1,0.9904,0.9904}15 & \cellcolor[rgb]{1,0.99232,0.99232}12 & \cellcolor[rgb]{1,0.9968,0.9968}5 & \cellcolor[rgb]{1,0.99872,0.99872}2 & \cellcolor[rgb]{1,0.99872,0.99872}2 & \cellcolor[rgb]{1,0.99872,0.99872}2 & \cellcolor[rgb]{1,0.99872,0.99872}2 & \cellcolor[rgb]{1,0.99872,0.99872}2 & \cellcolor[rgb]{1,0.99744,0.99744}4 \\ 
   \hline
 & 0.1 & \cellcolor[rgb]{1,0.99424,0.99424}9 & \cellcolor[rgb]{1,0.99424,0.99424}9 & \cellcolor[rgb]{1,0.99232,0.99232}12 & \cellcolor[rgb]{1,0.99488,0.99488}8 & \cellcolor[rgb]{1,0.99744,0.99744}4 & \cellcolor[rgb]{1,0.99808,0.99808}3 & \cellcolor[rgb]{1,0.99872,0.99872}2 & \cellcolor[rgb]{1,0.99872,0.99872}2 & \cellcolor[rgb]{1,0.99872,0.99872}2 \\ 
   & 0.2 & \cellcolor[rgb]{1,0.98528,0.98528}23 & \cellcolor[rgb]{1,0.9872,0.9872}20 & \cellcolor[rgb]{1,0.99296,0.99296}11 & \cellcolor[rgb]{1,0.9968,0.9968}5 & \cellcolor[rgb]{1,0.99872,0.99872}2 & \cellcolor[rgb]{1,0.99872,0.99872}2 & \cellcolor[rgb]{1,0.99872,0.99872}2 & \cellcolor[rgb]{1,0.99936,0.99936}1 & \cellcolor[rgb]{1,0.99936,0.99936}1 \\ 
  {\bf GIS-C} (3-block) & 0.3 & \cellcolor[rgb]{1,0.98592,0.98592}22 & \cellcolor[rgb]{1,0.9904,0.9904}15 & \cellcolor[rgb]{1,0.99424,0.99424}9 & \cellcolor[rgb]{1,0.99808,0.99808}3 & \cellcolor[rgb]{1,0.99872,0.99872}2 & \cellcolor[rgb]{1,0.99936,0.99936}1 & \cellcolor[rgb]{1,0.99936,0.99936}1 & \cellcolor[rgb]{1,0.99936,0.99936}1 & \cellcolor[rgb]{1,0.99872,0.99872}2 \\ 
  $T_\text{CPU}= 2.37 $ & 0.4 & \cellcolor[rgb]{1,0.98912,0.98912}17 & \cellcolor[rgb]{1,0.99168,0.99168}13 & \cellcolor[rgb]{1,0.99552,0.99552}7 & \cellcolor[rgb]{1,0.99808,0.99808}3 & \cellcolor[rgb]{1,0.99936,0.99936}1 & \cellcolor[rgb]{1,0.99936,0.99936}1 & \cellcolor[rgb]{1,0.99936,0.99936}1 & \cellcolor[rgb]{1,0.99936,0.99936}1 & \cellcolor[rgb]{1,0.99872,0.99872}2 \\ 
   & 0.5 & \cellcolor[rgb]{1,0.9904,0.9904}15 & \cellcolor[rgb]{1,0.99232,0.99232}12 & \cellcolor[rgb]{1,0.9968,0.9968}5 & \cellcolor[rgb]{1,0.99872,0.99872}2 & \cellcolor[rgb]{1,0.99936,0.99936}1 & \cellcolor[rgb]{1,0.99936,0.99936}1 & \cellcolor[rgb]{1,0.99936,0.99936}1 & \cellcolor[rgb]{1,0.99872,0.99872}2 & \cellcolor[rgb]{1,0.99872,0.99872}2 \\ 
   \hline
\hline
 & 0.1 & \cellcolor[rgb]{1,0.99424,0.99424}9 & \cellcolor[rgb]{1,0.99424,0.99424}9 & \cellcolor[rgb]{1,0.99296,0.99296}11 & \cellcolor[rgb]{1,0.99488,0.99488}8 & \cellcolor[rgb]{1,0.99744,0.99744}4 & \cellcolor[rgb]{1,0.99808,0.99808}3 & \cellcolor[rgb]{1,0.99808,0.99808}3 & \cellcolor[rgb]{1,0.99872,0.99872}2 & \cellcolor[rgb]{1,0.99808,0.99808}3 \\ 
   & 0.2 & \cellcolor[rgb]{1,0.98528,0.98528}23 & \cellcolor[rgb]{1,0.9872,0.9872}20 & \cellcolor[rgb]{1,0.99296,0.99296}11 & \cellcolor[rgb]{1,0.9968,0.9968}5 & \cellcolor[rgb]{1,0.99872,0.99872}2 & \cellcolor[rgb]{1,0.99872,0.99872}2 & \cellcolor[rgb]{1,0.99872,0.99872}2 & \cellcolor[rgb]{1,0.99872,0.99872}2 & \cellcolor[rgb]{1,0.99808,0.99808}3 \\ 
  {\bf GIS-NC} (2-block) & 0.3 & \cellcolor[rgb]{1,0.98592,0.98592}22 & \cellcolor[rgb]{1,0.98976,0.98976}16 & \cellcolor[rgb]{1,0.99424,0.99424}9 & \cellcolor[rgb]{1,0.99808,0.99808}3 & \cellcolor[rgb]{1,0.99872,0.99872}2 & \cellcolor[rgb]{1,0.99872,0.99872}2 & \cellcolor[rgb]{1,0.99872,0.99872}2 & \cellcolor[rgb]{1,0.99872,0.99872}2 & \cellcolor[rgb]{1,0.99808,0.99808}3 \\ 
  $T_\text{CPU}= 2.40 $ & 0.4 & \cellcolor[rgb]{1,0.98912,0.98912}17 & \cellcolor[rgb]{1,0.99168,0.99168}13 & \cellcolor[rgb]{1,0.99552,0.99552}7 & \cellcolor[rgb]{1,0.99808,0.99808}3 & \cellcolor[rgb]{1,0.99872,0.99872}2 & \cellcolor[rgb]{1,0.99872,0.99872}2 & \cellcolor[rgb]{1,0.99872,0.99872}2 & \cellcolor[rgb]{1,0.99872,0.99872}2 & \cellcolor[rgb]{1,0.99744,0.99744}4 \\ 
   & 0.5 & \cellcolor[rgb]{1,0.9904,0.9904}15 & \cellcolor[rgb]{1,0.99232,0.99232}12 & \cellcolor[rgb]{1,0.9968,0.9968}5 & \cellcolor[rgb]{1,0.99872,0.99872}2 & \cellcolor[rgb]{1,0.99872,0.99872}2 & \cellcolor[rgb]{1,0.99872,0.99872}2 & \cellcolor[rgb]{1,0.99872,0.99872}2 & \cellcolor[rgb]{1,0.99872,0.99872}2 & \cellcolor[rgb]{1,0.99744,0.99744}4 \\ 
   \hline
   \hline
\end{tabular}
\end{table}

 Median IFs obtained from draws from $p(\mu|\mathbf{y})$ in Table~\ref{ifmu} confirm clearly that the centered parameterization is quite capable of efficiently estimating the level $\mu$ of the latent process throughout a wide parameter range, no matter which blocking strategy is used. Only a combination of both small $\sigma_\text{true}$ and small $\phi_\text{true}$ leads to large inefficiency. As was to be expected, this is exactly the area where the non-centered parameterization performs comparably well; median IFs are small to moderately large. 
On the other end of the scale -- where we find both highly persistent and highly varying latent variables -- NC becomes close to useless with very large IFs of $1000$ and above for both blocking strategies.

The lower three panels of Table~\ref{ifmu} show the performance of the interwoven samplers with different baselines, GIS-C and GIS-NC. It stands out that in terms of simulation efficiency, both variants are always better than or en par with the ideal parameterization, while there are practically no differences between the sampler with baseline C and the one with baseline NC.
Comparing CPU time of the raw samplers with their interwoven counterparts reveals the computational cost of interweaving, which amounts to merely around $2\%$ in our setup. Thus, even when taking into account the extra cost, interweaving is hardly ever a bad choice. Also note that GIS-C is practically as fast as NC.

Table~\ref{gainmu} shows a direct comparison of the interwoven sampler with the raw parameterizations in terms of increase in effective sample size. All numbers are positive, showing that interweaving is more efficient than the ideal parameterization, but sometimes only slightly. Note that in comparison to the suboptimal parameterization, GIS is always at least twice as effective.

\begin{table}[ht!]
\caption{Percentage gains in effective sample size for the 2-block and the 3-block sampler. First and third table: $\text{ESS}_\text{GIS-C}$ vs.~$\max(\text{ESS}_\text{C},\text{ESS}_\text{NC})$. Second and fourth table: $\text{ESS}_\text{GIS-C}$ vs.~$\min(\text{ESS}_\text{C},\text{ESS}_\text{NC})$.\\[.5em] Shading: \begin{tabular}{|x{ 0.84 cm} x{0.84cm}x{0.84cm}x{0.84cm}x{0.84cm}x{0.84cm}x{0.84cm}x{0.84cm}x{0.84cm}x{0.84cm} x{ 0.84 cm}|}
  \hline
\cellcolor[rgb]{ 1 ,1, 1 } 0&\cellcolor[rgb]{ 0.95 ,1, 0.95 } 10&\cellcolor[rgb]{ 0.9 ,1, 0.9 } 20&\cellcolor[rgb]{ 0.85 ,1, 0.85 } 30&\cellcolor[rgb]{ 0.8 ,1, 0.8 } 40&\cellcolor[rgb]{ 0.75 ,1, 0.75 } 50&\cellcolor[rgb]{ 0.7 ,1, 0.7 } 60&\cellcolor[rgb]{ 0.65 ,1, 0.65 } 70&\cellcolor[rgb]{ 0.6 ,1, 0.6 } 80&\cellcolor[rgb]{ 0.55 ,1, 0.55 } 90&\cellcolor[rgb]{ 0.5 ,1, 0.5 } 100+ \tabularnewline\hline
  \end{tabular} (1 and 3)\\Shading: \begin{tabular}{|x{ 0.84 cm} x{0.84cm}x{0.84cm}x{0.84cm}x{0.84cm}x{0.84cm}x{0.84cm}x{0.84cm}x{0.84cm}x{0.84cm} x{ 0.84 cm}|}
  \hline
\cellcolor[rgb]{ 1 , 1 ,1} 0&\cellcolor[rgb]{ 0.977 , 0.977 ,1} 100&\cellcolor[rgb]{ 0.954 , 0.954 ,1} 200&\cellcolor[rgb]{ 0.931 , 0.931 ,1} 300&\cellcolor[rgb]{ 0.908 , 0.908 ,1} 400&\cellcolor[rgb]{ 0.885 , 0.885 ,1} 500&\cellcolor[rgb]{ 0.862 , 0.862 ,1} 600&\cellcolor[rgb]{ 0.839 , 0.839 ,1} 700&\cellcolor[rgb]{ 0.816 , 0.816 ,1} 800&\cellcolor[rgb]{ 0.793 , 0.793 ,1} 900&\cellcolor[rgb]{ 0.77 , 0.77 ,1} 1000+ \tabularnewline\hline
  \end{tabular} (2 and 4)}
\label{gainmu}
\begin{tabular}{ccrrrrrrrrr}
   \hline
$p(\mu|\mathbf{y})$ & \backslashbox{$\sigma_{\text{true}}$}{$\phi_{\text{true}}$} & 0 & 0.5 & 0.8 & 0.9 & 0.95 & 0.96 & 0.97 & 0.98 & 0.99 \\ 
   \hline
\hline
 & 0.1 & \cellcolor[rgb]{0.99,1,0.99}2 & \cellcolor[rgb]{0.99,1,0.99}2 & \cellcolor[rgb]{0.955,1,0.955}9 & \cellcolor[rgb]{0.755,1,0.755}49 & \cellcolor[rgb]{0.735,1,0.735}53 & \cellcolor[rgb]{0.83,1,0.83}34 & \cellcolor[rgb]{0.895,1,0.895}21 & \cellcolor[rgb]{0.93,1,0.93}14 & \cellcolor[rgb]{0.945,1,0.945}11 \\ 
  {\bf GIS-C} vs. & 0.2 & \cellcolor[rgb]{0.98,1,0.98}4 & \cellcolor[rgb]{0.98,1,0.98}4 & \cellcolor[rgb]{0.82,1,0.82}36 & \cellcolor[rgb]{0.695,1,0.695}61 & \cellcolor[rgb]{0.91,1,0.91}18 & \cellcolor[rgb]{0.935,1,0.935}13 & \cellcolor[rgb]{0.955,1,0.955}9 & \cellcolor[rgb]{0.97,1,0.97}6 & \cellcolor[rgb]{0.925,1,0.925}15 \\ 
  {\bf better} & 0.3 & \cellcolor[rgb]{0.99,1,0.99}2 & \cellcolor[rgb]{0.935,1,0.935}13 & \cellcolor[rgb]{0.605,1,0.605}79 & \cellcolor[rgb]{0.805,1,0.805}39 & \cellcolor[rgb]{0.945,1,0.945}11 & \cellcolor[rgb]{0.955,1,0.955}9 & \cellcolor[rgb]{0.97,1,0.97}6 & \cellcolor[rgb]{0.97,1,0.97}6 & \cellcolor[rgb]{0.895,1,0.895}21 \\ 
  (2-block) & 0.4 & \cellcolor[rgb]{0.96,1,0.96}8 & \cellcolor[rgb]{0.875,1,0.875}25 & \cellcolor[rgb]{0.69,1,0.69}62 & \cellcolor[rgb]{0.885,1,0.885}23 & \cellcolor[rgb]{0.965,1,0.965}7 & \cellcolor[rgb]{0.975,1,0.975}5 & \cellcolor[rgb]{0.975,1,0.975}5 & \cellcolor[rgb]{0.965,1,0.965}7 & \cellcolor[rgb]{0.86,1,0.86}28 \\ 
   & 0.5 & \cellcolor[rgb]{0.935,1,0.935}13 & \cellcolor[rgb]{0.795,1,0.795}41 & \cellcolor[rgb]{0.765,1,0.765}47 & \cellcolor[rgb]{0.925,1,0.925}15 & \cellcolor[rgb]{0.975,1,0.975}5 & \cellcolor[rgb]{0.97,1,0.97}6 & \cellcolor[rgb]{0.975,1,0.975}5 & \cellcolor[rgb]{0.96,1,0.96}8 & \cellcolor[rgb]{0.84,1,0.84}32 \\ 
   \hline
 & 0.1 & \cellcolor[rgb]{0.77,0.77,1}7404 & \cellcolor[rgb]{0.77,0.77,1}2264 & \cellcolor[rgb]{0.89443,0.89443,1}459 & \cellcolor[rgb]{0.97401,0.97401,1}113 & \cellcolor[rgb]{0.90593,0.90593,1}409 & \cellcolor[rgb]{0.81669,0.81669,1}797 & \cellcolor[rgb]{0.77,0.77,1}1825 & \cellcolor[rgb]{0.77,0.77,1}5290 & \cellcolor[rgb]{0.77,0.77,1}16547 \\ 
  {\bf GIS-C} vs. & 0.2 & \cellcolor[rgb]{0.77023,0.77023,1}999 & \cellcolor[rgb]{0.92295,0.92295,1}335 & \cellcolor[rgb]{0.97539,0.97539,1}107 & \cellcolor[rgb]{0.93146,0.93146,1}298 & \cellcolor[rgb]{0.77,0.77,1}2924 & \cellcolor[rgb]{0.77,0.77,1}5415 & \cellcolor[rgb]{0.77,0.77,1}10731 & \cellcolor[rgb]{0.77,0.77,1}24612 & \cellcolor[rgb]{0.77,0.77,1}64832 \\ 
  {\bf worse} & 0.3 & \cellcolor[rgb]{0.90455,0.90455,1}415 & \cellcolor[rgb]{0.9517,0.9517,1}210 & \cellcolor[rgb]{0.97654,0.97654,1}102 & \cellcolor[rgb]{0.77,0.77,1}1070 & \cellcolor[rgb]{0.77,0.77,1}8051 & \cellcolor[rgb]{0.77,0.77,1}14082 & \cellcolor[rgb]{0.77,0.77,1}25470 & \cellcolor[rgb]{0.77,0.77,1}54664 & \cellcolor[rgb]{0.77,0.77,1}114292 \\ 
  (2-block) & 0.4 & \cellcolor[rgb]{0.92433,0.92433,1}329 & \cellcolor[rgb]{0.96458,0.96458,1}154 & \cellcolor[rgb]{0.9425,0.9425,1}250 & \cellcolor[rgb]{0.77,0.77,1}2588 & \cellcolor[rgb]{0.77,0.77,1}16212 & \cellcolor[rgb]{0.77,0.77,1}26504 & \cellcolor[rgb]{0.77,0.77,1}45144 & \cellcolor[rgb]{0.77,0.77,1}88863 & \cellcolor[rgb]{0.77,0.77,1}164695 \\ 
   & 0.5 & \cellcolor[rgb]{0.94296,0.94296,1}248 & \cellcolor[rgb]{0.97378,0.97378,1}114 & \cellcolor[rgb]{0.88293,0.88293,1}509 & \cellcolor[rgb]{0.77,0.77,1}4754 & \cellcolor[rgb]{0.77,0.77,1}26553 & \cellcolor[rgb]{0.77,0.77,1}43176 & \cellcolor[rgb]{0.77,0.77,1}72398 & \cellcolor[rgb]{0.77,0.77,1}121917 & \cellcolor[rgb]{0.77,0.77,1}240626 \\ 
   \hline
\hline
 & 0.1 & \cellcolor[rgb]{0.99,1,0.99}2 & \cellcolor[rgb]{0.99,1,0.99}2 & \cellcolor[rgb]{0.955,1,0.955}9 & \cellcolor[rgb]{0.76,1,0.76}48 & \cellcolor[rgb]{0.71,1,0.71}58 & \cellcolor[rgb]{0.725,1,0.725}55 & \cellcolor[rgb]{0.73,1,0.73}54 & \cellcolor[rgb]{0.825,1,0.825}35 & \cellcolor[rgb]{0.895,1,0.895}21 \\ 
  {\bf GIS-C} vs. & 0.2 & \cellcolor[rgb]{0.995,1,0.995}1 & \cellcolor[rgb]{0.98,1,0.98}4 & \cellcolor[rgb]{0.835,1,0.835}33 & \cellcolor[rgb]{0.725,1,0.725}55 & \cellcolor[rgb]{0.85,1,0.85}30 & \cellcolor[rgb]{0.9,1,0.9}20 & \cellcolor[rgb]{0.94,1,0.94}12 & \cellcolor[rgb]{0.965,1,0.965}7 & \cellcolor[rgb]{0.94,1,0.94}12 \\ 
  {\bf better} & 0.3 & \cellcolor[rgb]{0.97,1,0.97}6 & \cellcolor[rgb]{0.93,1,0.93}14 & \cellcolor[rgb]{0.675,1,0.675}65 & \cellcolor[rgb]{0.75,1,0.75}50 & \cellcolor[rgb]{0.905,1,0.905}19 & \cellcolor[rgb]{0.935,1,0.935}13 & \cellcolor[rgb]{0.965,1,0.965}7 & \cellcolor[rgb]{0.97,1,0.97}6 & \cellcolor[rgb]{0.93,1,0.93}14 \\ 
  (3-block) & 0.4 & \cellcolor[rgb]{0.96,1,0.96}8 & \cellcolor[rgb]{0.87,1,0.87}26 & \cellcolor[rgb]{0.685,1,0.685}63 & \cellcolor[rgb]{0.765,1,0.765}47 & \cellcolor[rgb]{0.93,1,0.93}14 & \cellcolor[rgb]{0.96,1,0.96}8 & \cellcolor[rgb]{0.975,1,0.975}5 & \cellcolor[rgb]{0.98,1,0.98}4 & \cellcolor[rgb]{0.945,1,0.945}11 \\ 
   & 0.5 & \cellcolor[rgb]{0.94,1,0.94}12 & \cellcolor[rgb]{0.8,1,0.8}40 & \cellcolor[rgb]{0.68,1,0.68}64 & \cellcolor[rgb]{0.785,1,0.785}43 & \cellcolor[rgb]{0.945,1,0.945}11 & \cellcolor[rgb]{0.97,1,0.97}6 & \cellcolor[rgb]{0.975,1,0.975}5 & \cellcolor[rgb]{0.99,1,0.99}2 & \cellcolor[rgb]{0.955,1,0.955}9 \\ 
   \hline
 & 0.1 & \cellcolor[rgb]{0.77,0.77,1}6897 & \cellcolor[rgb]{0.77,0.77,1}2475 & \cellcolor[rgb]{0.89006,0.89006,1}478 & \cellcolor[rgb]{0.96826,0.96826,1}138 & \cellcolor[rgb]{0.90156,0.90156,1}428 & \cellcolor[rgb]{0.80312,0.80312,1}856 & \cellcolor[rgb]{0.77,0.77,1}2032 & \cellcolor[rgb]{0.77,0.77,1}6553 & \cellcolor[rgb]{0.77,0.77,1}34199 \\ 
  {\bf GIS-C} vs. & 0.2 & \cellcolor[rgb]{0.77621,0.77621,1}973 & \cellcolor[rgb]{0.92272,0.92272,1}336 & \cellcolor[rgb]{0.97447,0.97447,1}111 & \cellcolor[rgb]{0.93123,0.93123,1}299 & \cellcolor[rgb]{0.77,0.77,1}3156 & \cellcolor[rgb]{0.77,0.77,1}5959 & \cellcolor[rgb]{0.77,0.77,1}12427 & \cellcolor[rgb]{0.77,0.77,1}32023 & \cellcolor[rgb]{0.77,0.77,1}129984 \\ 
  {\bf worse} & 0.3 & \cellcolor[rgb]{0.90409,0.90409,1}417 & \cellcolor[rgb]{0.95239,0.95239,1}207 & \cellcolor[rgb]{0.97746,0.97746,1}98 & \cellcolor[rgb]{0.77,0.77,1}1106 & \cellcolor[rgb]{0.77,0.77,1}8806 & \cellcolor[rgb]{0.77,0.77,1}16048 & \cellcolor[rgb]{0.77,0.77,1}30069 & \cellcolor[rgb]{0.77,0.77,1}70163 & \cellcolor[rgb]{0.77,0.77,1}220819 \\ 
  (3-block) & 0.4 & \cellcolor[rgb]{0.92801,0.92801,1}313 & \cellcolor[rgb]{0.9655,0.9655,1}150 & \cellcolor[rgb]{0.9425,0.9425,1}250 & \cellcolor[rgb]{0.77,0.77,1}2679 & \cellcolor[rgb]{0.77,0.77,1}17978 & \cellcolor[rgb]{0.77,0.77,1}29935 & \cellcolor[rgb]{0.77,0.77,1}53886 & \cellcolor[rgb]{0.77,0.77,1}116617 & \cellcolor[rgb]{0.77,0.77,1}373912 \\ 
   & 0.5 & \cellcolor[rgb]{0.94411,0.94411,1}243 & \cellcolor[rgb]{0.97516,0.97516,1}108 & \cellcolor[rgb]{0.88201,0.88201,1}513 & \cellcolor[rgb]{0.77,0.77,1}4933 & \cellcolor[rgb]{0.77,0.77,1}29265 & \cellcolor[rgb]{0.77,0.77,1}48629 & \cellcolor[rgb]{0.77,0.77,1}85550 & \cellcolor[rgb]{0.77,0.77,1}167572 & \cellcolor[rgb]{0.77,0.77,1}533651 \\ 
   \hline
   \hline
\end{tabular}
\end{table}

\begin{table}[ht!]
\caption{Inefficiency factors for $100\,000$ draws from $p(\phi|\mathbf{y})$ in various parameterizations using different blocking strategies. Time series length $T=5000$, the values reported are medians of $500$ repetitions and $T_\text{CPU}$ denotes the median time to complete $1000$ iterations.\\[.5em] Shading: \begin{tabular}{|x{ 0.84 cm} x{0.84cm}x{0.84cm}x{0.84cm}x{0.84cm}x{0.84cm}x{0.84cm}x{0.84cm}x{0.84cm}x{0.84cm} x{ 0.84 cm}|}
   \hline
\cellcolor[rgb]{1, 1 , 1 } 0&\cellcolor[rgb]{1, 0.968 , 0.968 } 50&\cellcolor[rgb]{1, 0.936 , 0.936 } 100&\cellcolor[rgb]{1, 0.904 , 0.904 } 150&\cellcolor[rgb]{1, 0.872 , 0.872 } 200&\cellcolor[rgb]{1, 0.84 , 0.84 } 250&\cellcolor[rgb]{1, 0.808 , 0.808 } 300&\cellcolor[rgb]{1, 0.776 , 0.776 } 350&\cellcolor[rgb]{1, 0.744 , 0.744 } 400&\cellcolor[rgb]{1, 0.712 , 0.712 } 450&\cellcolor[rgb]{1, 0.68 , 0.68 } 500+ \tabularnewline\hline
   \end{tabular}}
\label{ifphi}
\begin{tabular}{ccrrrrrrrrr}
   \hline
$p(\phi|\mathbf{y})$ & \backslashbox{$\sigma_{\text{true}}$}{$\phi_{\text{true}}$} & 0 & 0.5 & 0.8 & 0.9 & 0.95 & 0.96 & 0.97 & 0.98 & 0.99 \\ 
   \hline
\hline
 & 0.1 & \cellcolor[rgb]{1,0.91168,0.91168}138 & \cellcolor[rgb]{1,0.87008,0.87008}203 & \cellcolor[rgb]{1,0.84768,0.84768}238 & \cellcolor[rgb]{1,0.79712,0.79712}317 & \cellcolor[rgb]{1,0.76512,0.76512}367 & \cellcolor[rgb]{1,0.78304,0.78304}339 & \cellcolor[rgb]{1,0.83232,0.83232}262 & \cellcolor[rgb]{1,0.8912,0.8912}170 & \cellcolor[rgb]{1,0.9488,0.9488}80 \\ 
   & 0.2 & \cellcolor[rgb]{1,0.91296,0.91296}136 & \cellcolor[rgb]{1,0.8752,0.8752}195 & \cellcolor[rgb]{1,0.81504,0.81504}289 & \cellcolor[rgb]{1,0.84576,0.84576}241 & \cellcolor[rgb]{1,0.92768,0.92768}113 & \cellcolor[rgb]{1,0.94304,0.94304}89 & \cellcolor[rgb]{1,0.95776,0.95776}66 & \cellcolor[rgb]{1,0.97248,0.97248}43 & \cellcolor[rgb]{1,0.98336,0.98336}26 \\ 
  {\bf C} (1-block) & 0.3 & \cellcolor[rgb]{1,0.91744,0.91744}129 & \cellcolor[rgb]{1,0.88992,0.88992}172 & \cellcolor[rgb]{1,0.86624,0.86624}209 & \cellcolor[rgb]{1,0.93344,0.93344}104 & \cellcolor[rgb]{1,0.96864,0.96864}49 & \cellcolor[rgb]{1,0.9744,0.9744}40 & \cellcolor[rgb]{1,0.9808,0.9808}30 & \cellcolor[rgb]{1,0.98592,0.98592}22 & \cellcolor[rgb]{1,0.9904,0.9904}15 \\ 
  $T_\text{CPU}= 2.30 $ & 0.4 & \cellcolor[rgb]{1,0.92768,0.92768}113 & \cellcolor[rgb]{1,0.90656,0.90656}146 & \cellcolor[rgb]{1,0.9264,0.9264}115 & \cellcolor[rgb]{1,0.96352,0.96352}57 & \cellcolor[rgb]{1,0.98208,0.98208}28 & \cellcolor[rgb]{1,0.98528,0.98528}23 & \cellcolor[rgb]{1,0.98784,0.98784}19 & \cellcolor[rgb]{1,0.99104,0.99104}14 & \cellcolor[rgb]{1,0.99168,0.99168}13 \\ 
   & 0.5 & \cellcolor[rgb]{1,0.93856,0.93856}96 & \cellcolor[rgb]{1,0.92832,0.92832}112 & \cellcolor[rgb]{1,0.95328,0.95328}73 & \cellcolor[rgb]{1,0.97696,0.97696}36 & \cellcolor[rgb]{1,0.98784,0.98784}19 & \cellcolor[rgb]{1,0.98976,0.98976}16 & \cellcolor[rgb]{1,0.99168,0.99168}13 & \cellcolor[rgb]{1,0.99296,0.99296}11 & \cellcolor[rgb]{1,0.99296,0.99296}11 \\ 
   \hline
 & 0.1 & \cellcolor[rgb]{1,0.91104,0.91104}139 & \cellcolor[rgb]{1,0.87136,0.87136}201 & \cellcolor[rgb]{1,0.85024,0.85024}234 & \cellcolor[rgb]{1,0.79648,0.79648}318 & \cellcolor[rgb]{1,0.7696,0.7696}360 & \cellcolor[rgb]{1,0.79584,0.79584}319 & \cellcolor[rgb]{1,0.84576,0.84576}241 & \cellcolor[rgb]{1,0.90272,0.90272}152 & \cellcolor[rgb]{1,0.95584,0.95584}69 \\ 
   & 0.2 & \cellcolor[rgb]{1,0.91232,0.91232}137 & \cellcolor[rgb]{1,0.87584,0.87584}194 & \cellcolor[rgb]{1,0.81568,0.81568}288 & \cellcolor[rgb]{1,0.85152,0.85152}232 & \cellcolor[rgb]{1,0.93216,0.93216}106 & \cellcolor[rgb]{1,0.94752,0.94752}82 & \cellcolor[rgb]{1,0.96224,0.96224}59 & \cellcolor[rgb]{1,0.97568,0.97568}38 & \cellcolor[rgb]{1,0.98592,0.98592}22 \\ 
  {\bf C} (2-block) & 0.3 & \cellcolor[rgb]{1,0.91744,0.91744}129 & \cellcolor[rgb]{1,0.89056,0.89056}171 & \cellcolor[rgb]{1,0.87008,0.87008}203 & \cellcolor[rgb]{1,0.936,0.936}100 & \cellcolor[rgb]{1,0.97056,0.97056}46 & \cellcolor[rgb]{1,0.97696,0.97696}36 & \cellcolor[rgb]{1,0.98272,0.98272}27 & \cellcolor[rgb]{1,0.98784,0.98784}19 & \cellcolor[rgb]{1,0.99168,0.99168}13 \\ 
  $T_\text{CPU}= 2.31 $ & 0.4 & \cellcolor[rgb]{1,0.92768,0.92768}113 & \cellcolor[rgb]{1,0.90528,0.90528}148 & \cellcolor[rgb]{1,0.92832,0.92832}112 & \cellcolor[rgb]{1,0.96544,0.96544}54 & \cellcolor[rgb]{1,0.98336,0.98336}26 & \cellcolor[rgb]{1,0.98592,0.98592}22 & \cellcolor[rgb]{1,0.98912,0.98912}17 & \cellcolor[rgb]{1,0.99232,0.99232}12 & \cellcolor[rgb]{1,0.99296,0.99296}11 \\ 
   & 0.5 & \cellcolor[rgb]{1,0.93856,0.93856}96 & \cellcolor[rgb]{1,0.92768,0.92768}113 & \cellcolor[rgb]{1,0.95456,0.95456}71 & \cellcolor[rgb]{1,0.9776,0.9776}35 & \cellcolor[rgb]{1,0.98848,0.98848}18 & \cellcolor[rgb]{1,0.9904,0.9904}15 & \cellcolor[rgb]{1,0.99232,0.99232}12 & \cellcolor[rgb]{1,0.99424,0.99424}9 & \cellcolor[rgb]{1,0.9936,0.9936}10 \\ 
   \hline
 & 0.1 & \cellcolor[rgb]{1,0.68,0.68}29900 & \cellcolor[rgb]{1,0.68,0.68}29980 & \cellcolor[rgb]{1,0.68,0.68}35443 & \cellcolor[rgb]{1,0.68,0.68}33223 & \cellcolor[rgb]{1,0.68,0.68}13461 & \cellcolor[rgb]{1,0.68,0.68}9054 & \cellcolor[rgb]{1,0.68,0.68}4862 & \cellcolor[rgb]{1,0.68,0.68}2347 & \cellcolor[rgb]{1,0.68,0.68}719 \\ 
   & 0.2 & \cellcolor[rgb]{1,0.68,0.68}29607 & \cellcolor[rgb]{1,0.68,0.68}30722 & \cellcolor[rgb]{1,0.68,0.68}27821 & \cellcolor[rgb]{1,0.68,0.68}8991 & \cellcolor[rgb]{1,0.68,0.68}1799 & \cellcolor[rgb]{1,0.68,0.68}1142 & \cellcolor[rgb]{1,0.68,0.68}685 & \cellcolor[rgb]{1,0.76768,0.76768}363 & \cellcolor[rgb]{1,0.90144,0.90144}154 \\ 
  {\bf C} (3-block) & 0.3 & \cellcolor[rgb]{1,0.68,0.68}26975 & \cellcolor[rgb]{1,0.68,0.68}26012 & \cellcolor[rgb]{1,0.68,0.68}11051 & \cellcolor[rgb]{1,0.68,0.68}2212 & \cellcolor[rgb]{1,0.68,0.68}564 & \cellcolor[rgb]{1,0.76192,0.76192}372 & \cellcolor[rgb]{1,0.84576,0.84576}241 & \cellcolor[rgb]{1,0.90976,0.90976}141 & \cellcolor[rgb]{1,0.95584,0.95584}69 \\ 
  $T_\text{CPU}= 2.31 $ & 0.4 & \cellcolor[rgb]{1,0.68,0.68}23672 & \cellcolor[rgb]{1,0.68,0.68}20661 & \cellcolor[rgb]{1,0.68,0.68}4382 & \cellcolor[rgb]{1,0.68,0.68}882 & \cellcolor[rgb]{1,0.84064,0.84064}249 & \cellcolor[rgb]{1,0.88416,0.88416}181 & \cellcolor[rgb]{1,0.92192,0.92192}122 & \cellcolor[rgb]{1,0.95136,0.95136}76 & \cellcolor[rgb]{1,0.97312,0.97312}42 \\ 
   & 0.5 & \cellcolor[rgb]{1,0.68,0.68}18800 & \cellcolor[rgb]{1,0.68,0.68}13056 & \cellcolor[rgb]{1,0.68,0.68}1990 & \cellcolor[rgb]{1,0.71264,0.71264}449 & \cellcolor[rgb]{1,0.90656,0.90656}146 & \cellcolor[rgb]{1,0.9328,0.9328}105 & \cellcolor[rgb]{1,0.95264,0.95264}74 & \cellcolor[rgb]{1,0.968,0.968}50 & \cellcolor[rgb]{1,0.9808,0.9808}30 \\ 
   \hline
\hline
 & 0.1 & \cellcolor[rgb]{1,0.91168,0.91168}138 & \cellcolor[rgb]{1,0.87136,0.87136}201 & \cellcolor[rgb]{1,0.85536,0.85536}226 & \cellcolor[rgb]{1,0.89184,0.89184}169 & \cellcolor[rgb]{1,0.9456,0.9456}85 & \cellcolor[rgb]{1,0.95264,0.95264}74 & \cellcolor[rgb]{1,0.96096,0.96096}61 & \cellcolor[rgb]{1,0.96608,0.96608}53 & \cellcolor[rgb]{1,0.96288,0.96288}58 \\ 
   & 0.2 & \cellcolor[rgb]{1,0.91296,0.91296}136 & \cellcolor[rgb]{1,0.87648,0.87648}193 & \cellcolor[rgb]{1,0.90784,0.90784}144 & \cellcolor[rgb]{1,0.9488,0.9488}80 & \cellcolor[rgb]{1,0.96672,0.96672}52 & \cellcolor[rgb]{1,0.96928,0.96928}48 & \cellcolor[rgb]{1,0.9712,0.9712}45 & \cellcolor[rgb]{1,0.9712,0.9712}45 & \cellcolor[rgb]{1,0.96288,0.96288}58 \\ 
  {\bf NC} (2-block) & 0.3 & \cellcolor[rgb]{1,0.9168,0.9168}130 & \cellcolor[rgb]{1,0.89824,0.89824}159 & \cellcolor[rgb]{1,0.94112,0.94112}92 & \cellcolor[rgb]{1,0.96416,0.96416}56 & \cellcolor[rgb]{1,0.97248,0.97248}43 & \cellcolor[rgb]{1,0.97312,0.97312}42 & \cellcolor[rgb]{1,0.97376,0.97376}41 & \cellcolor[rgb]{1,0.97184,0.97184}44 & \cellcolor[rgb]{1,0.96608,0.96608}53 \\ 
  $T_\text{CPU}= 2.34 $ & 0.4 & \cellcolor[rgb]{1,0.92768,0.92768}113 & \cellcolor[rgb]{1,0.92256,0.92256}121 & \cellcolor[rgb]{1,0.95712,0.95712}67 & \cellcolor[rgb]{1,0.96992,0.96992}47 & \cellcolor[rgb]{1,0.97504,0.97504}39 & \cellcolor[rgb]{1,0.9744,0.9744}40 & \cellcolor[rgb]{1,0.9744,0.9744}40 & \cellcolor[rgb]{1,0.97248,0.97248}43 & \cellcolor[rgb]{1,0.96928,0.96928}48 \\ 
   & 0.5 & \cellcolor[rgb]{1,0.93856,0.93856}96 & \cellcolor[rgb]{1,0.94176,0.94176}91 & \cellcolor[rgb]{1,0.9648,0.9648}55 & \cellcolor[rgb]{1,0.97312,0.97312}42 & \cellcolor[rgb]{1,0.97568,0.97568}38 & \cellcolor[rgb]{1,0.97632,0.97632}37 & \cellcolor[rgb]{1,0.97568,0.97568}38 & \cellcolor[rgb]{1,0.9744,0.9744}40 & \cellcolor[rgb]{1,0.97312,0.97312}42 \\ 
   \hline
 & 0.1 & \cellcolor[rgb]{1,0.91104,0.91104}139 & \cellcolor[rgb]{1,0.87072,0.87072}202 & \cellcolor[rgb]{1,0.85728,0.85728}223 & \cellcolor[rgb]{1,0.89248,0.89248}168 & \cellcolor[rgb]{1,0.9456,0.9456}85 & \cellcolor[rgb]{1,0.95328,0.95328}73 & \cellcolor[rgb]{1,0.9616,0.9616}60 & \cellcolor[rgb]{1,0.96736,0.96736}51 & \cellcolor[rgb]{1,0.96352,0.96352}57 \\ 
   & 0.2 & \cellcolor[rgb]{1,0.91232,0.91232}137 & \cellcolor[rgb]{1,0.87712,0.87712}192 & \cellcolor[rgb]{1,0.90784,0.90784}144 & \cellcolor[rgb]{1,0.94944,0.94944}79 & \cellcolor[rgb]{1,0.96672,0.96672}52 & \cellcolor[rgb]{1,0.96928,0.96928}48 & \cellcolor[rgb]{1,0.9712,0.9712}45 & \cellcolor[rgb]{1,0.9712,0.9712}45 & \cellcolor[rgb]{1,0.96288,0.96288}58 \\ 
  {\bf NC} (3-block) & 0.3 & \cellcolor[rgb]{1,0.91744,0.91744}129 & \cellcolor[rgb]{1,0.89696,0.89696}161 & \cellcolor[rgb]{1,0.94112,0.94112}92 & \cellcolor[rgb]{1,0.96416,0.96416}56 & \cellcolor[rgb]{1,0.97248,0.97248}43 & \cellcolor[rgb]{1,0.97312,0.97312}42 & \cellcolor[rgb]{1,0.97376,0.97376}41 & \cellcolor[rgb]{1,0.97248,0.97248}43 & \cellcolor[rgb]{1,0.96672,0.96672}52 \\ 
  $T_\text{CPU}= 2.35 $ & 0.4 & \cellcolor[rgb]{1,0.92768,0.92768}113 & \cellcolor[rgb]{1,0.9232,0.9232}120 & \cellcolor[rgb]{1,0.95712,0.95712}67 & \cellcolor[rgb]{1,0.96992,0.96992}47 & \cellcolor[rgb]{1,0.97504,0.97504}39 & \cellcolor[rgb]{1,0.97504,0.97504}39 & \cellcolor[rgb]{1,0.9744,0.9744}40 & \cellcolor[rgb]{1,0.97248,0.97248}43 & \cellcolor[rgb]{1,0.968,0.968}50 \\ 
   & 0.5 & \cellcolor[rgb]{1,0.93856,0.93856}96 & \cellcolor[rgb]{1,0.94176,0.94176}91 & \cellcolor[rgb]{1,0.96416,0.96416}56 & \cellcolor[rgb]{1,0.97312,0.97312}42 & \cellcolor[rgb]{1,0.97568,0.97568}38 & \cellcolor[rgb]{1,0.97568,0.97568}38 & \cellcolor[rgb]{1,0.97568,0.97568}38 & \cellcolor[rgb]{1,0.9744,0.9744}40 & \cellcolor[rgb]{1,0.9712,0.9712}45 \\ 
   \hline
\hline
 & 0.1 & \cellcolor[rgb]{1,0.91808,0.91808}128 & \cellcolor[rgb]{1,0.88096,0.88096}186 & \cellcolor[rgb]{1,0.86816,0.86816}206 & \cellcolor[rgb]{1,0.89952,0.89952}157 & \cellcolor[rgb]{1,0.95136,0.95136}76 & \cellcolor[rgb]{1,0.95776,0.95776}66 & \cellcolor[rgb]{1,0.96736,0.96736}51 & \cellcolor[rgb]{1,0.97504,0.97504}39 & \cellcolor[rgb]{1,0.98336,0.98336}26 \\ 
   & 0.2 & \cellcolor[rgb]{1,0.92,0.92}125 & \cellcolor[rgb]{1,0.8848,0.8848}180 & \cellcolor[rgb]{1,0.91424,0.91424}134 & \cellcolor[rgb]{1,0.9552,0.9552}70 & \cellcolor[rgb]{1,0.97504,0.97504}39 & \cellcolor[rgb]{1,0.97888,0.97888}33 & \cellcolor[rgb]{1,0.98208,0.98208}28 & \cellcolor[rgb]{1,0.98656,0.98656}21 & \cellcolor[rgb]{1,0.99104,0.99104}14 \\ 
  {\bf GIS-C} (2-block) & 0.3 & \cellcolor[rgb]{1,0.92384,0.92384}119 & \cellcolor[rgb]{1,0.90464,0.90464}149 & \cellcolor[rgb]{1,0.94752,0.94752}82 & \cellcolor[rgb]{1,0.97248,0.97248}43 & \cellcolor[rgb]{1,0.984,0.984}25 & \cellcolor[rgb]{1,0.98656,0.98656}21 & \cellcolor[rgb]{1,0.98912,0.98912}17 & \cellcolor[rgb]{1,0.99168,0.99168}13 & \cellcolor[rgb]{1,0.99424,0.99424}9 \\ 
  $T_\text{CPU}= 2.36 $ & 0.4 & \cellcolor[rgb]{1,0.9328,0.9328}105 & \cellcolor[rgb]{1,0.92768,0.92768}113 & \cellcolor[rgb]{1,0.9648,0.9648}55 & \cellcolor[rgb]{1,0.98016,0.98016}31 & \cellcolor[rgb]{1,0.98848,0.98848}18 & \cellcolor[rgb]{1,0.9904,0.9904}15 & \cellcolor[rgb]{1,0.99232,0.99232}12 & \cellcolor[rgb]{1,0.99424,0.99424}9 & \cellcolor[rgb]{1,0.99552,0.99552}7 \\ 
   & 0.5 & \cellcolor[rgb]{1,0.94368,0.94368}88 & \cellcolor[rgb]{1,0.94624,0.94624}84 & \cellcolor[rgb]{1,0.97376,0.97376}41 & \cellcolor[rgb]{1,0.98528,0.98528}23 & \cellcolor[rgb]{1,0.99104,0.99104}14 & \cellcolor[rgb]{1,0.99232,0.99232}12 & \cellcolor[rgb]{1,0.99424,0.99424}9 & \cellcolor[rgb]{1,0.99552,0.99552}7 & \cellcolor[rgb]{1,0.99616,0.99616}6 \\ 
   \hline
 & 0.1 & \cellcolor[rgb]{1,0.91104,0.91104}139 & \cellcolor[rgb]{1,0.87136,0.87136}201 & \cellcolor[rgb]{1,0.85728,0.85728}223 & \cellcolor[rgb]{1,0.89248,0.89248}168 & \cellcolor[rgb]{1,0.94944,0.94944}79 & \cellcolor[rgb]{1,0.95712,0.95712}67 & \cellcolor[rgb]{1,0.96672,0.96672}52 & \cellcolor[rgb]{1,0.9744,0.9744}40 & \cellcolor[rgb]{1,0.98208,0.98208}28 \\ 
   & 0.2 & \cellcolor[rgb]{1,0.91232,0.91232}137 & \cellcolor[rgb]{1,0.87584,0.87584}194 & \cellcolor[rgb]{1,0.91104,0.91104}139 & \cellcolor[rgb]{1,0.95392,0.95392}72 & \cellcolor[rgb]{1,0.9744,0.9744}40 & \cellcolor[rgb]{1,0.97824,0.97824}34 & \cellcolor[rgb]{1,0.98208,0.98208}28 & \cellcolor[rgb]{1,0.98592,0.98592}22 & \cellcolor[rgb]{1,0.99104,0.99104}14 \\ 
  {\bf GIS-C} (3-block) & 0.3 & \cellcolor[rgb]{1,0.9168,0.9168}130 & \cellcolor[rgb]{1,0.8976,0.8976}160 & \cellcolor[rgb]{1,0.94624,0.94624}84 & \cellcolor[rgb]{1,0.97184,0.97184}44 & \cellcolor[rgb]{1,0.98336,0.98336}26 & \cellcolor[rgb]{1,0.98592,0.98592}22 & \cellcolor[rgb]{1,0.98848,0.98848}18 & \cellcolor[rgb]{1,0.99104,0.99104}14 & \cellcolor[rgb]{1,0.99424,0.99424}9 \\ 
  $T_\text{CPU}= 2.37 $ & 0.4 & \cellcolor[rgb]{1,0.92704,0.92704}114 & \cellcolor[rgb]{1,0.9232,0.9232}120 & \cellcolor[rgb]{1,0.96416,0.96416}56 & \cellcolor[rgb]{1,0.98016,0.98016}31 & \cellcolor[rgb]{1,0.98848,0.98848}18 & \cellcolor[rgb]{1,0.98976,0.98976}16 & \cellcolor[rgb]{1,0.99168,0.99168}13 & \cellcolor[rgb]{1,0.9936,0.9936}10 & \cellcolor[rgb]{1,0.99552,0.99552}7 \\ 
   & 0.5 & \cellcolor[rgb]{1,0.9392,0.9392}95 & \cellcolor[rgb]{1,0.94432,0.94432}87 & \cellcolor[rgb]{1,0.97312,0.97312}42 & \cellcolor[rgb]{1,0.98464,0.98464}24 & \cellcolor[rgb]{1,0.99104,0.99104}14 & \cellcolor[rgb]{1,0.99232,0.99232}12 & \cellcolor[rgb]{1,0.9936,0.9936}10 & \cellcolor[rgb]{1,0.99488,0.99488}8 & \cellcolor[rgb]{1,0.99616,0.99616}6 \\ 
   \hline
\hline
 & 0.1 & \cellcolor[rgb]{1,0.91872,0.91872}127 & \cellcolor[rgb]{1,0.88032,0.88032}187 & \cellcolor[rgb]{1,0.86688,0.86688}208 & \cellcolor[rgb]{1,0.9008,0.9008}155 & \cellcolor[rgb]{1,0.95136,0.95136}76 & \cellcolor[rgb]{1,0.95776,0.95776}66 & \cellcolor[rgb]{1,0.96736,0.96736}51 & \cellcolor[rgb]{1,0.97504,0.97504}39 & \cellcolor[rgb]{1,0.98272,0.98272}27 \\ 
   & 0.2 & \cellcolor[rgb]{1,0.91936,0.91936}126 & \cellcolor[rgb]{1,0.8848,0.8848}180 & \cellcolor[rgb]{1,0.91424,0.91424}134 & \cellcolor[rgb]{1,0.9552,0.9552}70 & \cellcolor[rgb]{1,0.97504,0.97504}39 & \cellcolor[rgb]{1,0.97888,0.97888}33 & \cellcolor[rgb]{1,0.98272,0.98272}27 & \cellcolor[rgb]{1,0.98656,0.98656}21 & \cellcolor[rgb]{1,0.99104,0.99104}14 \\ 
  {\bf GIS-NC} (2-block) & 0.3 & \cellcolor[rgb]{1,0.92384,0.92384}119 & \cellcolor[rgb]{1,0.904,0.904}150 & \cellcolor[rgb]{1,0.94752,0.94752}82 & \cellcolor[rgb]{1,0.97248,0.97248}43 & \cellcolor[rgb]{1,0.984,0.984}25 & \cellcolor[rgb]{1,0.98592,0.98592}22 & \cellcolor[rgb]{1,0.98912,0.98912}17 & \cellcolor[rgb]{1,0.99168,0.99168}13 & \cellcolor[rgb]{1,0.99424,0.99424}9 \\ 
  $T_\text{CPU}= 2.40 $ & 0.4 & \cellcolor[rgb]{1,0.9328,0.9328}105 & \cellcolor[rgb]{1,0.92768,0.92768}113 & \cellcolor[rgb]{1,0.9648,0.9648}55 & \cellcolor[rgb]{1,0.98016,0.98016}31 & \cellcolor[rgb]{1,0.98848,0.98848}18 & \cellcolor[rgb]{1,0.9904,0.9904}15 & \cellcolor[rgb]{1,0.99232,0.99232}12 & \cellcolor[rgb]{1,0.99424,0.99424}9 & \cellcolor[rgb]{1,0.99552,0.99552}7 \\ 
   & 0.5 & \cellcolor[rgb]{1,0.94304,0.94304}89 & \cellcolor[rgb]{1,0.94624,0.94624}84 & \cellcolor[rgb]{1,0.97376,0.97376}41 & \cellcolor[rgb]{1,0.98528,0.98528}23 & \cellcolor[rgb]{1,0.99104,0.99104}14 & \cellcolor[rgb]{1,0.99296,0.99296}11 & \cellcolor[rgb]{1,0.99424,0.99424}9 & \cellcolor[rgb]{1,0.99552,0.99552}7 & \cellcolor[rgb]{1,0.99616,0.99616}6 \\ 
   \hline
   \hline
\end{tabular}
\end{table}

Next, we turn to assessing simulation efficiency for the persistence parameter $\phi$, summarized in Table~\ref{ifphi}.
Because $\phi$ is not involved in the reparameterization, the differences between C and NC (and consequently also between the raw and the interwoven samplers) are much less pronounced. For a summary of efficiency gains, see Table~\ref{gainphi}.
It stands out that one- and two-block samplers show very similar IFs, whereas the three-block sampler deteriorates due to massive overconditioning for moderate and small $\phipar_\text{true}$ or $\sigmapar_\text{true}$. Note however that again the interwoven sampler is exempt from these defects due to the fact that NC performs solidly. 
Results not reported here show that for shorter time series with $T=500$, sampling inefficiency is uniformly smaller for all parameterizations. The interwoven 2-block samplers for instance show IFs of 30 or below for all underlying true parameter values.

\begin{table}[ht!]
\caption{Percentage gains in effective sample size for the 2-block and the 3-block sampler. First and third table: $\text{ESS}_\text{GIS-C}$ vs.~$\max(\text{ESS}_\text{C},\text{ESS}_\text{NC})$. Second and fourth table: $\text{ESS}_\text{GIS-C}$ vs.~$\min(\text{ESS}_\text{C},\text{ESS}_\text{NC})$.\\[.5em] Shading: \begin{tabular}{|x{ 0.84 cm} x{0.84cm}x{0.84cm}x{0.84cm}x{0.84cm}x{0.84cm}x{0.84cm}x{0.84cm}x{0.84cm}x{0.84cm} x{ 0.84 cm}|}
  \hline
\cellcolor[rgb]{ 1 ,1, 1 } 0&\cellcolor[rgb]{ 0.95 ,1, 0.95 } 10&\cellcolor[rgb]{ 0.9 ,1, 0.9 } 20&\cellcolor[rgb]{ 0.85 ,1, 0.85 } 30&\cellcolor[rgb]{ 0.8 ,1, 0.8 } 40&\cellcolor[rgb]{ 0.75 ,1, 0.75 } 50&\cellcolor[rgb]{ 0.7 ,1, 0.7 } 60&\cellcolor[rgb]{ 0.65 ,1, 0.65 } 70&\cellcolor[rgb]{ 0.6 ,1, 0.6 } 80&\cellcolor[rgb]{ 0.55 ,1, 0.55 } 90&\cellcolor[rgb]{ 0.5 ,1, 0.5 } 100+ \tabularnewline\hline
  \end{tabular} (1 and 3)\\Shading: \begin{tabular}{|x{ 0.84 cm} x{0.84cm}x{0.84cm}x{0.84cm}x{0.84cm}x{0.84cm}x{0.84cm}x{0.84cm}x{0.84cm}x{0.84cm} x{ 0.84 cm}|}
  \hline
\cellcolor[rgb]{ 1 , 1 ,1} 0&\cellcolor[rgb]{ 0.977 , 0.977 ,1} 100&\cellcolor[rgb]{ 0.954 , 0.954 ,1} 200&\cellcolor[rgb]{ 0.931 , 0.931 ,1} 300&\cellcolor[rgb]{ 0.908 , 0.908 ,1} 400&\cellcolor[rgb]{ 0.885 , 0.885 ,1} 500&\cellcolor[rgb]{ 0.862 , 0.862 ,1} 600&\cellcolor[rgb]{ 0.839 , 0.839 ,1} 700&\cellcolor[rgb]{ 0.816 , 0.816 ,1} 800&\cellcolor[rgb]{ 0.793 , 0.793 ,1} 900&\cellcolor[rgb]{ 0.77 , 0.77 ,1} 1000+ \tabularnewline\hline
  \end{tabular} (2 and 4)}
\label{gainphi}
\begin{tabular}{ccrrrrrrrrr}
   \hline
$p(\phi|\mathbf{y})$ & \backslashbox{$\sigma_{\text{true}}$}{$\phi_{\text{true}}$} & 0 & 0.5 & 0.8 & 0.9 & 0.95 & 0.96 & 0.97 & 0.98 & 0.99 \\ 
   \hline
\hline
 & 0.1 & \cellcolor[rgb]{0.96,1,0.96}8 & \cellcolor[rgb]{0.96,1,0.96}8 & \cellcolor[rgb]{0.95,1,0.95}10 & \cellcolor[rgb]{0.965,1,0.965}7 & \cellcolor[rgb]{0.94,1,0.94}12 & \cellcolor[rgb]{0.94,1,0.94}12 & \cellcolor[rgb]{0.9,1,0.9}20 & \cellcolor[rgb]{0.82,1,0.82}36 & \cellcolor[rgb]{0.5,1,0.5}119 \\ 
  {\bf GIS-C} vs. & 0.2 & \cellcolor[rgb]{0.955,1,0.955}9 & \cellcolor[rgb]{0.965,1,0.965}7 & \cellcolor[rgb]{0.96,1,0.96}8 & \cellcolor[rgb]{0.93,1,0.93}14 & \cellcolor[rgb]{0.83,1,0.83}34 & \cellcolor[rgb]{0.78,1,0.78}44 & \cellcolor[rgb]{0.68,1,0.68}64 & \cellcolor[rgb]{0.59,1,0.59}82 & \cellcolor[rgb]{0.715,1,0.715}57 \\ 
  {\bf better} & 0.3 & \cellcolor[rgb]{0.955,1,0.955}9 & \cellcolor[rgb]{0.965,1,0.965}7 & \cellcolor[rgb]{0.94,1,0.94}12 & \cellcolor[rgb]{0.845,1,0.845}31 & \cellcolor[rgb]{0.645,1,0.645}71 & \cellcolor[rgb]{0.655,1,0.655}69 & \cellcolor[rgb]{0.74,1,0.74}52 & \cellcolor[rgb]{0.79,1,0.79}42 & \cellcolor[rgb]{0.765,1,0.765}47 \\ 
  (2-block) & 0.4 & \cellcolor[rgb]{0.965,1,0.965}7 & \cellcolor[rgb]{0.965,1,0.965}7 & \cellcolor[rgb]{0.89,1,0.89}22 & \cellcolor[rgb]{0.74,1,0.74}52 & \cellcolor[rgb]{0.77,1,0.77}46 & \cellcolor[rgb]{0.795,1,0.795}41 & \cellcolor[rgb]{0.83,1,0.83}34 & \cellcolor[rgb]{0.84,1,0.84}32 & \cellcolor[rgb]{0.675,1,0.675}65 \\ 
   & 0.5 & \cellcolor[rgb]{0.96,1,0.96}8 & \cellcolor[rgb]{0.96,1,0.96}8 & \cellcolor[rgb]{0.825,1,0.825}35 & \cellcolor[rgb]{0.75,1,0.75}50 & \cellcolor[rgb]{0.845,1,0.845}31 & \cellcolor[rgb]{0.86,1,0.86}28 & \cellcolor[rgb]{0.86,1,0.86}28 & \cellcolor[rgb]{0.86,1,0.86}28 & \cellcolor[rgb]{0.595,1,0.595}81 \\ 
   \hline
 & 0.1 & \cellcolor[rgb]{0.99793,0.99793,1}9 & \cellcolor[rgb]{0.99793,0.99793,1}9 & \cellcolor[rgb]{0.99678,0.99678,1}14 & \cellcolor[rgb]{0.97654,0.97654,1}102 & \cellcolor[rgb]{0.91467,0.91467,1}371 & \cellcolor[rgb]{0.91237,0.91237,1}381 & \cellcolor[rgb]{0.91352,0.91352,1}376 & \cellcolor[rgb]{0.93261,0.93261,1}293 & \cellcolor[rgb]{0.96274,0.96274,1}162 \\ 
  {\bf GIS-C} vs. & 0.2 & \cellcolor[rgb]{0.99793,0.99793,1}9 & \cellcolor[rgb]{0.99816,0.99816,1}8 & \cellcolor[rgb]{0.97355,0.97355,1}115 & \cellcolor[rgb]{0.94687,0.94687,1}231 & \cellcolor[rgb]{0.96021,0.96021,1}173 & \cellcolor[rgb]{0.96642,0.96642,1}146 & \cellcolor[rgb]{0.97378,0.97378,1}114 & \cellcolor[rgb]{0.97355,0.97355,1}115 & \cellcolor[rgb]{0.92755,0.92755,1}315 \\ 
  {\bf worse} & 0.3 & \cellcolor[rgb]{0.99793,0.99793,1}9 & \cellcolor[rgb]{0.99655,0.99655,1}15 & \cellcolor[rgb]{0.96619,0.96619,1}147 & \cellcolor[rgb]{0.96987,0.96987,1}131 & \cellcolor[rgb]{0.98137,0.98137,1}81 & \cellcolor[rgb]{0.97861,0.97861,1}93 & \cellcolor[rgb]{0.96872,0.96872,1}136 & \cellcolor[rgb]{0.94641,0.94641,1}233 & \cellcolor[rgb]{0.8896,0.8896,1}480 \\ 
  (2-block) & 0.4 & \cellcolor[rgb]{0.99816,0.99816,1}8 & \cellcolor[rgb]{0.99287,0.99287,1}31 & \cellcolor[rgb]{0.97608,0.97608,1}104 & \cellcolor[rgb]{0.98252,0.98252,1}76 & \cellcolor[rgb]{0.97286,0.97286,1}118 & \cellcolor[rgb]{0.96366,0.96366,1}158 & \cellcolor[rgb]{0.9494,0.9494,1}220 & \cellcolor[rgb]{0.91766,0.91766,1}358 & \cellcolor[rgb]{0.862,0.862,1}600 \\ 
   & 0.5 & \cellcolor[rgb]{0.99816,0.99816,1}8 & \cellcolor[rgb]{0.99218,0.99218,1}34 & \cellcolor[rgb]{0.98344,0.98344,1}72 & \cellcolor[rgb]{0.98183,0.98183,1}79 & \cellcolor[rgb]{0.95998,0.95998,1}174 & \cellcolor[rgb]{0.94825,0.94825,1}225 & \cellcolor[rgb]{0.92709,0.92709,1}317 & \cellcolor[rgb]{0.89857,0.89857,1}441 & \cellcolor[rgb]{0.85372,0.85372,1}636 \\ 
   \hline
\hline
 & 0.1 & \cellcolor[rgb]{1,1,1}0 & \cellcolor[rgb]{1,1,1}0 & \cellcolor[rgb]{1,1,1}0 & \cellcolor[rgb]{1,1,1}0 & \cellcolor[rgb]{0.965,1,0.965}7 & \cellcolor[rgb]{0.955,1,0.955}9 & \cellcolor[rgb]{0.93,1,0.93}14 & \cellcolor[rgb]{0.86,1,0.86}28 & \cellcolor[rgb]{0.5,1,0.5}107 \\ 
  {\bf GIS-C} vs. & 0.2 & \cellcolor[rgb]{1,1,1}0 & \cellcolor[rgb]{1,1,1}0 & \cellcolor[rgb]{0.985,1,0.985}3 & \cellcolor[rgb]{0.945,1,0.945}11 & \cellcolor[rgb]{0.85,1,0.85}30 & \cellcolor[rgb]{0.79,1,0.79}42 & \cellcolor[rgb]{0.7,1,0.7}60 & \cellcolor[rgb]{0.5,1,0.5}105 & \cellcolor[rgb]{0.5,1,0.5}301 \\ 
  {\bf better} & 0.3 & \cellcolor[rgb]{1,1,1}0 & \cellcolor[rgb]{0.995,1,0.995}1 & \cellcolor[rgb]{0.955,1,0.955}9 & \cellcolor[rgb]{0.86,1,0.86}28 & \cellcolor[rgb]{0.665,1,0.665}67 & \cellcolor[rgb]{0.55,1,0.55}90 & \cellcolor[rgb]{0.5,1,0.5}129 & \cellcolor[rgb]{0.5,1,0.5}213 & \cellcolor[rgb]{0.5,1,0.5}455 \\ 
  (3-block) & 0.4 & \cellcolor[rgb]{1,1,1}0 & \cellcolor[rgb]{1,1,1}0 & \cellcolor[rgb]{0.905,1,0.905}19 & \cellcolor[rgb]{0.745,1,0.745}51 & \cellcolor[rgb]{0.5,1,0.5}116 & \cellcolor[rgb]{0.5,1,0.5}147 & \cellcolor[rgb]{0.5,1,0.5}211 & \cellcolor[rgb]{0.5,1,0.5}332 & \cellcolor[rgb]{0.5,1,0.5}486 \\ 
   & 0.5 & \cellcolor[rgb]{1,1,1}0 & \cellcolor[rgb]{0.98,1,0.98}4 & \cellcolor[rgb]{0.835,1,0.835}33 & \cellcolor[rgb]{0.62,1,0.62}76 & \cellcolor[rgb]{0.5,1,0.5}167 & \cellcolor[rgb]{0.5,1,0.5}214 & \cellcolor[rgb]{0.5,1,0.5}297 & \cellcolor[rgb]{0.5,1,0.5}420 & \cellcolor[rgb]{0.5,1,0.5}404 \\ 
   \hline
 & 0.1 & \cellcolor[rgb]{0.77,0.77,1}21444 & \cellcolor[rgb]{0.77,0.77,1}14828 & \cellcolor[rgb]{0.77,0.77,1}15826 & \cellcolor[rgb]{0.77,0.77,1}19673 & \cellcolor[rgb]{0.77,0.77,1}16944 & \cellcolor[rgb]{0.77,0.77,1}13337 & \cellcolor[rgb]{0.77,0.77,1}9245 & \cellcolor[rgb]{0.77,0.77,1}5750 & \cellcolor[rgb]{0.77,0.77,1}2505 \\ 
  {\bf GIS-C} vs. & 0.2 & \cellcolor[rgb]{0.77,0.77,1}21498 & \cellcolor[rgb]{0.77,0.77,1}15775 & \cellcolor[rgb]{0.77,0.77,1}19897 & \cellcolor[rgb]{0.77,0.77,1}12448 & \cellcolor[rgb]{0.77,0.77,1}4413 & \cellcolor[rgb]{0.77,0.77,1}3304 & \cellcolor[rgb]{0.77,0.77,1}2347 & \cellcolor[rgb]{0.77,0.77,1}1571 & \cellcolor[rgb]{0.77782,0.77782,1}966 \\ 
  {\bf worse} & 0.3 & \cellcolor[rgb]{0.77,0.77,1}20674 & \cellcolor[rgb]{0.77,0.77,1}16207 & \cellcolor[rgb]{0.77,0.77,1}12989 & \cellcolor[rgb]{0.77,0.77,1}4984 & \cellcolor[rgb]{0.77,0.77,1}2081 & \cellcolor[rgb]{0.77,0.77,1}1593 & \cellcolor[rgb]{0.77,0.77,1}1252 & \cellcolor[rgb]{0.78725,0.78725,1}925 & \cellcolor[rgb]{0.85487,0.85487,1}631 \\ 
  (3-block) & 0.4 & \cellcolor[rgb]{0.77,0.77,1}20651 & \cellcolor[rgb]{0.77,0.77,1}17183 & \cellcolor[rgb]{0.77,0.77,1}7693 & \cellcolor[rgb]{0.77,0.77,1}2735 & \cellcolor[rgb]{0.77,0.77,1}1267 & \cellcolor[rgb]{0.77,0.77,1}1044 & \cellcolor[rgb]{0.80266,0.80266,1}858 & \cellcolor[rgb]{0.84659,0.84659,1}667 & \cellcolor[rgb]{0.86476,0.86476,1}588 \\ 
   & 0.5 & \cellcolor[rgb]{0.77,0.77,1}19595 & \cellcolor[rgb]{0.77,0.77,1}14829 & \cellcolor[rgb]{0.77,0.77,1}4642 & \cellcolor[rgb]{0.77,0.77,1}1788 & \cellcolor[rgb]{0.78518,0.78518,1}934 & \cellcolor[rgb]{0.82014,0.82014,1}782 & \cellcolor[rgb]{0.8482,0.8482,1}660 & \cellcolor[rgb]{0.87419,0.87419,1}547 & \cellcolor[rgb]{0.84866,0.84866,1}658 \\ 
   \hline
   \hline
\end{tabular}
\end{table}



Finally, we investigate sampling efficiency for $\sigma$. Table~\ref{ifsigma} summarizes median IFs for draws from $p(\sigma|\mathbf{y})$. We observe a similar overall picture to the one presented in Table~\ref{ifmu}: C performs poorly when $\sigma_\text{true}$ and $\phi_\text{true}$ are small, and NC performs poorly when $\sigma_\text{true}$ and $\phi_\text{true}$ are large, while interweaving strategies perform well for all underlying parameter values.
This result partially contrasts the conclusions of \cite{str-etal:par}, who associate better mixing with larger $|\phi|$ for all parameterizations and recommend the non-centered parameterization in any setup. It should be noted, however, that these authors use a different sampling algorithm that does not rely on Gaussian mixture approximation. Moreover, the parameter range investigated in their paper
does not span the range of parameters examined in our paper. For a summary of percentage gains in terms of effective sample size, see Table~\ref{gainsigma}.

\begin{table}[ht!]
\caption{Inefficiency factors for $100\,000$ draws from $p(\sigma|\mathbf{y})$ in various parameterizations using different blocking strategies. Time series length $T=5000$, the values reported are medians of $500$ repetitions and $T_\text{CPU}$ denotes the median time to complete $1000$ iterations.\\[.5em] Shading: \begin{tabular}{|x{ 0.84 cm} x{0.84cm}x{0.84cm}x{0.84cm}x{0.84cm}x{0.84cm}x{0.84cm}x{0.84cm}x{0.84cm}x{0.84cm} x{ 0.84 cm}|}
   \hline
\cellcolor[rgb]{1, 1 , 1 } 0&\cellcolor[rgb]{1, 0.968 , 0.968 } 50&\cellcolor[rgb]{1, 0.936 , 0.936 } 100&\cellcolor[rgb]{1, 0.904 , 0.904 } 150&\cellcolor[rgb]{1, 0.872 , 0.872 } 200&\cellcolor[rgb]{1, 0.84 , 0.84 } 250&\cellcolor[rgb]{1, 0.808 , 0.808 } 300&\cellcolor[rgb]{1, 0.776 , 0.776 } 350&\cellcolor[rgb]{1, 0.744 , 0.744 } 400&\cellcolor[rgb]{1, 0.712 , 0.712 } 450&\cellcolor[rgb]{1, 0.68 , 0.68 } 500+ \tabularnewline\hline
   \end{tabular}}
\label{ifsigma}
\begin{tabular}{ccrrrrrrrrr}
   \hline
$p(\sigma|\mathbf{y})$ & \backslashbox{$\sigma_{\text{true}}$}{$\phi_{\text{true}}$} & 0 & 0.5 & 0.8 & 0.9 & 0.95 & 0.96 & 0.97 & 0.98 & 0.99 \\ 
   \hline
\hline
 & 0.1 & \cellcolor[rgb]{1,0.68,0.68}5403 & \cellcolor[rgb]{1,0.68,0.68}5125 & \cellcolor[rgb]{1,0.68,0.68}3634 & \cellcolor[rgb]{1,0.68,0.68}1907 & \cellcolor[rgb]{1,0.68,0.68}829 & \cellcolor[rgb]{1,0.68,0.68}679 & \cellcolor[rgb]{1,0.68192,0.68192}497 & \cellcolor[rgb]{1,0.77728,0.77728}348 & \cellcolor[rgb]{1,0.84192,0.84192}247 \\ 
   & 0.2 & \cellcolor[rgb]{1,0.68,0.68}3067 & \cellcolor[rgb]{1,0.68,0.68}1938 & \cellcolor[rgb]{1,0.68,0.68}809 & \cellcolor[rgb]{1,0.74528,0.74528}398 & \cellcolor[rgb]{1,0.87456,0.87456}196 & \cellcolor[rgb]{1,0.89312,0.89312}167 & \cellcolor[rgb]{1,0.9104,0.9104}140 & \cellcolor[rgb]{1,0.92512,0.92512}117 & \cellcolor[rgb]{1,0.93152,0.93152}107 \\ 
  {\bf C} (1-block) & 0.3 & \cellcolor[rgb]{1,0.68,0.68}687 & \cellcolor[rgb]{1,0.68,0.68}505 & \cellcolor[rgb]{1,0.79008,0.79008}328 & \cellcolor[rgb]{1,0.89504,0.89504}164 & \cellcolor[rgb]{1,0.93728,0.93728}98 & \cellcolor[rgb]{1,0.94368,0.94368}88 & \cellcolor[rgb]{1,0.95008,0.95008}78 & \cellcolor[rgb]{1,0.95456,0.95456}71 & \cellcolor[rgb]{1,0.95328,0.95328}73 \\ 
  $T_\text{CPU}= 2.30 $ & 0.4 & \cellcolor[rgb]{1,0.84256,0.84256}246 & \cellcolor[rgb]{1,0.82784,0.82784}269 & \cellcolor[rgb]{1,0.89184,0.89184}169 & \cellcolor[rgb]{1,0.93856,0.93856}96 & \cellcolor[rgb]{1,0.95904,0.95904}64 & \cellcolor[rgb]{1,0.9616,0.9616}60 & \cellcolor[rgb]{1,0.96416,0.96416}56 & \cellcolor[rgb]{1,0.96608,0.96608}53 & \cellcolor[rgb]{1,0.9616,0.9616}60 \\ 
   & 0.5 & \cellcolor[rgb]{1,0.9104,0.9104}140 & \cellcolor[rgb]{1,0.89184,0.89184}169 & \cellcolor[rgb]{1,0.93024,0.93024}109 & \cellcolor[rgb]{1,0.95776,0.95776}66 & \cellcolor[rgb]{1,0.96864,0.96864}49 & \cellcolor[rgb]{1,0.97056,0.97056}46 & \cellcolor[rgb]{1,0.97248,0.97248}43 & \cellcolor[rgb]{1,0.97248,0.97248}43 & \cellcolor[rgb]{1,0.968,0.968}50 \\ 
   \hline
 & 0.1 & \cellcolor[rgb]{1,0.68,0.68}5440 & \cellcolor[rgb]{1,0.68,0.68}4899 & \cellcolor[rgb]{1,0.68,0.68}3608 & \cellcolor[rgb]{1,0.68,0.68}1845 & \cellcolor[rgb]{1,0.68,0.68}768 & \cellcolor[rgb]{1,0.68,0.68}604 & \cellcolor[rgb]{1,0.72416,0.72416}431 & \cellcolor[rgb]{1,0.808,0.808}300 & \cellcolor[rgb]{1,0.87584,0.87584}194 \\ 
   & 0.2 & \cellcolor[rgb]{1,0.68,0.68}3001 & \cellcolor[rgb]{1,0.68,0.68}1813 & \cellcolor[rgb]{1,0.68,0.68}779 & \cellcolor[rgb]{1,0.76064,0.76064}374 & \cellcolor[rgb]{1,0.88416,0.88416}181 & \cellcolor[rgb]{1,0.90336,0.90336}151 & \cellcolor[rgb]{1,0.92064,0.92064}124 & \cellcolor[rgb]{1,0.936,0.936}100 & \cellcolor[rgb]{1,0.9488,0.9488}80 \\ 
  {\bf C} (2-block) & 0.3 & \cellcolor[rgb]{1,0.68,0.68}663 & \cellcolor[rgb]{1,0.6864,0.6864}490 & \cellcolor[rgb]{1,0.79776,0.79776}316 & \cellcolor[rgb]{1,0.9008,0.9008}155 & \cellcolor[rgb]{1,0.9424,0.9424}90 & \cellcolor[rgb]{1,0.94944,0.94944}79 & \cellcolor[rgb]{1,0.95648,0.95648}68 & \cellcolor[rgb]{1,0.96224,0.96224}59 & \cellcolor[rgb]{1,0.968,0.968}50 \\ 
  $T_\text{CPU}= 2.31 $ & 0.4 & \cellcolor[rgb]{1,0.84768,0.84768}238 & \cellcolor[rgb]{1,0.8336,0.8336}260 & \cellcolor[rgb]{1,0.89568,0.89568}163 & \cellcolor[rgb]{1,0.94176,0.94176}91 & \cellcolor[rgb]{1,0.96288,0.96288}58 & \cellcolor[rgb]{1,0.96544,0.96544}54 & \cellcolor[rgb]{1,0.96928,0.96928}48 & \cellcolor[rgb]{1,0.97248,0.97248}43 & \cellcolor[rgb]{1,0.97568,0.97568}38 \\ 
   & 0.5 & \cellcolor[rgb]{1,0.9136,0.9136}135 & \cellcolor[rgb]{1,0.89376,0.89376}166 & \cellcolor[rgb]{1,0.9328,0.9328}105 & \cellcolor[rgb]{1,0.95968,0.95968}63 & \cellcolor[rgb]{1,0.97184,0.97184}44 & \cellcolor[rgb]{1,0.97376,0.97376}41 & \cellcolor[rgb]{1,0.97632,0.97632}37 & \cellcolor[rgb]{1,0.97824,0.97824}34 & \cellcolor[rgb]{1,0.98016,0.98016}31 \\ 
   \hline
 & 0.1 & \cellcolor[rgb]{1,0.68,0.68}5347 & \cellcolor[rgb]{1,0.68,0.68}5083 & \cellcolor[rgb]{1,0.68,0.68}3274 & \cellcolor[rgb]{1,0.68,0.68}1265 & \cellcolor[rgb]{1,0.68,0.68}701 & \cellcolor[rgb]{1,0.68,0.68}617 & \cellcolor[rgb]{1,0.6832,0.6832}495 & \cellcolor[rgb]{1,0.75168,0.75168}388 & \cellcolor[rgb]{1,0.8592,0.8592}220 \\ 
   & 0.2 & \cellcolor[rgb]{1,0.68,0.68}3089 & \cellcolor[rgb]{1,0.68,0.68}1774 & \cellcolor[rgb]{1,0.68,0.68}559 & \cellcolor[rgb]{1,0.68,0.68}565 & \cellcolor[rgb]{1,0.78112,0.78112}342 & \cellcolor[rgb]{1,0.8208,0.8208}280 & \cellcolor[rgb]{1,0.86432,0.86432}212 & \cellcolor[rgb]{1,0.904,0.904}150 & \cellcolor[rgb]{1,0.94112,0.94112}92 \\ 
  {\bf C} (3-block) & 0.3 & \cellcolor[rgb]{1,0.68,0.68}675 & \cellcolor[rgb]{1,0.74528,0.74528}398 & \cellcolor[rgb]{1,0.6992,0.6992}470 & \cellcolor[rgb]{1,0.77152,0.77152}357 & \cellcolor[rgb]{1,0.87904,0.87904}189 & \cellcolor[rgb]{1,0.90592,0.90592}147 & \cellcolor[rgb]{1,0.92832,0.92832}112 & \cellcolor[rgb]{1,0.95008,0.95008}78 & \cellcolor[rgb]{1,0.9648,0.9648}55 \\ 
  $T_\text{CPU}= 2.31 $ & 0.4 & \cellcolor[rgb]{1,0.84704,0.84704}239 & \cellcolor[rgb]{1,0.85536,0.85536}226 & \cellcolor[rgb]{1,0.7568,0.7568}380 & \cellcolor[rgb]{1,0.84512,0.84512}242 & \cellcolor[rgb]{1,0.9264,0.9264}115 & \cellcolor[rgb]{1,0.94112,0.94112}92 & \cellcolor[rgb]{1,0.95584,0.95584}69 & \cellcolor[rgb]{1,0.96736,0.96736}51 & \cellcolor[rgb]{1,0.97376,0.97376}41 \\ 
   & 0.5 & \cellcolor[rgb]{1,0.91488,0.91488}133 & \cellcolor[rgb]{1,0.8656,0.8656}210 & \cellcolor[rgb]{1,0.80224,0.80224}309 & \cellcolor[rgb]{1,0.89376,0.89376}166 & \cellcolor[rgb]{1,0.95072,0.95072}77 & \cellcolor[rgb]{1,0.96096,0.96096}61 & \cellcolor[rgb]{1,0.96992,0.96992}47 & \cellcolor[rgb]{1,0.97504,0.97504}39 & \cellcolor[rgb]{1,0.97952,0.97952}32 \\ 
   \hline
\hline
 & 0.1 & \cellcolor[rgb]{1,0.96352,0.96352}57 & \cellcolor[rgb]{1,0.95904,0.95904}64 & \cellcolor[rgb]{1,0.94176,0.94176}91 & \cellcolor[rgb]{1,0.9168,0.9168}130 & \cellcolor[rgb]{1,0.9424,0.9424}90 & \cellcolor[rgb]{1,0.94688,0.94688}83 & \cellcolor[rgb]{1,0.95328,0.95328}73 & \cellcolor[rgb]{1,0.95456,0.95456}71 & \cellcolor[rgb]{1,0.94432,0.94432}87 \\ 
   & 0.2 & \cellcolor[rgb]{1,0.93664,0.93664}99 & \cellcolor[rgb]{1,0.9392,0.9392}95 & \cellcolor[rgb]{1,0.92256,0.92256}121 & \cellcolor[rgb]{1,0.94432,0.94432}87 & \cellcolor[rgb]{1,0.9552,0.9552}70 & \cellcolor[rgb]{1,0.9552,0.9552}70 & \cellcolor[rgb]{1,0.952,0.952}75 & \cellcolor[rgb]{1,0.94304,0.94304}89 & \cellcolor[rgb]{1,0.91232,0.91232}137 \\ 
  {\bf NC} (2-block) & 0.3 & \cellcolor[rgb]{1,0.96096,0.96096}61 & \cellcolor[rgb]{1,0.95072,0.95072}77 & \cellcolor[rgb]{1,0.93984,0.93984}94 & \cellcolor[rgb]{1,0.9552,0.9552}70 & \cellcolor[rgb]{1,0.95456,0.95456}71 & \cellcolor[rgb]{1,0.95072,0.95072}77 & \cellcolor[rgb]{1,0.94304,0.94304}89 & \cellcolor[rgb]{1,0.92704,0.92704}114 & \cellcolor[rgb]{1,0.87904,0.87904}189 \\ 
  $T_\text{CPU}= 2.34 $ & 0.4 & \cellcolor[rgb]{1,0.97568,0.97568}38 & \cellcolor[rgb]{1,0.952,0.952}75 & \cellcolor[rgb]{1,0.952,0.952}75 & \cellcolor[rgb]{1,0.95776,0.95776}66 & \cellcolor[rgb]{1,0.94944,0.94944}79 & \cellcolor[rgb]{1,0.9424,0.9424}90 & \cellcolor[rgb]{1,0.93216,0.93216}106 & \cellcolor[rgb]{1,0.90784,0.90784}144 & \cellcolor[rgb]{1,0.83552,0.83552}257 \\ 
   & 0.5 & \cellcolor[rgb]{1,0.98016,0.98016}31 & \cellcolor[rgb]{1,0.95648,0.95648}68 & \cellcolor[rgb]{1,0.95712,0.95712}67 & \cellcolor[rgb]{1,0.95712,0.95712}67 & \cellcolor[rgb]{1,0.9424,0.9424}90 & \cellcolor[rgb]{1,0.93408,0.93408}103 & \cellcolor[rgb]{1,0.91936,0.91936}126 & \cellcolor[rgb]{1,0.88928,0.88928}173 & \cellcolor[rgb]{1,0.79776,0.79776}316 \\ 
   \hline
 & 0.1 & \cellcolor[rgb]{1,0.96352,0.96352}57 & \cellcolor[rgb]{1,0.95904,0.95904}64 & \cellcolor[rgb]{1,0.9424,0.9424}90 & \cellcolor[rgb]{1,0.9168,0.9168}130 & \cellcolor[rgb]{1,0.9424,0.9424}90 & \cellcolor[rgb]{1,0.94688,0.94688}83 & \cellcolor[rgb]{1,0.95328,0.95328}73 & \cellcolor[rgb]{1,0.95392,0.95392}72 & \cellcolor[rgb]{1,0.94112,0.94112}92 \\ 
   & 0.2 & \cellcolor[rgb]{1,0.93664,0.93664}99 & \cellcolor[rgb]{1,0.93856,0.93856}96 & \cellcolor[rgb]{1,0.92128,0.92128}123 & \cellcolor[rgb]{1,0.94432,0.94432}87 & \cellcolor[rgb]{1,0.9552,0.9552}70 & \cellcolor[rgb]{1,0.95456,0.95456}71 & \cellcolor[rgb]{1,0.95136,0.95136}76 & \cellcolor[rgb]{1,0.94176,0.94176}91 & \cellcolor[rgb]{1,0.9072,0.9072}145 \\ 
  {\bf NC} (3-block) & 0.3 & \cellcolor[rgb]{1,0.9616,0.9616}60 & \cellcolor[rgb]{1,0.95072,0.95072}77 & \cellcolor[rgb]{1,0.9392,0.9392}95 & \cellcolor[rgb]{1,0.9552,0.9552}70 & \cellcolor[rgb]{1,0.95392,0.95392}72 & \cellcolor[rgb]{1,0.95008,0.95008}78 & \cellcolor[rgb]{1,0.9424,0.9424}90 & \cellcolor[rgb]{1,0.9264,0.9264}115 & \cellcolor[rgb]{1,0.87264,0.87264}199 \\ 
  $T_\text{CPU}= 2.35 $ & 0.4 & \cellcolor[rgb]{1,0.97568,0.97568}38 & \cellcolor[rgb]{1,0.952,0.952}75 & \cellcolor[rgb]{1,0.952,0.952}75 & \cellcolor[rgb]{1,0.95776,0.95776}66 & \cellcolor[rgb]{1,0.94816,0.94816}81 & \cellcolor[rgb]{1,0.94176,0.94176}91 & \cellcolor[rgb]{1,0.93088,0.93088}108 & \cellcolor[rgb]{1,0.90656,0.90656}146 & \cellcolor[rgb]{1,0.82528,0.82528}273 \\ 
   & 0.5 & \cellcolor[rgb]{1,0.98016,0.98016}31 & \cellcolor[rgb]{1,0.95648,0.95648}68 & \cellcolor[rgb]{1,0.95712,0.95712}67 & \cellcolor[rgb]{1,0.95712,0.95712}67 & \cellcolor[rgb]{1,0.9424,0.9424}90 & \cellcolor[rgb]{1,0.9328,0.9328}105 & \cellcolor[rgb]{1,0.91744,0.91744}129 & \cellcolor[rgb]{1,0.88736,0.88736}176 & \cellcolor[rgb]{1,0.7856,0.7856}335 \\ 
   \hline
\hline
 & 0.1 & \cellcolor[rgb]{1,0.96416,0.96416}56 & \cellcolor[rgb]{1,0.95904,0.95904}64 & \cellcolor[rgb]{1,0.94304,0.94304}89 & \cellcolor[rgb]{1,0.92192,0.92192}122 & \cellcolor[rgb]{1,0.94752,0.94752}82 & \cellcolor[rgb]{1,0.952,0.952}75 & \cellcolor[rgb]{1,0.95904,0.95904}64 & \cellcolor[rgb]{1,0.96288,0.96288}58 & \cellcolor[rgb]{1,0.96096,0.96096}61 \\ 
   & 0.2 & \cellcolor[rgb]{1,0.93792,0.93792}97 & \cellcolor[rgb]{1,0.94048,0.94048}93 & \cellcolor[rgb]{1,0.92704,0.92704}114 & \cellcolor[rgb]{1,0.95136,0.95136}76 & \cellcolor[rgb]{1,0.96608,0.96608}53 & \cellcolor[rgb]{1,0.968,0.968}50 & \cellcolor[rgb]{1,0.96928,0.96928}48 & \cellcolor[rgb]{1,0.96928,0.96928}48 & \cellcolor[rgb]{1,0.96736,0.96736}51 \\ 
  {\bf GIS-C} (2-block) & 0.3 & \cellcolor[rgb]{1,0.96288,0.96288}58 & \cellcolor[rgb]{1,0.95392,0.95392}72 & \cellcolor[rgb]{1,0.94752,0.94752}82 & \cellcolor[rgb]{1,0.96608,0.96608}53 & \cellcolor[rgb]{1,0.97312,0.97312}42 & \cellcolor[rgb]{1,0.97376,0.97376}41 & \cellcolor[rgb]{1,0.9744,0.9744}40 & \cellcolor[rgb]{1,0.97504,0.97504}39 & \cellcolor[rgb]{1,0.9744,0.9744}40 \\ 
  $T_\text{CPU}= 2.36 $ & 0.4 & \cellcolor[rgb]{1,0.9776,0.9776}35 & \cellcolor[rgb]{1,0.95584,0.95584}69 & \cellcolor[rgb]{1,0.96288,0.96288}58 & \cellcolor[rgb]{1,0.97312,0.97312}42 & \cellcolor[rgb]{1,0.9776,0.9776}35 & \cellcolor[rgb]{1,0.9776,0.9776}35 & \cellcolor[rgb]{1,0.97824,0.97824}34 & \cellcolor[rgb]{1,0.97888,0.97888}33 & \cellcolor[rgb]{1,0.97888,0.97888}33 \\ 
   & 0.5 & \cellcolor[rgb]{1,0.98208,0.98208}28 & \cellcolor[rgb]{1,0.96224,0.96224}59 & \cellcolor[rgb]{1,0.96992,0.96992}47 & \cellcolor[rgb]{1,0.9776,0.9776}35 & \cellcolor[rgb]{1,0.98016,0.98016}31 & \cellcolor[rgb]{1,0.9808,0.9808}30 & \cellcolor[rgb]{1,0.98144,0.98144}29 & \cellcolor[rgb]{1,0.98144,0.98144}29 & \cellcolor[rgb]{1,0.98208,0.98208}28 \\ 
   \hline
 & 0.1 & \cellcolor[rgb]{1,0.96416,0.96416}56 & \cellcolor[rgb]{1,0.95904,0.95904}64 & \cellcolor[rgb]{1,0.94304,0.94304}89 & \cellcolor[rgb]{1,0.92,0.92}125 & \cellcolor[rgb]{1,0.9456,0.9456}85 & \cellcolor[rgb]{1,0.95136,0.95136}76 & \cellcolor[rgb]{1,0.9584,0.9584}65 & \cellcolor[rgb]{1,0.9616,0.9616}60 & \cellcolor[rgb]{1,0.95968,0.95968}63 \\ 
   & 0.2 & \cellcolor[rgb]{1,0.93792,0.93792}97 & \cellcolor[rgb]{1,0.94048,0.94048}93 & \cellcolor[rgb]{1,0.92512,0.92512}117 & \cellcolor[rgb]{1,0.95136,0.95136}76 & \cellcolor[rgb]{1,0.96544,0.96544}54 & \cellcolor[rgb]{1,0.96736,0.96736}51 & \cellcolor[rgb]{1,0.96864,0.96864}49 & \cellcolor[rgb]{1,0.96864,0.96864}49 & \cellcolor[rgb]{1,0.96672,0.96672}52 \\ 
  {\bf GIS-C} (3-block) & 0.3 & \cellcolor[rgb]{1,0.96352,0.96352}57 & \cellcolor[rgb]{1,0.95328,0.95328}73 & \cellcolor[rgb]{1,0.94688,0.94688}83 & \cellcolor[rgb]{1,0.96608,0.96608}53 & \cellcolor[rgb]{1,0.97248,0.97248}43 & \cellcolor[rgb]{1,0.97376,0.97376}41 & \cellcolor[rgb]{1,0.9744,0.9744}40 & \cellcolor[rgb]{1,0.9744,0.9744}40 & \cellcolor[rgb]{1,0.97376,0.97376}41 \\ 
  $T_\text{CPU}= 2.37 $ & 0.4 & \cellcolor[rgb]{1,0.97824,0.97824}34 & \cellcolor[rgb]{1,0.9552,0.9552}70 & \cellcolor[rgb]{1,0.96224,0.96224}59 & \cellcolor[rgb]{1,0.97248,0.97248}43 & \cellcolor[rgb]{1,0.9776,0.9776}35 & \cellcolor[rgb]{1,0.9776,0.9776}35 & \cellcolor[rgb]{1,0.97824,0.97824}34 & \cellcolor[rgb]{1,0.97888,0.97888}33 & \cellcolor[rgb]{1,0.97888,0.97888}33 \\ 
   & 0.5 & \cellcolor[rgb]{1,0.98208,0.98208}28 & \cellcolor[rgb]{1,0.9616,0.9616}60 & \cellcolor[rgb]{1,0.96928,0.96928}48 & \cellcolor[rgb]{1,0.97696,0.97696}36 & \cellcolor[rgb]{1,0.98016,0.98016}31 & \cellcolor[rgb]{1,0.9808,0.9808}30 & \cellcolor[rgb]{1,0.98144,0.98144}29 & \cellcolor[rgb]{1,0.98144,0.98144}29 & \cellcolor[rgb]{1,0.98208,0.98208}28 \\ 
   \hline
\hline
 & 0.1 & \cellcolor[rgb]{1,0.96416,0.96416}56 & \cellcolor[rgb]{1,0.95904,0.95904}64 & \cellcolor[rgb]{1,0.94368,0.94368}88 & \cellcolor[rgb]{1,0.92128,0.92128}123 & \cellcolor[rgb]{1,0.94688,0.94688}83 & \cellcolor[rgb]{1,0.95264,0.95264}74 & \cellcolor[rgb]{1,0.95904,0.95904}64 & \cellcolor[rgb]{1,0.96288,0.96288}58 & \cellcolor[rgb]{1,0.96096,0.96096}61 \\ 
   & 0.2 & \cellcolor[rgb]{1,0.93856,0.93856}96 & \cellcolor[rgb]{1,0.94048,0.94048}93 & \cellcolor[rgb]{1,0.92768,0.92768}113 & \cellcolor[rgb]{1,0.952,0.952}75 & \cellcolor[rgb]{1,0.96608,0.96608}53 & \cellcolor[rgb]{1,0.968,0.968}50 & \cellcolor[rgb]{1,0.96928,0.96928}48 & \cellcolor[rgb]{1,0.96928,0.96928}48 & \cellcolor[rgb]{1,0.968,0.968}50 \\ 
  {\bf GIS-NC} (2-block) & 0.3 & \cellcolor[rgb]{1,0.96288,0.96288}58 & \cellcolor[rgb]{1,0.95392,0.95392}72 & \cellcolor[rgb]{1,0.94816,0.94816}81 & \cellcolor[rgb]{1,0.96672,0.96672}52 & \cellcolor[rgb]{1,0.97312,0.97312}42 & \cellcolor[rgb]{1,0.97376,0.97376}41 & \cellcolor[rgb]{1,0.9744,0.9744}40 & \cellcolor[rgb]{1,0.97504,0.97504}39 & \cellcolor[rgb]{1,0.9744,0.9744}40 \\ 
  $T_\text{CPU}= 2.40 $ & 0.4 & \cellcolor[rgb]{1,0.9776,0.9776}35 & \cellcolor[rgb]{1,0.95584,0.95584}69 & \cellcolor[rgb]{1,0.96224,0.96224}59 & \cellcolor[rgb]{1,0.97312,0.97312}42 & \cellcolor[rgb]{1,0.9776,0.9776}35 & \cellcolor[rgb]{1,0.9776,0.9776}35 & \cellcolor[rgb]{1,0.97824,0.97824}34 & \cellcolor[rgb]{1,0.97888,0.97888}33 & \cellcolor[rgb]{1,0.97888,0.97888}33 \\ 
   & 0.5 & \cellcolor[rgb]{1,0.98208,0.98208}28 & \cellcolor[rgb]{1,0.96288,0.96288}58 & \cellcolor[rgb]{1,0.96992,0.96992}47 & \cellcolor[rgb]{1,0.9776,0.9776}35 & \cellcolor[rgb]{1,0.98016,0.98016}31 & \cellcolor[rgb]{1,0.9808,0.9808}30 & \cellcolor[rgb]{1,0.98144,0.98144}29 & \cellcolor[rgb]{1,0.98144,0.98144}29 & \cellcolor[rgb]{1,0.98208,0.98208}28 \\ 
   \hline
   \hline
\end{tabular}
\end{table}

\begin{table}[ht!]
\caption{Percentage gains in effective sample size for the 2-block and the 3-block sampler. First and third table: $\text{ESS}_\text{GIS-C}$ vs.~$\max(\text{ESS}_\text{C},\text{ESS}_\text{NC})$. Second and fourth table: $\text{ESS}_\text{GIS-C}$ vs.~$\min(\text{ESS}_\text{C},\text{ESS}_\text{NC})$.\\[.5em] Shading: \begin{tabular}{|x{ 0.84 cm} x{0.84cm}x{0.84cm}x{0.84cm}x{0.84cm}x{0.84cm}x{0.84cm}x{0.84cm}x{0.84cm}x{0.84cm} x{ 0.84 cm}|}
  \hline
\cellcolor[rgb]{ 1 ,1, 1 } 0&\cellcolor[rgb]{ 0.95 ,1, 0.95 } 10&\cellcolor[rgb]{ 0.9 ,1, 0.9 } 20&\cellcolor[rgb]{ 0.85 ,1, 0.85 } 30&\cellcolor[rgb]{ 0.8 ,1, 0.8 } 40&\cellcolor[rgb]{ 0.75 ,1, 0.75 } 50&\cellcolor[rgb]{ 0.7 ,1, 0.7 } 60&\cellcolor[rgb]{ 0.65 ,1, 0.65 } 70&\cellcolor[rgb]{ 0.6 ,1, 0.6 } 80&\cellcolor[rgb]{ 0.55 ,1, 0.55 } 90&\cellcolor[rgb]{ 0.5 ,1, 0.5 } 100+ \tabularnewline\hline
  \end{tabular} (1 and 3)\\Shading: \begin{tabular}{|x{ 0.84 cm} x{0.84cm}x{0.84cm}x{0.84cm}x{0.84cm}x{0.84cm}x{0.84cm}x{0.84cm}x{0.84cm}x{0.84cm} x{ 0.84 cm}|}
  \hline
\cellcolor[rgb]{ 1 , 1 ,1} 0&\cellcolor[rgb]{ 0.977 , 0.977 ,1} 100&\cellcolor[rgb]{ 0.954 , 0.954 ,1} 200&\cellcolor[rgb]{ 0.931 , 0.931 ,1} 300&\cellcolor[rgb]{ 0.908 , 0.908 ,1} 400&\cellcolor[rgb]{ 0.885 , 0.885 ,1} 500&\cellcolor[rgb]{ 0.862 , 0.862 ,1} 600&\cellcolor[rgb]{ 0.839 , 0.839 ,1} 700&\cellcolor[rgb]{ 0.816 , 0.816 ,1} 800&\cellcolor[rgb]{ 0.793 , 0.793 ,1} 900&\cellcolor[rgb]{ 0.77 , 0.77 ,1} 1000+ \tabularnewline\hline
  \end{tabular} (2 and 4)}
\label{gainsigma}
\begin{tabular}{ccrrrrrrrrr}
   \hline
$p(\sigma|\mathbf{y})$ & \backslashbox{$\sigma_{\text{true}}$}{$\phi_{\text{true}}$} & 0 & 0.5 & 0.8 & 0.9 & 0.95 & 0.96 & 0.97 & 0.98 & 0.99 \\ 
   \hline
\hline
 & 0.1 & \cellcolor[rgb]{0.995,1,0.995}1 & \cellcolor[rgb]{1,1,1}0 & \cellcolor[rgb]{0.99,1,0.99}2 & \cellcolor[rgb]{0.97,1,0.97}6 & \cellcolor[rgb]{0.95,1,0.95}10 & \cellcolor[rgb]{0.945,1,0.945}11 & \cellcolor[rgb]{0.925,1,0.925}15 & \cellcolor[rgb]{0.895,1,0.895}21 & \cellcolor[rgb]{0.785,1,0.785}43 \\ 
  {\bf GIS-C} vs. & 0.2 & \cellcolor[rgb]{0.99,1,0.99}2 & \cellcolor[rgb]{0.99,1,0.99}2 & \cellcolor[rgb]{0.97,1,0.97}6 & \cellcolor[rgb]{0.925,1,0.925}15 & \cellcolor[rgb]{0.835,1,0.835}33 & \cellcolor[rgb]{0.805,1,0.805}39 & \cellcolor[rgb]{0.725,1,0.725}55 & \cellcolor[rgb]{0.575,1,0.575}85 & \cellcolor[rgb]{0.71,1,0.71}58 \\ 
  {\bf better} & 0.3 & \cellcolor[rgb]{0.975,1,0.975}5 & \cellcolor[rgb]{0.965,1,0.965}7 & \cellcolor[rgb]{0.925,1,0.925}15 & \cellcolor[rgb]{0.835,1,0.835}33 & \cellcolor[rgb]{0.655,1,0.655}69 & \cellcolor[rgb]{0.555,1,0.555}89 & \cellcolor[rgb]{0.635,1,0.635}73 & \cellcolor[rgb]{0.755,1,0.755}49 & \cellcolor[rgb]{0.87,1,0.87}26 \\ 
  (2-block) & 0.4 & \cellcolor[rgb]{0.96,1,0.96}8 & \cellcolor[rgb]{0.955,1,0.955}9 & \cellcolor[rgb]{0.86,1,0.86}28 & \cellcolor[rgb]{0.715,1,0.715}57 & \cellcolor[rgb]{0.67,1,0.67}66 & \cellcolor[rgb]{0.725,1,0.725}55 & \cellcolor[rgb]{0.79,1,0.79}42 & \cellcolor[rgb]{0.855,1,0.855}29 & \cellcolor[rgb]{0.925,1,0.925}15 \\ 
   & 0.5 & \cellcolor[rgb]{0.935,1,0.935}13 & \cellcolor[rgb]{0.92,1,0.92}16 & \cellcolor[rgb]{0.795,1,0.795}41 & \cellcolor[rgb]{0.62,1,0.62}76 & \cellcolor[rgb]{0.785,1,0.785}43 & \cellcolor[rgb]{0.82,1,0.82}36 & \cellcolor[rgb]{0.865,1,0.865}27 & \cellcolor[rgb]{0.905,1,0.905}19 & \cellcolor[rgb]{0.95,1,0.95}10 \\ 
   \hline
 & 0.1 & \cellcolor[rgb]{0.77,0.77,1}9550 & \cellcolor[rgb]{0.77,0.77,1}7593 & \cellcolor[rgb]{0.77,0.77,1}3947 & \cellcolor[rgb]{0.77,0.77,1}1409 & \cellcolor[rgb]{0.80818,0.80818,1}834 & \cellcolor[rgb]{0.83716,0.83716,1}708 & \cellcolor[rgb]{0.86798,0.86798,1}574 & \cellcolor[rgb]{0.90478,0.90478,1}414 & \cellcolor[rgb]{0.94986,0.94986,1}218 \\ 
  {\bf GIS-C} vs. & 0.2 & \cellcolor[rgb]{0.77,0.77,1}2999 & \cellcolor[rgb]{0.77,0.77,1}1842 & \cellcolor[rgb]{0.86568,0.86568,1}584 & \cellcolor[rgb]{0.90961,0.90961,1}393 & \cellcolor[rgb]{0.94434,0.94434,1}242 & \cellcolor[rgb]{0.954,0.954,1}200 & \cellcolor[rgb]{0.96389,0.96389,1}157 & \cellcolor[rgb]{0.97516,0.97516,1}108 & \cellcolor[rgb]{0.9609,0.9609,1}170 \\ 
  {\bf worse} & 0.3 & \cellcolor[rgb]{0.77,0.77,1}1041 & \cellcolor[rgb]{0.86591,0.86591,1}583 & \cellcolor[rgb]{0.93422,0.93422,1}286 & \cellcolor[rgb]{0.95515,0.95515,1}195 & \cellcolor[rgb]{0.97378,0.97378,1}114 & \cellcolor[rgb]{0.97861,0.97861,1}93 & \cellcolor[rgb]{0.97148,0.97148,1}124 & \cellcolor[rgb]{0.9563,0.9563,1}190 & \cellcolor[rgb]{0.91398,0.91398,1}374 \\ 
  (2-block) & 0.4 & \cellcolor[rgb]{0.86476,0.86476,1}588 & \cellcolor[rgb]{0.93629,0.93629,1}277 & \cellcolor[rgb]{0.9586,0.9586,1}180 & \cellcolor[rgb]{0.97355,0.97355,1}115 & \cellcolor[rgb]{0.97102,0.97102,1}126 & \cellcolor[rgb]{0.9632,0.9632,1}160 & \cellcolor[rgb]{0.95055,0.95055,1}215 & \cellcolor[rgb]{0.92364,0.92364,1}332 & \cellcolor[rgb]{0.84429,0.84429,1}677 \\ 
   & 0.5 & \cellcolor[rgb]{0.91214,0.91214,1}382 & \cellcolor[rgb]{0.95791,0.95791,1}183 & \cellcolor[rgb]{0.97171,0.97171,1}123 & \cellcolor[rgb]{0.97999,0.97999,1}87 & \cellcolor[rgb]{0.95607,0.95607,1}191 & \cellcolor[rgb]{0.94342,0.94342,1}246 & \cellcolor[rgb]{0.92364,0.92364,1}332 & \cellcolor[rgb]{0.88362,0.88362,1}506 & \cellcolor[rgb]{0.77,0.77,1}1031 \\ 
   \hline
\hline
 & 0.1 & \cellcolor[rgb]{0.99,1,0.99}2 & \cellcolor[rgb]{1,1,1}0 & \cellcolor[rgb]{0.99,1,0.99}2 & \cellcolor[rgb]{0.985,1,0.985}3 & \cellcolor[rgb]{0.965,1,0.965}7 & \cellcolor[rgb]{0.955,1,0.955}9 & \cellcolor[rgb]{0.935,1,0.935}13 & \cellcolor[rgb]{0.895,1,0.895}21 & \cellcolor[rgb]{0.775,1,0.775}45 \\ 
  {\bf GIS-C} vs. & 0.2 & \cellcolor[rgb]{0.99,1,0.99}2 & \cellcolor[rgb]{0.98,1,0.98}4 & \cellcolor[rgb]{0.975,1,0.975}5 & \cellcolor[rgb]{0.93,1,0.93}14 & \cellcolor[rgb]{0.85,1,0.85}30 & \cellcolor[rgb]{0.8,1,0.8}40 & \cellcolor[rgb]{0.725,1,0.725}55 & \cellcolor[rgb]{0.575,1,0.575}85 & \cellcolor[rgb]{0.615,1,0.615}77 \\ 
  {\bf better} & 0.3 & \cellcolor[rgb]{0.975,1,0.975}5 & \cellcolor[rgb]{0.975,1,0.975}5 & \cellcolor[rgb]{0.93,1,0.93}14 & \cellcolor[rgb]{0.845,1,0.845}31 & \cellcolor[rgb]{0.655,1,0.655}69 & \cellcolor[rgb]{0.55,1,0.55}90 & \cellcolor[rgb]{0.5,1,0.5}126 & \cellcolor[rgb]{0.515,1,0.515}97 & \cellcolor[rgb]{0.82,1,0.82}36 \\ 
  (3-block) & 0.4 & \cellcolor[rgb]{0.955,1,0.955}9 & \cellcolor[rgb]{0.97,1,0.97}6 & \cellcolor[rgb]{0.87,1,0.87}26 & \cellcolor[rgb]{0.72,1,0.72}56 & \cellcolor[rgb]{0.5,1,0.5}127 & \cellcolor[rgb]{0.5,1,0.5}160 & \cellcolor[rgb]{0.5,1,0.5}103 & \cellcolor[rgb]{0.74,1,0.74}52 & \cellcolor[rgb]{0.885,1,0.885}23 \\ 
   & 0.5 & \cellcolor[rgb]{0.94,1,0.94}12 & \cellcolor[rgb]{0.935,1,0.935}13 & \cellcolor[rgb]{0.795,1,0.795}41 & \cellcolor[rgb]{0.56,1,0.56}88 & \cellcolor[rgb]{0.5,1,0.5}150 & \cellcolor[rgb]{0.5,1,0.5}102 & \cellcolor[rgb]{0.695,1,0.695}61 & \cellcolor[rgb]{0.82,1,0.82}36 & \cellcolor[rgb]{0.92,1,0.92}16 \\ 
   \hline
 & 0.1 & \cellcolor[rgb]{0.77,0.77,1}9449 & \cellcolor[rgb]{0.77,0.77,1}7825 & \cellcolor[rgb]{0.77,0.77,1}3575 & \cellcolor[rgb]{0.79093,0.79093,1}909 & \cellcolor[rgb]{0.83279,0.83279,1}727 & \cellcolor[rgb]{0.83578,0.83578,1}714 & \cellcolor[rgb]{0.84751,0.84751,1}663 & \cellcolor[rgb]{0.87304,0.87304,1}552 & \cellcolor[rgb]{0.94296,0.94296,1}248 \\ 
  {\bf GIS-C} vs. & 0.2 & \cellcolor[rgb]{0.77,0.77,1}3086 & \cellcolor[rgb]{0.77,0.77,1}1810 & \cellcolor[rgb]{0.91283,0.91283,1}379 & \cellcolor[rgb]{0.85234,0.85234,1}642 & \cellcolor[rgb]{0.87603,0.87603,1}539 & \cellcolor[rgb]{0.89581,0.89581,1}453 & \cellcolor[rgb]{0.92341,0.92341,1}333 & \cellcolor[rgb]{0.95262,0.95262,1}206 & \cellcolor[rgb]{0.9586,0.9586,1}180 \\ 
  {\bf worse} & 0.3 & \cellcolor[rgb]{0.77,0.77,1}1075 & \cellcolor[rgb]{0.89788,0.89788,1}444 & \cellcolor[rgb]{0.89351,0.89351,1}463 & \cellcolor[rgb]{0.86936,0.86936,1}568 & \cellcolor[rgb]{0.92065,0.92065,1}345 & \cellcolor[rgb]{0.94112,0.94112,1}256 & \cellcolor[rgb]{0.95837,0.95837,1}181 & \cellcolor[rgb]{0.95607,0.95607,1}191 & \cellcolor[rgb]{0.91007,0.91007,1}391 \\ 
  (3-block) & 0.4 & \cellcolor[rgb]{0.86338,0.86338,1}594 & \cellcolor[rgb]{0.94894,0.94894,1}222 & \cellcolor[rgb]{0.87534,0.87534,1}542 & \cellcolor[rgb]{0.89213,0.89213,1}469 & \cellcolor[rgb]{0.94848,0.94848,1}224 & \cellcolor[rgb]{0.96274,0.96274,1}162 & \cellcolor[rgb]{0.94986,0.94986,1}218 & \cellcolor[rgb]{0.92226,0.92226,1}338 & \cellcolor[rgb]{0.8344,0.8344,1}720 \\ 
   & 0.5 & \cellcolor[rgb]{0.91398,0.91398,1}374 & \cellcolor[rgb]{0.94296,0.94296,1}248 & \cellcolor[rgb]{0.87396,0.87396,1}548 & \cellcolor[rgb]{0.91605,0.91605,1}365 & \cellcolor[rgb]{0.95584,0.95584,1}192 & \cellcolor[rgb]{0.94296,0.94296,1}248 & \cellcolor[rgb]{0.92157,0.92157,1}341 & \cellcolor[rgb]{0.88201,0.88201,1}513 & \cellcolor[rgb]{0.77,0.77,1}1103 \\ 
   \hline
   \hline
\end{tabular}
\end{table}


\section{Application to Exchange Rate Data}
\label{sec:app}

We apply our estimation methodology to daily Euro exchange rates. The data stems from the European Central Bank's Statistical Data Warehouse and comprises 3140 observations of 23 currencies ranging from January 3, 2000 to April 4, 2012. In choosing the prior for $\phi$ we follow \citet{kim-etal:sto}, i.e.~$(\phi+1)/2 \sim \Betadis{20, 1.5}$, and for the other parameters we pick rather vague priors: $\mu \sim \Normal{-10, 100}$ and $\sigma^2 \sim
\Gammad{\frac{1}{2},\frac{1}{2}}$. After a burn-in of $10\ 000$, we use $1\ 000\ 000$ draws from the respective distributions in each parameterization for posterior inference.

\begin{figure}[h!t]
\centering
  \includegraphics[width=\textwidth]{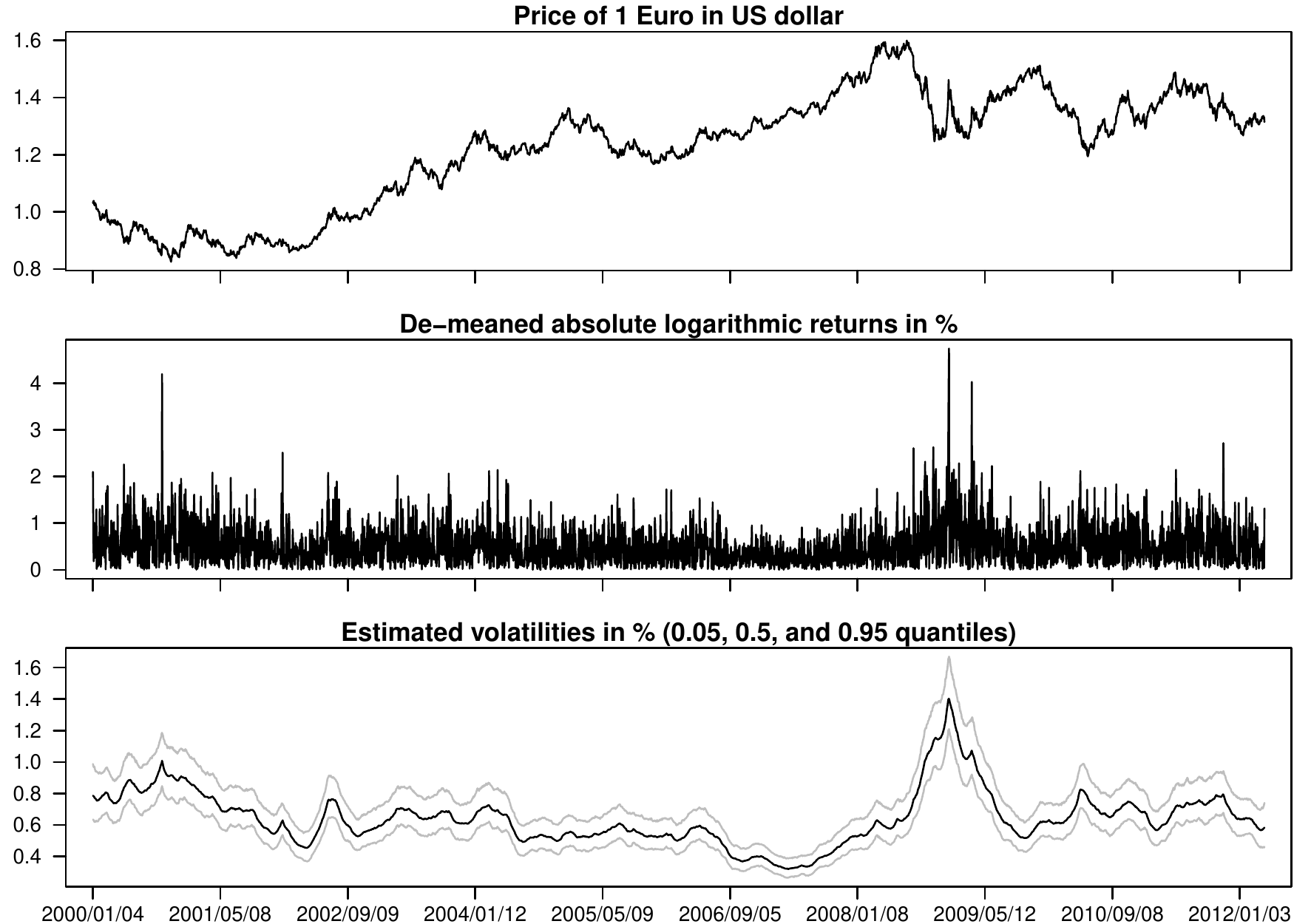}
 \caption{Indirectly quoted EUR/US\$ exchange rates (top) with de-meaned absolute log-returns (middle) and estimated instantaneous volatilities (bottom) based on $1000\,000$ draws via GIS-C.}
 \label{us}
\end{figure}

\begin{figure}[h!t]
\centering
  \includegraphics[width=\textwidth]{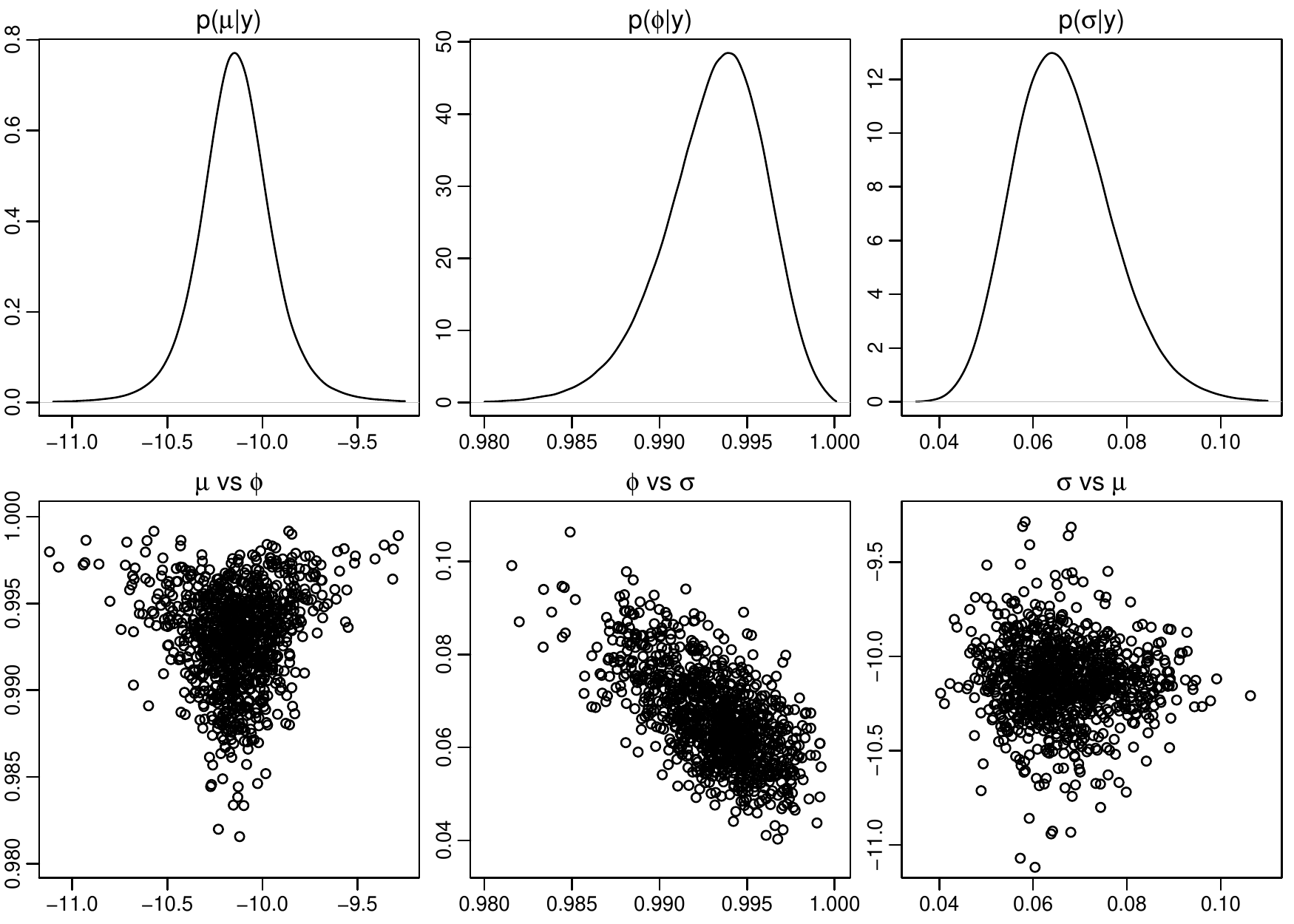}
 \caption{Marginal posterior density estimates and bivariate scatterplots of $1000$ thinned draws for the EUR/US\$ exchange rate data, based on $1000\,000$ samples obtained via GIS-C.}
 \label{usdens}
\end{figure}

To exemplify, Figure~\ref{us} shows exchange rates of EUR/US\$ along with absolute de-meaned log-returns, which are then used to estimate the time-varying volatilities displayed below. The transformed latent process and the absolute log-returns exhibit a similar overall pattern. Nevertheless, the volatility path is much smoother, which is due to the highly persistent autoregressive process (the posterior mean of $\phi$ is $0.993$, the posterior mean of $\sigma$ is $0.07$). Marginal posterior density estimates and two-way scatterplots can be found in Figure~\ref{usdens}. Note that only $p(\mu|\mathbf{y})$ is symmetric, while both $p(\phi|\mathbf{y})$ and $p(\sigma|\mathbf{y})$ are skewed. Moreover, the parameter draws are (sometimes nonlinearly) correlated.

Results for all 23 examined exchange rates are displayed in Table~\ref{exratesummary}. It stands out that for currencies which are closely tied to the Euro, posterior parameter means differ substantially to those found above. Most notably, the Danish krone exhibits very low overall level of volatility ($\mu_\text{mean} = -18$), paired with moderate persistence ($\phi_\text{mean}=0.916$) and moderately high volatility of volatility ($\sigma_\text{mean}=0.38$). Looking at the inefficiency factors for the raw parameterizations, one observes striking superiority of C in terms of sampling efficiency of $\mu$, while NC usually performs better in terms of sampling efficiency of $\sigma$. Again, interweaving overcomes these problems by showing lowest IFs uniformly for all parameters and all time series. Even though not reported here in detail due to space constraints, the choice of the baseline (GIS-C vs. GIS-NC) is negligible.

\begin{table}[ht]
{\small
\centering
\begin{tabular}{r@{\hspace{.4cm}}rrr@{\hspace{.7cm}}rrr@{\hspace{.75cm}}rrr@{\hspace{.7cm}}rrr}
\hline
 
&\multicolumn{3}{c}{Posterior means}&\multicolumn{3}{c}{$\text{IF}_\text{C}$}&\multicolumn{3}{c}{$\text{IF}_\text{NC}$}&\multicolumn{3}{c}{$\text{IF}_\text{GIS-C}$} \\
& $\mu$ & $\phi$ & $\sigma$ & $\mu$ & $\phi$ & $\sigma$& $\mu$ & $\phi$ & $\sigma$& $\mu$ & $\phi$ & $\sigma$ \\

   \hline
\hline
Australian dollar & -10.3 & 0.976 & 0.17 & 3 & 167 & 256 & 216 & 120 & 149 & 2 & 68 & 97 \\ 
  Canadian dollar & -10.1 & 0.987 & 0.09 & 3 & 350 & 539 & 211 & 128 & 159 & 3 & 89 & 120 \\ 
  Swiss franc & -12.0 & 0.985 & 0.21 & 3 & 55 & 129 & 807 & 99 & 162 & 3 & 33 & 73 \\ 
  Czech koruna & -11.5 & 0.953 & 0.28 & 3 & 140 & 200 & 151 & 121 & 154 & 3 & 72 & 96 \\ 
  Danish krone & -18.0 & 0.916 & 0.38 & 5 & 102 & 145 & 85 & 90 & 115 & 4 & 57 & 72 \\ 
   \hline
UK pound sterling & -10.8 & 0.992 & 0.10 & 2 & 95 & 233 & 802 & 120 & 150 & 2 & 39 & 87 \\ 
  Hong Kong dollar & -10.2 & 0.993 & 0.07 & 2 & 128 & 309 & 526 & 89 & 96 & 2 & 36 & 75 \\ 
  Indonesian rupiah & -9.9 & 0.966 & 0.23 & 4 & 234 & 313 & 216 & 201 & 242 & 3 & 114 & 142 \\ 
  Japanese yen & -10.0 & 0.989 & 0.12 & 3 & 110 & 227 & 513 & 98 & 140 & 3 & 47 & 91 \\ 
  Korean won & -10.0 & 0.987 & 0.14 & 2 & 92 & 192 & 501 & 96 & 134 & 2 & 40 & 79 \\ 
   \hline
Mexican peso & -9.8 & 0.977 & 0.16 & 2 & 137 & 221 & 220 & 101 & 128 & 2 & 60 & 86 \\ 
  Malaysian ringgit & -10.3 & 0.990 & 0.08 & 2 & 220 & 390 & 404 & 104 & 124 & 2 & 59 & 90 \\ 
  Norwegian krone & -11.1 & 0.970 & 0.18 & 2 & 139 & 217 & 150 & 83 & 107 & 2 & 53 & 76 \\ 
  New Zealand dollar & -10.0 & 0.963 & 0.17 & 4 & 344 & 432 & 105 & 153 & 177 & 3 & 114 & 135 \\ 
  Philippine peso & -10.1 & 0.981 & 0.12 & 3 & 376 & 529 & 209 & 189 & 220 & 2 & 123 & 160 \\ 
   \hline
Polish zloty & -10.4 & 0.975 & 0.19 & 2 & 96 & 171 & 261 & 85 & 117 & 2 & 43 & 69 \\ 
  Romanian leu & -11.1 & 0.970 & 0.31 & 2 & 51 & 100 & 446 & 87 & 137 & 2 & 32 & 60 \\ 
  Russian rouble & -10.6 & 0.988 & 0.15 & 4 & 75 & 172 & 891 & 121 & 151 & 3 & 38 & 82 \\ 
  Swedish krona & -11.3 & 0.991 & 0.11 & 1 & 52 & 156 & 752 & 75 & 99 & 1 & 23 & 60 \\ 
  Singapore dollar & -10.6 & 0.995 & 0.07 & 4 & 138 & 348 & 998 & 126 & 132 & 4 & 47 & 100 \\ 
   \hline
Thai bhat & -10.2 & 0.980 & 0.13 & 3 & 202 & 314 & 207 & 102 & 125 & 3 & 64 & 90 \\ 
  Turkish lira & -9.8 & 0.966 & 0.27 & 2 & 72 & 127 & 259 & 84 & 127 & 2 & 42 & 69 \\ 
  US dollar & -10.1 & 0.993 & 0.07 & 2 & 126 & 308 & 504 & 87 & 99 & 2 & 37 & 74 \\ 
   \hline
   \hline
\end{tabular}
\caption{Posterior means and inefficiency factors for various estimation methods of the SV model, applied to EUR exchange rate data.}
\label{exratesummary}
}
\end{table}

\section{Concluding Remarks}
\label{sec:con}

Previous studies have shown that simple reparameterizations often turn out to have substantial impact on MCMC simulation efficiency in state-space specifications. This paper contributes to the literature by exploring the influence of choosing between two selected parameterizations for Bayesian estimation of SV models. Moreover, it provides evidence that inefficiency factors obtained from simulation experiments can heavily depend on the realization of the data generating process. Through the findings of this paper it becomes clear that employing an ancillarity-sufficiency interweaving strategy (ASIS) introduced by \cite{yu-men:cen} helps to overcome shortcomings of either the centered or the non-centered parameterization by outperforming those in terms of sampling efficiency with respect to all parameters at very little extra computational cost, whereas the baseline of the interweaving strategy is of minor influence.

The concept of interweaving different parameterizations of state-space models is clearly very general, and there is good reason to hope for similar magic when applying ASIS to extension of the basic SV model such as more general innovation distributions \citep[e.g.][]{lie-jun:sto, del-gri:bay}, asymmetry \citep[e.g.][]{yu:on, omo-etal:sto} or both \citep[e.g.][]{chi-etal:mar, wan-etal:sto, tsi-on, ish-omo:eff, nak-omo:sto}. Preliminary results for an SV model with leverage, where a centered parameterization from \cite{yu:on} is compared with a non-centered version based on transforming $\hsv_t$ into $\tilde \hsv_t = (\hsv_t-\mu)/\sigma$ as in the present paper, show that this hope is in fact an actual possibility. A thorough investigation of this issue is however beyond the scope of this article.

\section{Acknowledgments}
\label{sec:ack}
The authors would like to thank the editor and two referees for their perspicacious comments on an earlier draft of this paper, and Stefan Theu\ss l for helpful advice concerning coding and implementation.

\bibliographystyle{biometrika}
\bibliography{bibfile}

\end{document}